\documentclass[a4paper,12pt]{article}
\usepackage{amsmath}
\usepackage{graphicx}
\usepackage{bm}
\usepackage{bbold}
\usepackage{xcolor}
\usepackage{authblk}

\usepackage[colorlinks=true,linkcolor=black,citecolor=magenta]{hyperref}

\allowdisplaybreaks  
\newcommand\TT{\rule{0pt}{2.5ex}}        
\newcommand\BB{\rule[-1.0ex]{0pt}{0pt}}  
\newcommand{\be}{\begin{equation}}
\newcommand{\en}{\end{equation}}
\newcommand{\lbl}[1]{\label{eq:#1}}

\newcommand{\rf}[1]{(\ref{eq:#1})}
\newcommand{\fig}[1]{\ref{fig:#1}}
\newcommand{\sect}[1]{\ref{sec:#1}}
\newcommand{\braque}[1]{{\langle #1 \rangle}}
\newcommand{\ket}[1]{{\vert #1 \rangle}}
\newcommand{\bt}{\begin{tabular}}
\newcommand{\et}{\end{tabular}}
\newcommand{\ba}{\begin{array}}
\newcommand{\ea}{\end{array}}
\newcommand{\bp}{\begin{pmatrix}}
\newcommand{\ep}{\end{pmatrix}}
\newcommand{\disc}{\hbox{disc}}
\newcommand{\lapprox}{\mathrel{%
\setbox0=\hbox{$<$}\raise0.6ex\copy0\kern-\wd0\lower0.65ex\hbox{$\sim$}}}

\newcommand{\mpid}{m_\pi^2}
\newcommand{\mpi}{m_\pi}
\newcommand{\mkd}{m_K^2}
\newcommand{\mk}{m_K}
\newcommand{\Kbar}{\bar{K}}

\newcommand{\braques}[1]{{\langle #1 \rangle}_s}
\newcommand{\braquet}[1]{{\langle #1 \rangle}_t} 
\newcommand{\Dp}{{D^+}}
\newcommand{\Dz}{{D^0}} 
\newcommand{\Kp}{{K^+}} 
\newcommand{\Km}{{K^-}}
\newcommand{\Kz}{{K^0}} 
\newcommand{\Kzb}{{\bar{K}^0}}
\newcommand{\piz}{{\pi^0}} 
\newcommand{\pip}{{\pi^+}}
\newcommand{\pim}{{\pi^-}}
\newcommand{\skpi}{{(m_K+m_\pi)^2}}
\newcommand{\spipi}{{4m_\pi^2}}
\newcommand{\mdd}{{m^2_D}}
\newcommand{\md}{{m_D}}

\newcommand{\fpi}{{F_\pi}}
\newcommand{\undemi}{{1/2}}
\newcommand{\trdemi}{{3/2}}

\newcommand{\unun}{{{1\over2}{1\over2}}}
\newcommand{\unmun}{{{1\over2}{-1\over2}}}
\newcommand{\trtr}{{{3\over2}{3\over2}}}
\newcommand{\untr}{{{1\over2}{3\over2}}}
\newcommand{\trun}{{{3\over2}{1\over2}}}
\newcommand{\CF}{{\cal F}}
\newcommand{\CG}{{\cal G}}
\newcommand{\CH}{{\cal H}}
\newcommand{\CA}{{\cal A}}
 
\newcommand{\zs}{Z_s}
\newcommand{\zu}{Z_u}
\newcommand{\umin}{{u_-(s)}}
\renewcommand{\uplus}{{u_+(s)}}
\newcommand{\tmin}{{t_-(s)}}
\newcommand{\tplus}{{t_+(s)}}
\newcommand{\smin}{{s_-(t)}}
\newcommand{\splus}{{s_+(t)}}
\newcommand{\sDmin}{{s_D^-}}

\newcommand{\Zu}{Z_u}
\newcommand{\su}{Z_t}
\newcommand{\Zs}{Z_s}

\makeatletter
\@addtoreset{equation}{section}
\makeatother

\DeclareMathSymbol{\widetildesym}{\mathord}{largesymbols}{"65}
\DeclareMathSymbol{\widehatsym}{\mathord}{largesymbols}{"62}
\newcommand{\TG}{%
\mathrel{%
\setbox0=\hbox{$G_1^1$}%
\setbox1=\hbox{$\widetildesym$}%
\copy0\kern-0.9\wd0\raise0.28\ht0\copy1\copy2\kern0.5\wd1}}

\newcommand{\TTG}{%
\mathrel{%
\setbox0=\hbox{$G_1^1$}%
\setbox1=\hbox{$\widetildesym$}%
\setbox2=\hbox{$\widehatsym$}%
\copy0\kern-0.9\wd0\raise0.25\ht0\copy1\kern-\wd1\raise0.5\ht0\copy2\kern\wd1}}

\title{Isospin symmetry and analyticity in $D\to\bar{K}\pi\pi$ decays}
\author[a]{Emi Kou}

\author[b]{Tetiana Moskalets}

\author[a]{Bachir Moussallam}

\affil[a]{\small Universit\'e
  Paris-Saclay, Laboratoire de physique des 2
  infinis Ir\`ene Joliot-Curie (CNRS/IN2P3,UMR9012), 91405 Orsay, France.}  

\affil[b]{\small Physikalisches Institut,
  Albert-Ludwigs-Universit\"{a}t Freiburg, Freiburg, Germany.} 

\begin{document}

\maketitle

\begin{abstract}
We perform a detailed study of the consequences of isospin symmetry in the
Cabibbo favoured $D\to\Kbar\pi\pi$ decays. These processes are important for
precision testing of the Standard Model and for hadronic physics. Combining
isospin symmetry with a dispersive reconstruction theorem we derive a
representation in terms of one-variable functions which allows one to predict 
all the $D\to\Kbar\pi\pi$ amplitudes given inputs from one $D^+$ mode
and one $D^0$ mode. From this, using dispersion relations and unitarity, we
derive a set of 6+6 Khuri-Treiman type integral equations which enable to take
three-body rescattering effects into account. A first test of this approach is
presented using experimental results on the $D^+\to K_S\piz\pip$ and the
$D^0\to K_S\pim\pip$ modes.

\end{abstract}

\tableofcontents

\reversemarginpar

\section{Introduction}
Precision measurements of non-leptonic weak decays of the $K$, $D$ and $B$
mesons provide crucial information on the origin of the violation of CP
symmetry. In the last two decades, studies of CP violation in B decays by the
Babar, Belle and LHCb collaborations have led to a considerable improvement in
the precision of the determination of the three CKM angles
$\alpha,\beta,\gamma$ (also called $\phi_1,\phi_2,\phi_3$). This effort is due
to be pursued at Belle II~\cite{Belle-II:2018jsg}.

Non-leptonic decays of flavoured mesons probe dimension six operators
as well as the dynamics of strong final-state rescattering. The $D$
mesons differ from the $K$ and the $B$ ones in that, for the $c$ quark,
neither the light-quark nor the heavy-quark expansions can be
justified. This difference could manifest itself in three-body
decays. Indeed, final-state rescattering which involve the three
particles are suppressed in $K$ decays (they vanish at LO and NLO in
the chiral expansion\footnote{The $K\to3\pi$ amplitude at NLO has
been computed in refs.~\cite{Kambor:1991ah,Bijnens:2002vr}, the two-loop
rescattering contributions have been evaluated in
ref.\cite{Kampf:2019bkf}.}) and are generally also
  suppressed in $B$ decays within the QCD factorisation
framework (see\cite{Krankl:2015fha} and references
  therein).
$D$ decays, in contrast, could well be sensitive to three-body rescattering
effects and it is still an open question whether such effects can be clearly
identified in the data.

In this paper we reconsider the theoretical description of the family of
three-body decays: $D^+,D^0\to \Kbar\pi\pi$. One motivation is related to the
method proposed in refs.~\cite{Bondar:2002xx,Giri:2003ty,Bondar:2005ki}
(BPGGSZ) for measuring the angle $\gamma$ (or $\phi_3$), which is the most
efficient one for that purpose. It uses the chain decay $B^\pm \to K^\pm D,
D\to K_S\pim\pip$ and requires knowledge of the behaviour of both the strong
phase and the modulus of the $D^0\to K_S\pim\pip$ amplitude\footnote{The same
inputs can also be used in measurements of the $D^0-\bar{D}^0$ mixing
parameters~\cite{CLEO:2005qym}.}. While averages of 
these quantities can actually be measured (see below) a reliable modelling
would further improve the accuracy of this approach. Another, different,
motivation is provided by the experimental observation, some time ago, of
several interesting features in the amplitude $D^+\to\Km\pip\pip$.
Based on an isobar model description of the Dalitz plot, a first clear
observation of the broad scalar $\kappa$ resonance was reported by the E791
collaboration~\cite{E791:2002xlc}. Moreover, this amplitude shows evidence of
a significant dynamical role of {\it repulsive} interactions, as between a
$\pi{K}$ pair with $I=3/2$~\cite{FOCUS:2007mcb} (see also~\cite{Edera:2005dk})
or a pion pair with isospin $I=2$~\cite{Bonvicini:2008jw}, which are 
often ignored. Several theoretical models have been proposed for
describing the $D^+\to \Km\pip\pip$
amplitude~\cite{Diakonou:1989sf,Oller:2004xm,Boito:2008gy,Boito:2009qd,Magalhaes:2011sh,Guimaraes:2014kor,Nakamura:2015qga,Niecknig:2015ija}.

We will combine two theoretical tools: isospin symmetry and dispersion
relations. While not exact, of course, isospin symmetry is still a very
precise dynamical constraint since isospin breaking effects are of the order
$\sim0.5-1$\%. Isospin decompositions of the four independent $K\to3\pi$
amplitudes have been implemented a long time
ago~\cite{Zemach:1963bc,Devlin:1978ye}. Analogous application to the $D\to
\Kbar\pi\pi$ amplitudes could be performed as well, and is simplified by the
fact that the relevant weak Hamiltonian, in that case, is a pure $I=1$
operator.
This was exploited previously in the work of Niecknig and
Kubis~\cite{Niecknig:2015ija,Niecknig:2017ylb} who performed a
combined study of the two modes $D^+\to
\Km\pip\pip,\Kzb\pip\piz$. They have also derived for the $S$- and the
$P$-wave amplitudes a set of Khuri-Treiman (KT) dispersive
equations~\cite{Khuri:1960zz,Sawyer:1960hrc}. This formalism allows one to take
three-body rescattering effects into account.  The mode $D^+\to
\Km\pip\pip$ has been studied in the analogous framework of
Faddeev-type equations in
refs.~\cite{Guimaraes:2014kor,Nakamura:2015qga}.

We extend here the application of isospin symmetry and of the KT equations to
{\it all} the three-body decay modes of the form $D\to\Kbar\pi\pi$. That is,
in addition to the two modes of the $D^+$ we will consider the three decay
modes of the $D^0$: $D^0\to \Km\pip\piz, \Kzb\pim\pip, \Kzb\piz\piz$. The
isospin relations which we provide are applicable, as well, to simpler models
like the isobar model. The possibility of performing combined fits of several
different amplitudes puts constraints on the theoretical models and
could also lead
to a better determination of the small double-Cabibbo suppressed amplitudes
which are present when a $K_S$ (or a $K_L$) is detected in the final
state. This would be useful in the search for a CP violating phase in this
sector as proposed in ref.~\cite{Bigi:1994aw}.

We perform a first test of our model using two sets of
experimental data which we could access. The first set concerns 
measurements of the $D^+\to K_S\piz\pip$ mode performed by the BESIII
collaboration~\cite{Ablikim:2014cea}. Their results on the distribution of
events are provided by the collaboration over a dense binning of the Dalitz plot
(with 1342 bins). The second set of data refer to measurements by
CLEO~\cite{Libby:2010nu} and by BESIII~\cite{BESIII:2020khq} of the $D^0\to
K_S\pim\pip$ mode. The data, in this case, can be retrieved from tables in the
publications. They are given on a limited number of bins (8 bins) but contain
information both on the modulus squared of the amplitude and on its phase
$\delta_D$ through averages $\bar{c}_i$, $\bar{s}_i$ of the cosine and sine of
the phase differences
\be\lbl{Deltadeltadef} \Delta\delta_D(s,u)\equiv
\delta_D(s,t,u)-\delta_D(u,t,s)
\en
($s,t,u$ being Mandelstam variables) performed over bin $i$. Remarkably,
measuring such phases is possible in $D^0$ decays by making use of correlated
$D$ pairs, e.g from $\psi(3770)\to D^0\bar{D^0}$, and tagging one of the $D's$
in a CP even/odd state~\cite{Giri:2003ty,Poluektov:2006ia}.

The plan of the paper is as follows. After introducing some notation
and reviewing the structure of the relevant weak Hamiltonian we
implement the Wigner-Eckart theorem on a number of $2\to2$ matrix
elements.  A rather predictive structure emerges when this is combined
with the dispersive ``reconstruction
theorem''~\cite{Khuri:1960zz,Stern:1993rg,Kambor:1995yc} which we
first limit to $j=0,1$ partial-waves. The two $D^+$ decay amplitudes
get expressed in terms of six functions of one variable (the
$F$-functions), as was found
in~\cite{Niecknig:2015ija,Niecknig:2017ylb}. The new results concern
the $D^0$ amplitudes for which we provide expressions involving the
same $F$-functions plus an additional set of six functions (the
$H$-functions).

Using these results two sets of KT integral equations are derived, from which
the $F$- and the $H$-functions can be computed numerically, depending in a
linear way respectively on 6 and on 6+7 complex polynomial parameters. These
form the core of our model for describing the amplitudes in the whole Dalitz
plot region.
The determination of the parameters based on the input experimental data
mentioned above is then discussed. In addition, we use qualitative information
which derives from four independent soft pion theorems and we consider the
decay widths of the five $D\to \Kbar\pi\pi$ modes. These widths are very
different from each other and provide good probes of the interference patterns
between the isospin $F$- and $H$-functions. We finally comment on the
difference between the $F$ and the $H$ I=1/2 $S$-waves near the $\pi{K}$
threshold.

\section{Isospin decomposition of  the $D\to\bar{K}\pi\pi$ amplitudes}
\begin{figure}
\centering
\includegraphics[width=0.5\linewidth]{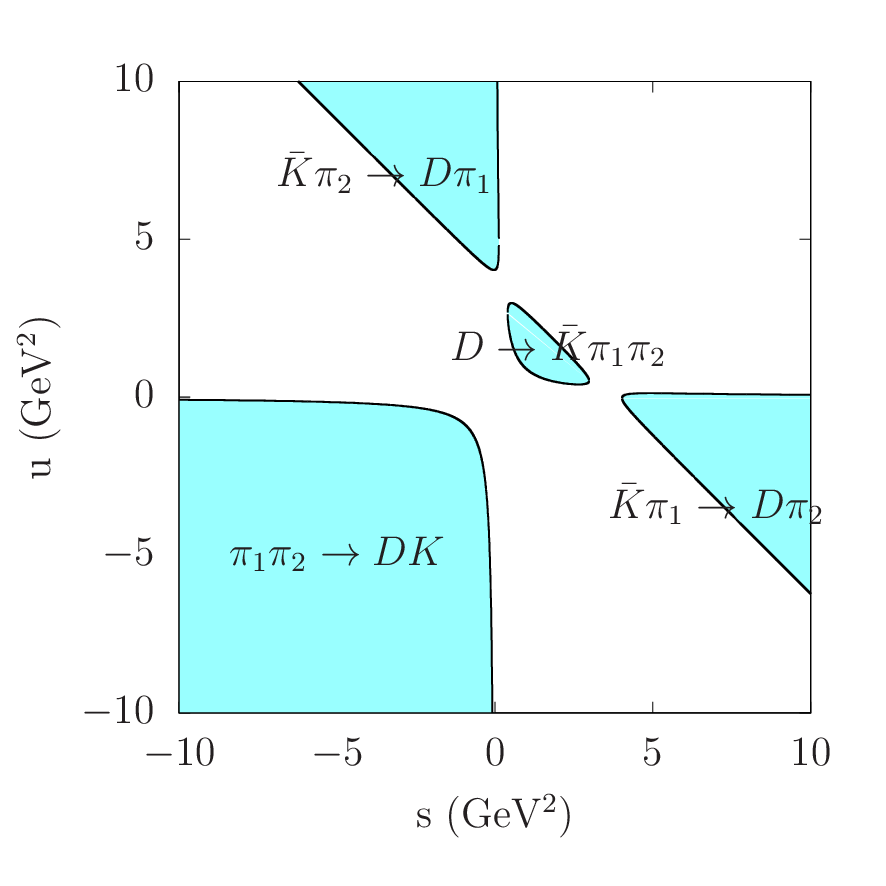}
\caption{\small  Physical regions for decay
  and for scattering as a function of the variables $u$, $s$.}
\label{fig:mandelstam}
\end{figure}
\subsection{Kinematics}
We consider amplitudes $D(p_D)\to \bar{K}(p_K)
\pi_1(p_{\pi_1})\pi_2(p_{\pi_2})$.
The Mandelstam variables $s$,  $u$, $t$ are defined as follows
\be
s=(p_K + p_{\pi_1})^2,\quad
u=(p_K + p_{\pi_2})^2,\quad
t=(p_{\pi_1}+ p_{\pi_2})^2\ .
\en
and satisfy
\be\lbl{s+t+u}
s+t+u=\Sigma,\ \Sigma=m_D^2+m_K^2+2m_\pi^2\ .
\en
The decay amplitudes are called $\CA_{\bar{K}\pi_1\pi_2}(s,t,u)$ and the
differential decay width is given by
\be
\frac{d^2\Gamma_{\bar{K}\pi_1\pi_2}}{ds dt}= \frac{1}{256\pi^3 m_D^3}\vert
\CA_{\bar{K}\pi_1\pi_2}(s,t,u)\vert^2\ .
\en
A symmetry factor $S_{12}=1/2$ must be appended when integrating
over $t$ when the two pions are identical. By analytical continuation,
the amplitude functions $\CA_{\bar{K}\pi_1\pi_2}(s,t,u)$ also describe
scattering processes, the physical regions are shown in
fig.~\fig{mandelstam}. Next, we recall the definitions of the angular
variables $\theta_s$ and $\theta_t$ associated with the $\bar{K}\pi_1$
and $\pi_1\pi_2$ reference frames.

\begin{figure}
\centering
\includegraphics[width=0.40\linewidth]{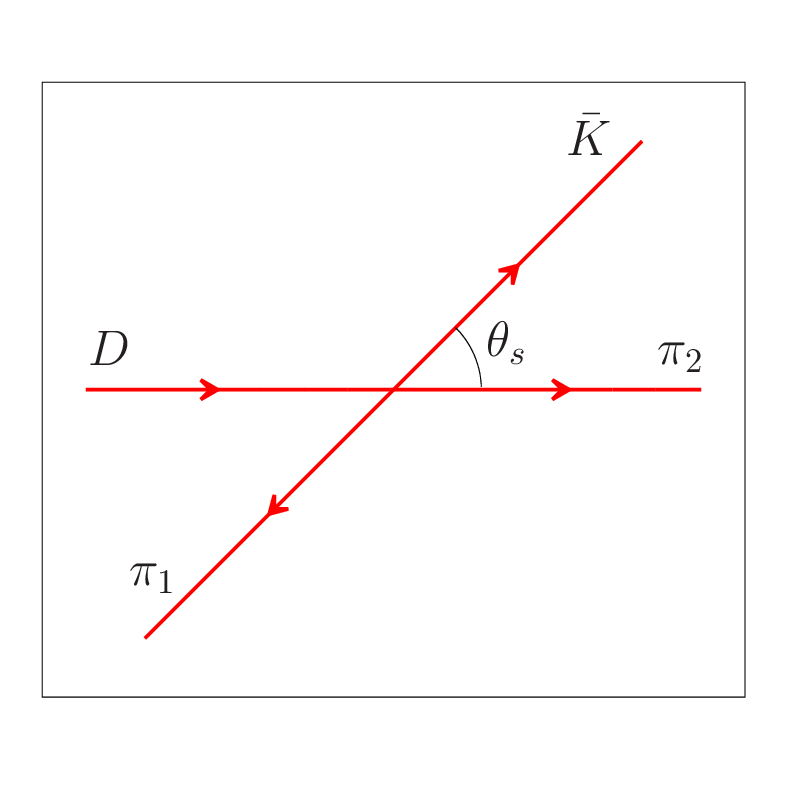}%
\includegraphics[width=0.40\linewidth]{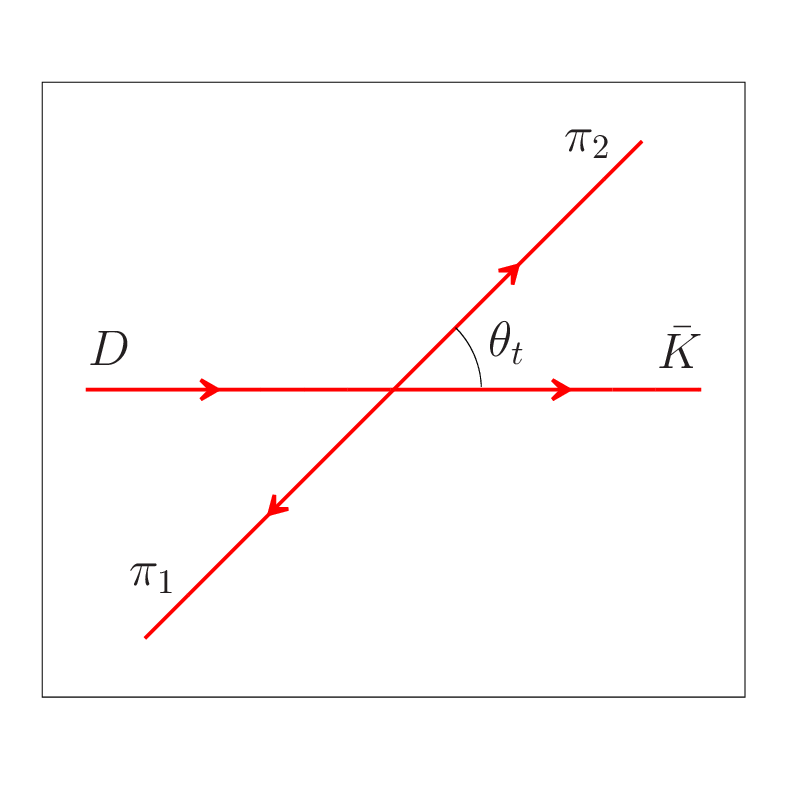}
\caption{\small Kinematics of $D\to\bar{K}\pi_1\pi_2$ in the $\bar{K}\pi_1$
(left) and the $\pi_1\pi_2$ centre-of-mass systems showing the
  definitions of the angles $\theta_s$ and $\theta_t$. } 
\label{fig:kinematics}
\end{figure}
\noindent{\bf 1) $\bar{K}\pi_1$ CMS} \\
We first consider the $\bar{K}\pi_1$ centre-of-mass system
(fig.~\fig{kinematics}, left). The  4-momenta in this frame read
\be
p_D=\dfrac{1}{2\sqrt{s}}
\begin{pmatrix}
{s+\Delta_{D\pi}}\\[0.2cm]
{\lambda_{D\pi}^\undemi(s)}\,\hat{u}\\
\end{pmatrix},\quad
p_{\pi_2}=\dfrac{1}{2\sqrt{s}}
\begin{pmatrix}
{\Delta_{D\pi}-s}\\[0.2cm]
{\lambda_{D\pi}^\undemi(s)}\,\hat{u}\\
\end{pmatrix}
\en
and
\be\lbl{DK-frame}
p_{K}=\dfrac{1}{2\sqrt{s}}
\begin{pmatrix}
{s+\Delta_{K\pi}}\\[0.2cm]
{\lambda_{K\pi}^\undemi(s)}\,\hat{v}\\
\end{pmatrix},\quad
p_{\pi_1}=\dfrac{1}{2\sqrt{s}}
\begin{pmatrix}
{s-\Delta_{K\pi}}\\[0.2cm]
-{\lambda_{K\pi}^\undemi(s)}\,\hat{v}\\
\end{pmatrix}
\en
with the definitions
\be
\Delta_{AB}=m^2_A-m^2_B,\quad
\lambda_{AB}(s)=(s-(m_A+m_B)^2)(s-(m_A-m_B)^2)\ .
\en
In eq.~\rf{DK-frame} $\hat{u}$ and $\hat{v}$ are unit 3-vectors and we define
the cosine of the scattering angle in this frame as
\be
 z_s\equiv \cos(\theta_s)=\hat{u}\cdot\hat{v}
=\dfrac{s(t-u)+\Delta_{D\pi}\Delta_{K\pi}}
{\sqrt{\lambda_{D\pi}(s)\lambda_{K\pi}(s)}}\ .
\en
From this, $t$ and $u$ can be written in terms of $s$ and $z_s$ as,
\be\lbl{tufunczs}
\ba{l}
t(s,z_s)=\dfrac{1}{2}\left(\Sigma-s-\dfrac{\Delta}{s} 
+{\kappa_s(s)} z_s  \right)\\
u(s,z_s)=\dfrac{1}{2}\left(\Sigma-s+\dfrac{\Delta}{s} 
-{\kappa_s(s)} z_s   \right)\\
\ea\en
with
\be\lbl{kappa_sdef}
\Delta=\Delta_{D\pi}\Delta_{K\pi}\ ,\quad
\kappa_s(s)=\dfrac{\sqrt{\lambda_{D\pi}(s)\lambda_{K\pi}(s)}}{s}\ .
\en

\medskip
\noindent{\bf 2) $\pi_1\pi_2$ CMS:} \\
In the $\pi_1\pi_2$ centre-of-mass system (fig.~\fig{kinematics}, right) the
momenta read
\be
p_D=\dfrac{1}{2\sqrt{t}}
\begin{pmatrix}
t+\Delta_{DK}\\[0.2cm]
{\lambda^\undemi_{DK}}(t)\hat{u}\\
\end{pmatrix},\quad
p_K=\dfrac{1}{2\sqrt{t}}
\begin{pmatrix}
\Delta_{DK}-t\\[0.2cm]
{\lambda^\undemi_{DK}}(t)\hat{u}\\
\end{pmatrix}
\en
and
\be
p_{\pi_1}=\dfrac{1}{2\sqrt{t}}
\begin{pmatrix}
t\\[0.2cm]
-{\lambda^\undemi_{\pi\pi}(t)} \hat{v}\\
\end{pmatrix},\quad
p_{\pi_2}=\dfrac{1}{2\sqrt{t}}
\begin{pmatrix}
t\\[0.2cm]
{\lambda^\undemi_{\pi\pi}(t)} \hat{v}
\end{pmatrix}\ .
\en
The cosine of the  scattering angle $z_t$ in this frame is then 
given by
\be
z_t(s,t,u)\equiv \cos(\theta_t)
= \frac{s-u}{\sqrt{\lambda_{D\Kz}(t)\,(1-4\mpid/t)}}\ 
\en
and one can  express $s$ and $u$ in terms of $t$ and $z_t$
\be\lbl{sufunczt}
\ba{l}
s(t,z_t)=\dfrac{1}{2}\left(\Sigma-t +\kappa_t(t)\, z_t\right)\\[0.2cm]
u(t,z_t)=\dfrac{1}{2}\left(\Sigma-t -\kappa_t(t)\, z_t\right)\\
\ea\en
with
\be\lbl{kappa_tdef}
\kappa_t(t)=\sqrt{\lambda_{DK}(t) (1-4\mpid/t)}\ .
\en

\subsection{Weak Hamiltonian}
In the Standard Model the $D\to \bar{K}\pi\pi$ decays are driven by
the following weak Hamiltonian 
\be
H_W^{(CF)}= \frac{G_F}{\sqrt2} \left[V^*_{cs}\,V_{ud} \left(
C_1 O_1 +C_2 O_2\right) + h.c. \right]
\en
with
\be\lbl{O1O2}
O_1=(\bar{s}_ic_j)_{V-A}\, (\bar{u}_jd_i)_{V-A}\ ,\quad
O_2=(\bar{s}_ic_i)_{V-A}\, (\bar{u}_jd_j)_{V-A}\
\en
(e.g. ref.~\cite{Buchalla:1995vs}). It is
Cabibbo favoured, i.e. $O(\lambda^0)$ in the Wolfenstein
parametrisation~\cite{Wolfenstein:1983yz},
with~\cite{ParticleDataGroup:2022pth}  
\be
\vert V^*_{cs} V_{ud}\vert= 0.969\pm0.016\ .
\en
An important property, as can be seen immediately from~\rf{O1O2} is that
the $H_W^{(CF)}$  transforms under isospin as an $I=1$ operator 
\be
H_W^{(CF)}\sim T_{II_z},\quad I=1, I_z=-1
\en
which will enable us to make use of the Wigner-Eckart theorem
(e.g.~\cite{Edmonds:1957}) 
\be\lbl{wignereckart}
\braque{I_1m_1\vert T_{I,I_z}\vert I_2m_2}= \braque{I I_z;I_2 m_2\vert I_1 m_1}
\,{\cal F}_I^{I_1I_2}
\en
when evaluating matrix elements.

In experimental situations where $K_S\pi_1\pi_2$ or $K_L\pi_1\pi_2$
are detected, contributions from $D\to K\pi_1\pi_2$ amplitudes must be
considered, which are driven by the Hamiltonian
\be
H_W^{(DCS)}= \frac{G_F}{\sqrt2}\left[ V^*_{us}\,V_{cd} \left(
C_1 \tilde{O}_1 +C_2 \tilde{O}_2\right)+ h.c. \right]
\en
where
\be
\tilde{O}_1=(\bar{d}_ic_j)_{V-A}\, (\bar{u}_js_i)_{V-A}\ ,\quad
\tilde{O}_2=(\bar{d}_ic_i)_{V-A}\, (\bar{u}_js_j)_{V-A}\ .
\en
$H_W^{(DCS)}$ is doubly Cabibbo suppressed, i.e. $O(\lambda^2)$, with
\be
\vert V^*_{us} V_{cd}\vert= 0.049\pm0.001\ .
\en
From the point of view of isospin, it has both an $I=0$ and a $I=1$
component
\be
H_W^{(DCS)}\sim \frac{1}{\sqrt2}T_{00} +  \frac{1}{\sqrt2}T_{10} \ .
\en
In that situation, isospin symmetry is less predictive than in the case of
the Cabibbo favoured operator. Performing an expansion of the DCS
amplitudes in terms of isospin one-variable functions analogous to the
ones presented below in eqs.~\rf{A1A2express} to~\rf{A7express} for the
CF amplitudes would involve 18 such functions instead of 12.

\subsection{Isospin amplitudes and one-variable functions}
It proves convenient to think in terms of scattering,
rather than decay, amplitudes i.e. to consider the processes
\be
\bar{K}\pi_1\to D \pi_2\ ,
\bar{K}\pi_2\to D \pi_1\ ,
\pi_1\pi_2\to D K \     
\en
which are associated with the matrix elements $\braque{D\pi_i\vert
  H_W\vert\bar{K}\pi_j}$ and $\braque{D K\vert H_W\vert \pi_1\pi_2}$ (dropping
the $CF$ index). The first ones will be expressed in terms of isospin
amplitudes labelled with a pair of half-integer isospins, referring to the
initial and final state, and the latter ones will be labelled with a pair of
integer isospins
\be\lbl{isoamplit}
\ba{ll}
\braque{D\pi_i\vert H_W\vert\bar{K}\pi_j} : &
\CF^{\trtr}(s,t,u),\ \CF^{\untr}(s,t,u),\ \CF^{\trun}(s,t,u),\ \CF^{\unun}(s,t,u)\\
\braque{D K\vert H_W\vert \pi_1\pi_2} : &
\CG^{10}(s,t,u),\ \CG^{12}(s,t,u),\ \CG^{01}(s,t,u),\  \CG^{11}(s,t,u)\ .
\ea\en
The explicit relations between the physical matrix elements and the
isospin amplitudes are derived in a straightforward way by applying
the Wigner-Eckart theorem~\rf{wignereckart}: they are given in
table~\ref{table:isoexpress} in appendix~\ref{sec:isodetails}. 
Using these, one can derive four relations among the five physical $D$ decay
amplitudes (e.g.~\cite{Matsuda:1977hf,Kaptanoglu:1978sr}). The first one
relates the two $D^+$ amplitudes
\be\lbl{generaliso1}
\CA_{\Km\pip\pip}(s,t,u)=-\sqrt2\big(
\CA_{\Kzb\piz\pip}(s,t,u)+ \CA_{\Kzb\piz\pip}(u,t,s) \big)
\en
and the next two relations involve $D^+$ and $D^0$ amplitudes and are
symmetric under $s$, $u$ interchange
\be\lbl{generaliso2}
\ba{l}\CA_{\Km\pip\pip}(s,t,u)= \sqrt2\big(
\CA_{\Km\piz\pip}(s,t,u)+ \CA_{\Km\piz\pip}(u,t,s) \big) \\[0.25cm]
\CA_{\Km\pip\pip}(s,t,u)= \CA_{\Kzb\pim\pip}(s,t,u)+\CA_{\Kzb\pim\pip}(u,t,s)
-2\CA_{\Kzb\piz\piz}(s,t,u)
\ea\en
while the last relation involves antisymmetrised amplitudes
\be\lbl{generaliso3}
\ba{l}
\CA_{\Kzb\piz\pip}(s,t,u)-\CA_{\Kzb\piz\pip}(u,t,s)=
\CA_{\Km\piz\pip}(s,t,u)-\CA_{\Km\piz\pip}(u,t,s)\\[0.20cm]
\phantom{\CA_{\Kzb\piz\pip}(s,t,u)-\CA_{\Kzb\piz\pip}(u,t,s)}
-\sqrt2\big(\CA_{\Kzb\pim\pip}(s,t,u)-\CA_{\Kzb\pim\pip}(u,t,s)\big)\ .
\ea\en

In order to go beyond these general results it is convenient, at
first, to make use of a redefined set of isospin amplitudes which
leads to simpler expressions for the $D^+$ amplitudes: this is
explained in appendix~\ref{sec:isodetails}. This set is labelled
with the isospins of the $\Kbar\pi$ or $\pi\pi$ systems
only as it mixes the isospins of the $D\pi$ or $DK$ states and we collect them
into vectors $\vec{\CF}(s,t,u)$ and $\vec{\CG}(s,t,u)$
\be\lbl{vecFvecG}
\vec{\CF}(s,t,u)\equiv\bp
\CF^\trdemi(s,t,u)\\
\CF^\undemi(s,t,u)\\
\CH^\trdemi(s,t,u)\\
\CH^\undemi(s,t,u)\\
\ep\ ,\
\vec{\CG}(s,t,u)\equiv\bp
\CG^0(s,t,u)\\
\CG^2(s,t,u)\\
\CG^1(s,t,u)\\
\tilde{\CG}^1(s,t,u)\\ \ep
\en
These vectors satisfy the following crossing symmetry relations
\be\lbl{isocrossrels}
\vec{\CG}(t,s,u)
= \bm{C_{st}}\vec{\CF}(s,t,u)
 ,\quad
\vec{\CF}(u,t,s)
=\bm{C_{us}} \vec{\CF}(s,t,u)\ , 
\en
the matrices $\bm{C_{st}}$, $\bm{C_{us}}$ are given in eq.~\rf{crossmat} in
appendix~\sect{crossingsym}. 
We next use an approximation in which these isospin amplitudes are
written as a finite sum of terms involving a function of one of the
Mandelstam variables times an angular factor. This structure can be
derived from dispersion theory, upon truncating the partial-wave
expansion in the absorptive parts to
$j=0,1$~\cite{Khuri:1960zz,Stern:1993rg,Kambor:1995yc}. The derivation
shows that the one-variable functions which are involved are analytic
with a right-hand cut. It has been adapted to the $D\to\Kbar\pi\pi$
amplitudes (also including $j=2$) in ref.~\cite{Niecknig:2016fva}. We
will assume here that the contributions generated by the $j\ge2$
partial-waves in the absorptive parts can be approximated by an
isobar-type model. The relevance of this description for
$D\to\Kbar\pi\pi$ is motivated by the experimental observation of a
large dominance of the $S$ and $P$ waves in the Dalitz plot region.
Restricting to $j=0,1$ for the moment, there are twelve one-variable
functions which are involved: $F_j^K$, $H_j^K$, $G_0^I$, $G_1^1$,
$\widetilde{G}_1^1$. Collecting them into vectors as follows 
\be\lbl{onevarvectors}
\vec{F}_0(z)=\bp
F_0^\trdemi(z)\\
F_0^\undemi(z)\\
H_0^\trdemi(z)\\
H_0^\undemi(z)\\ \ep\, ,\
\vec{F}_1(z)=\bp
F_1^\trdemi(z)\\
F_1^\undemi(z)\\
H_1^\trdemi(z)\\
H_1^\undemi(z)\\ \ep\, ,\ 
\vec{G}_0(z)=\bp
G_0^0(z)\\
G_0^2(z)\\
0\\
0\\
\ep\, ,\
\vec{G}_1(z)=\bp
0\\
0\\
G_1^1(z)\\
\TG(z)\\
\ep\ 
\en
and taking into account the crossing relations~\rf{isocrossrels}, one easily
arrives at the following representations for the $D\to\Kbar\pi\pi$ isospin
amplitudes $\vec{\CF}$ and $\vec{\CG}$ 
\be\lbl{isosvrep1}
\ba{ll}
\vec{\CF}(s,t,u)
= & \vec{F}_0(s)+(s(t-u)+\Delta) \vec{F}_1(s)
+\bm{C_{us}} [\vec{F}_0(u)+(u(t-s)+\Delta) \vec{F}_1(u)]\\[0.2cm]
\ &+\bm{C_{st}}^{-1} [\vec{G}_0(t) +(s-u)\vec{G}_1(t)]\\[0.2cm]
\vec{\CG}(t,s,u)= & \vec{G}_0(t) +(s-u)\vec{G}_1(t)]
 +\bm{C_{st}}[\vec{F}_0(s)+(s(t-u)+\Delta) \vec{F}_1(s)]\\[0.2cm]
\ &+ \bm{C_{st}}\bm{C_{us}} [\vec{F}_0(u)+(u(t-s)+\Delta)
  \vec{F}_1(u)]\ .
\ea\en

\subsection{Physical amplitudes in terms of $j=0,1$ one-variable functions} 
Using eqs.~\rf{isosvrep1},~\rf{onevarvectors} together with
table~\ref{table:newisoexpress} in appendix~\ref{sec:isodetails} one can
express the physical amplitudes in terms of the twelve one-variable isospin
functions introduced above.  The $D^+$ amplitudes involve only six of them:
$F_0^\undemi$, $F_0^\trdemi$, $F_1^\undemi$, $F_1^\trdemi$, $G_0^2$ and
$G_1^1$ (called collectively the $F$-functions).
The $D^0$ amplitudes 
involve the same six $F$-functions plus six additional ones:
$H_0^\undemi$, $H_0^\trdemi$, $H_1^\undemi$, $H_1^\trdemi$, $G_0^0$
and $\TG$ (called $H$-functions in the following). The
expressions of all the physical amplitudes are listed below, starting
with the $D^+$ amplitudes.
\begin{itemize}
\item[{\bf 1-}]{\bf $D^+$ amplitudes:}\\
The two amplitudes $\Dp\to \Km \pip\pip$ and $\Dp\to \Kzb \piz
\pip$ have the following expressions
\be\lbl{A1A2express}
\ba{l@{}l}
{\CA}_{\Km \pip\pip}(s,t,u)= & 
-{\sqrt2}\bigg[ F_0^\trdemi(s)+ F_0^\undemi(s) 
       +Z_s\big(F_1^\trdemi(s)+ F_1^\undemi(s)\big)\\
\ & +(s\leftrightarrow{u})\bigg]+G_0^2(t) \\
{\CA}_{\Kzb \piz\pip}(s,t,u)= &  -2F_0^\trdemi(s)+F_0^\undemi(s) 
          +Z_s\big(-2F_1^\trdemi(s)+F_1^\undemi(s)\big) \\[0.2cm]
\ &+3F_0^\trdemi(u)+ 3\,\zu F_1^\trdemi(u)
-\frac{\sqrt2}{4} G_0^2(t)+(s-u) G_1^1(t)\\
\ea\en
where
\be
\zs=s(t-u)+\Delta,\quad
\zu=u(t-s)+\Delta\ 
\en
are the $j=1$ angular factors.
The above formulae are in agreement with the result
of ref.~\cite{Niecknig:2015ija}\footnote{The functions used here are
  proportional to those of ref.~\cite{Niecknig:2015ija}:
  $F_j^\trdemi=(-1)^j{\cal F}_j^\trdemi/\sqrt{15}$,
  $F_j^\undemi=(-1)^{j+1} {\cal F}_j^\undemi/\sqrt6$, 
  $G_0^2={\cal  F}_0^2$, $G_1^1={\cal F}_1^1\sqrt{3/8}$.}.

\item[{\bf 2)}] {\bf $D^0$ amplitudes:}\\
The $\Dz\to \Km\piz\pip$ amplitude reads
\be\lbl{A4express}
\ba{l@{}l}
{\CA}_{\Km\piz\pip}(s,t,u)=
\ & -{2}\,F^\trdemi_0(s)+F^\trdemi_0(u)-F^\undemi_0(u)\\
\ &-{2}Z_s\,F^\trdemi_1(s)+\zu\big(F^\trdemi_1(u)-F^\undemi_1(u)\big)\\
\ &+\frac{\sqrt2}{4}\,{G}_0^2(t)+(s-u)G_1^1(t)\\
\ &+\sqrt2\Big( H^\trdemi_0(s)-H^\trdemi_0(u)
+ Z_s H^\trdemi_1(s)-\zu H^\trdemi_1(u)\Big)\\
\ &-\frac{\sqrt2}{2}\Big(H^\undemi_0(s)-H^\undemi_0(u)
+  Z_s H^\undemi_1(s)-  \zu H^\undemi_1(u)\Big)\\
\ &-2(s-u)\TG(t)\ .
\ea\en
The contributions from the $H$-functions in this amplitude are
antisymmetric under $s,u$ interchange, they cancel when the amplitude
is symmetrised in $s$, $u$, which is expected from the first isospin
symmetry relation in eq.~\rf{generaliso2}.

The $\Dz\to \Kzb\pim\pip$ amplitude reads
\be\lbl{A6express}
\ba{l@{}l}
{\CA}_{\Kzb\pim \pip}(s,t,u)= & {\sqrt2}\Big( F^\trdemi_0(s)+\zs F^\trdemi_1(s)\Big)
+\frac{1}{6}\,{G}_0^2(t)\\
\ & -\Big( H^\trdemi_0(s)+H^\undemi_0(s)
+\zs\big( H^\trdemi_1(s)+H^\undemi_1(s)\big) \Big)\\
\ & -3\Big(H^\trdemi_0(u)+\zu H^\trdemi_1(u)\Big)
-{G}_0^0(t)-\sqrt2\,(s-u)\,\TG(t)\ .
\ea\en

Finally, the $\Dz\to \Kzb\piz\piz$ amplitude's expression is
\be\lbl{A7express}
\ba{ll}
{\CA}_{\Kzb\piz\piz}(s,t,u)= & 
\sqrt2\Big[F^\trdemi_0(s)+\frac{1}{2} F^\undemi_0(s)
+\zs\big(F^\trdemi_1(s)+\frac{1}{2} F^\undemi_1(s)\big) 
+(s\leftrightarrow u)\Big]\\
\ & -\frac{1}{3}\,{G}_0^2(t)
-\Big[2H^\trdemi_0(s)+ \frac{1}{2}H^\undemi_0(s)
+\zs\big(2H^\trdemi_1(s)+ \frac{1}{2}H^\undemi_1(s) 
\big)\\
\ & +(s\leftrightarrow u)\Big] -{G}_0^0(t)\ .
\ea\en
\end{itemize}

We discuss below how to derive the functions $F_j^K$, $H_j^K$, $G_j^I$ with
$j=0,1$ from a set of Khuri-Treiman equations. The formulas above can be used,
if one wishes, within simpler isobar model approaches.  In order to properly
describe the physical amplitudes it proves necessary to introduce some $j=2$
contributions. It is not difficult to extend the above formulas to include $j=2$
resonances consistently with isospin symmetry. One must simply perform the
following replacements,
\be\lbl{j=2extension}
\ba{l}
Z_w F_1^K(w) \to  Z_w F_1^K(w) + Z_{2w} F_2^K(w)\\
Z_w H_1^K(w) \to  Z_w H_1^K(w) + Z_{2w} H_2^K(w)\\
G_0^0(t)    \to G_0^0(t)+Z_{2t} G_2^0(t)
\ea\en
where $w=s,u$ and one can take for the $j=2$ angular functions
\be\lbl{Z2s}
\ba{l}
Z_{2w}= 3Z_s^2-\lambda_{D\pi}(w)\lambda_{K\pi}(w)\\
Z_{2t}=3t(s-u)^2-\lambda_{DK}(t)(t-4\mpid)\ .
\ea\en

\section{Unitarity, dispersion relations and Khuri-Treiman equations}
In this section we formulate a set of Khuri-Treiman type equations
for the isospin one-variable functions.
Let us first consider the asymptotic
behaviour of the amplitudes, which conditions the number of subtractions needed
in their dispersive representations.

\subsection{Asymptotic behaviour and uniqueness conditions}

Letting the variable $s$ go to infinity while keeping $t$ fixed and small
the leading asymptotic behaviour of the isospin amplitudes $\CH^\undemi$, 
$\CH^\trdemi$ is given by Pomeron exchange, while the behaviour of 
the amplitudes $\CF^\undemi$, $\CF^\trdemi$ is given by the $\rho$
Reggeon exchange. Therefore, a dispersion relation in the variable $s$
for the amplitudes $\CH^I$ requires two subtractions while a single
subtraction is needed in the case of $\CF^I$.
Now, when $t$ goes to infinity and the variable $s$ is kept fixed, 
the asymptotic behaviour is given by the $K^*$, $K^*_2$ Reggeon exchanges. A
dispersion relation in $t$ thus requires one subtraction.

Here, in order to improve the description in the finite
energy region of interest, we use a number of subtractions larger than the
minimal one and assume twice-subtracted dispersion relations for all
amplitudes. This assumption, which must be considered as part of the
model, leads to the following asymptotic behaviour  
\be\lbl{regge}
\ba{l}
\lim_{s\to\infty} \left.\CA_i(s,t,u)\right|_{t\,{fixed}}\sim s\\
\lim_{t\to\infty} \left.\CA_i(s,t,u)\right|_{s\,{fixed}}\sim  t\ .\\
\ea\en
The corresponding asymptotic behaviour of the one-variable functions is then
as follows 
\be\lbl{asysingle}
\ba{l}
F_0^\undemi(s),\ F_0^\trdemi,\ H_0^\undemi(s),\ H_0^\trdemi\sim s\\
F_1^\undemi(s),\ F_1^\trdemi,\ H_1^\undemi(s),\ H_1^\trdemi\sim s^{-1}\\
G_0^2(t),\ G_0^0(t) \sim t\\
G_1^1(t),\ \widetilde{G}_1^1(t) \sim t^0\ .
\ea\en
Because there are only two independent Mandelstam variables, there is a family
of re-definitions of the one-variable functions which leave the physical
amplitudes invariant~\cite{Stern:1993rg}. Using the vectorial
notation~\rf{onevarvectors} linear translations consistent with the asymptotic
behaviour are as follows
\be\lbl{lineartrans}
\ba{l}
\vec{F}_0(z)\to \vec{F}_0(z) +\vec{a}_0 + \vec{b}_0 z\\
\vec{G}_0(z)\to \vec{G}_0(z) +\vec{a}_{0G} +\vec{b}_{0G} z\\
\vec{G}_1(z)\to \vec{G}_1(z)+\vec{a}_{1G}\\
\ea\en
where the vectors $\vec{a}_{0G}$, $\vec{b}_{0G}$, $\vec{a}_{1G}$ have two
vanishing components. Inserting this into eq.~\rf{isosvrep1} for the full
amplitude it is easy to verify, using eq.~\rf{s+t+u}, that it is
left invariant under the family of translations~\rf{lineartrans} in which the
14 parameters satisfy the conditions
\be
\ba{l}
\vec{a}_{0G}+diag(1,1,0,0)\bm{C_{st}}(2\vec{a}_0+\Sigma\vec{b}_0)=0\\[0.1cm]
\vec{b}_{0G}-diag(1,1,0,0)\bm{C_{st}}\vec{b}_0=0\\[0.1cm]
\vec{a}_{1G}+diag(0,0,1,1)\bm{C_{st}}\vec{b}_0=0\ \\
\ea\en
such that 8 parameters can be chosen arbitrarily. In order to define uniquely
the one-variable functions we must therefore impose eight conditions. In the
following, we will set
\be\lbl{condFH}
\ba{l}
F_0^\trdemi(0)=\dot{F}_0^\trdemi(0)=H_0^\trdemi(0)=\dot{H}_0^\trdemi(0)=0\\
G_0^2(0)=\dot{G}_0^2(0)=G_0^0(0)=\dot{G}_0^0(0)=0\ .\\
\ea
\en
\subsection{Partial-waves and unitarity}
The one-variable functions introduced above satisfy dispersion representations
as a function of their discontinuities across the right-hand cuts (see
eqs.~\rf{ordinaryDRF},~\rf{ordinaryDRH} in appendix~\sect{kernelang}). These
discontinuities can be derived from unitarity relations of $2\to2$
partial-wave amplitudes~\cite{Khuri:1960zz}. In our case, these amplitudes are
$\Kbar\pi\to{D}\pi$ ($s$-channel) and $\pi\pi\to{D}K$ ($t$-channel).
We define the partial-waves in the $s$-channel as 
\be\lbl{pw-schannel}
\vec{\CF}_j(s)=\frac{2j+1}{2}\frac{1}{(s\kappa_s(s))^j}\int_{-1}^1 dz_s\,
P_j(z_s) \vec{\CF}(s,t,u)\  
\en
where $P_j$ is a Legendre polynomial. In the integrand of
eq.~\rf{pw-schannel}, we have to express $t$, $u$ in terms of $s$, $z_s$ as
given in eq.~\rf{tufunczs}. Similarly, in the $t$-channel, the partial-waves
are defined as
\be\lbl{pw-tchannel}
\vec{\CG}_j(t)=\frac{2j+1}{2}
\frac{1}{(\kappa_t(t))^j}\int_{-1}^1 dz_t P_j(z_t)\, \vec{\CG}(t,s,u)\ 
\en
and the variables $s$, $u$ in the integrand must be expressed in terms of $t$
and $z_t$  using eq.~\rf{sufunczt}.
\begin{figure}
\centering
\includegraphics[width=0.5\linewidth]{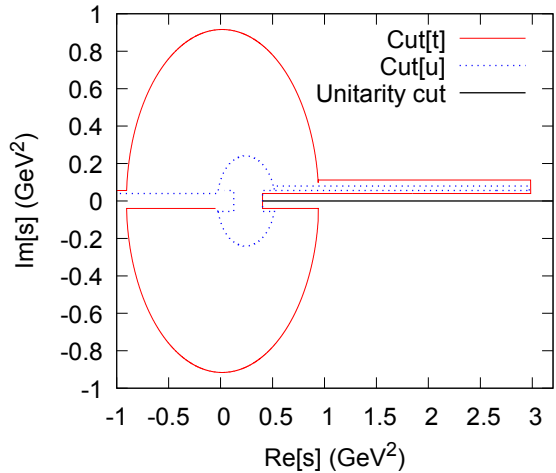}
\caption{\small Complex cuts of the hat-functions $\hat{F}_j^K(s)$,
  $\hat{H}_j^K(s)$ illustrating the $m^2_D+i\epsilon$ prescription. The solid
  red curve is generated from the $z_s$ angular integrations of the functions
  of $t$ and the blue dotted curve is generated from the integrations of the
  functions of $u$. The thick black line is the unitarity cut.}
\label{fig:cut_s}
\end{figure}

%
Inserting the representation~\rf{isosvrep1} of the amplitudes
$\vec{\CF}(s,t,u)$ and $\vec{\CG}(t,s,u)$ in terms of single-variable
functions into the angular integrals~\rf{pw-schannel},~\rf{pw-tchannel}, the
$s$-channel $j=0,1$ partial-waves get expressed as a sum of two functions
\be\lbl{Fhatdef}
\ba{l}
{\CF}^K_0(s)={F}^K_0(s)+ {\widehat{F}}^K_0(s)\\
{\CH}^K_0(s)={H}^K_0(s)+ {\widehat{H}}^K_0(s)\\
{\CF}^K_1(s)={F}^K_1(s)+ \widehat{F}^K_1(s)\\
{\CH}^K_1(s)={H}^K_1(s)+ \widehat{H}^K_1(s)\\
\ea\en
and the $j=0,1$ partial-waves in the $t-$channel have a similar form,
\be\lbl{Ghatdef}
\ba{l}
\CG_0^0(t)= G_0^0(t)+ \widehat{G}_0^0(t)\\
\CG^2_0(t)= G_0^2(t)+ \widehat{G}_0^2(t)\\
\CG_1^1(t)= G_1^1(t)+ \widehat{G}_1^1(t)\\
\widetilde{\CG}^1_1(t)=\widetilde{G}_1^1(t)+ \TTG(t)\ .\\
\ea\en 
The ``hat-functions'' which appear in the expressions of the
partial-waves are composed of a sum of angular integrals involving the
one-variable functions, they are given in
appendix~\ref{sec:Hat-functions}. The hat-functions carry the
left-hand cuts of the partial-wave amplitudes. These cuts contain the
negative real axis and have a complex component because of the unequal
masses of the participating particles~\cite{Kennedy:1962}. In
addition, because $m_D > m_K+2m_\pi$ the left-hand cuts have
components on the positive real axis which overlap with the unitarity
cut: in the region $(m_K+\mpi)^2\le s\le (m_D-m_\pi)^2$ for the $s$
variable and $4\mpid\le{t}\le(m_D-m_K)^2$ for the $t$ variable. The
prescription for separating the unitarity cut from the left-hand cut
in these regions is well known (e.g.~\cite{Kambor:1995yc} and
references therein): one must add an infinitesimal imaginary part to
the $D$ mass, $m^2_D\to{m^2_D+i\epsilon}$, with $\epsilon>0$, this is
illustrated in fig.~\fig{cut_s}.
The discontinuities of the one-variable functions across the unitarity cut are
then simply equal to those of the partial-wave amplitudes given by unitarity.
Keeping the elastic contribution to unitarity for the $\Kbar\pi\to D\pi$
partial-wave amplitude one has
\be\lbl{unitarity-s}
\disc[F_j^K(s)]=
\disc[\CF^K_j(s)]=\exp(-i\delta^K_j(s))\sin(\delta^K_j(s))\,\CF^K_j(s)\ , 
\en
in which $\delta^K_j$ are the $\Kbar\pi\to \Kbar\pi$ scattering
phase-shifts with isospin $K=1/2,3/2$. This relation gives the
discontinuity across the unitarity cut, using analyticity, defining
the discontinuity as 
\be
\disc[\CF^K_j(s)]\equiv
\frac{\CF^K_j(s+i\epsilon)-\CF^K_j(s-i\epsilon)}{2i}\ .
\en
The same unitarity relations hold for the amplitudes $\CH^K_j(s)$.
For the $t-$channel amplitudes  $D K\to \pi\pi$, keeping
the elastic contribution to unitarity one has
\be\lbl{unitarity-t}
\disc[G_j^I(t)]=
\disc[\mathcal{G}_j^I(t)]=\exp(-i\delta_j^I(t))\sin(\delta_j^I(t))\,
\mathcal{G}_j^I(t)
\en 
which involves the $\pi\pi\to \pi\pi$ phase-shifts $\delta_j^I$, with $I=0,1,2$.

\subsection{Khuri-Treiman equations for the $D^+$ and $D^0$ functions}
The form of the discontinuities given in
eqs.~\rf{unitarity-s},~\rf{unitarity-t} (which are based on elastic
unitarity) allows one to express the one-variable functions with
Muskhelishvili-Omn\`es (MO) dispersion
relations~\cite{Muskhelishvili,Omnes:1958hv,Neveu:1970tn} which
exhibit the dependence on the scattering phase-shifts explicitly. A key
ingredient is the Omn\`es function which is expressed as follows in
terms of a given phase-shift $\phi$,
\be
\Omega(w)=\exp\left(\frac{w}{\pi}
\int_{w_{th}}^\infty dw'\frac{\phi(w')}{w'(w'-w)}\right)
\en
where the threshold value $w_{th}$ is $4\mpid$ for $\pi\pi$ phases and
$(m_K+m_\pi)^2$ for $K\pi$ and it has been assumed that the phase is bounded
by a constant at infinity.
We will further assume that this constant is a
multiple of $\pi$. As is well known, the asymptotic behaviour of the Omn\`es
function is related to that of the phase by,
\be\lbl{asyomnes}
\left.\phi(w)\right\vert_{w\to\infty} =  m\pi,\qquad 
\left.\Omega(w)\right\vert_{w\to\infty} \sim \frac{1}{w^m}\ .
\en
Outside of the energy region where scattering is
elastic the value of the phase to be used in the Omn\`es function is, in fact,
arbitrary. We will assume the following asymptotic values, chosen to be
a multiple of $\pi$ close to the value of the phase-shift at the border of the
inelasticity region\footnote{We differ here from
  ref.~\cite{Niecknig:2015ija} who take $\delta_0^{1/2}(\infty)=2\pi$.},
\be\lbl{asymptoticphases}
\ba{l}
\delta_0^{1/2}(\infty)=\pi,\ 
\delta_1^{1/2}(\infty)=\pi,\
\delta_0^{3/2}(\infty)=0,\ 
\delta_1^{3/2}(\infty)=0\\
\delta_0^0(\infty)=\pi,\ 
\delta_1^1(\infty)=\pi,\ 
\delta_0^2(\infty)=0\ .
\ea\en
Implementing the uniqueness conditions~\rf{condFH} as well as the
assumed linear asymptotic behaviour, one can write the
MO dispersive representations for the six $D^+$ functions as follows
\be\lbl{MOrepres}
\ba{l}
F_0^{3/2}(w)=\Omega_0^{3/2}(w)\, w^2\,
\widehat{I}_{0F}^{3/2}(2,w)\\[0.2cm]
F_0^{1/2}(w)=\Omega_0^{1/2}(w)\,\Big[c_0 +c_1\,w+c_2\,w^2 + w^3\,
\widehat{I}_{0F}^{1/2}(3,w) \Big]\\[0.2cm]
F_1^{3/2}(w)=\Omega_{1}^{3/2}(w)\,
\widehat{I}_{1F}^{3/2}(0,w)\\[0.2cm]
F_1^{1/2}(w)=\Omega_1^{1/2}(w)\Big[c_3+ w\,
\widehat{I}_{1F}^{1/2}(1,w) \Big]\\[0.2cm]
G_0^2(t)= \Omega_0^2(t)\,t^2\,
\widehat{I}_{0G}^2(2,t)\\[0.2cm]
G_1^1(t)= \Omega_1^1(t)\Big[c_4+ c_5\,t+ t^2\,
\widehat{I}_{1G}^1(2,t) \Big]
\ea\en
involving six polynomial parameters $c_i$.  
In these formulae the functions $\widehat{I}(n,z)$ are dispersive integrals
with $n$ subtractions involving the hat-functions (see
eqs.~\rf{Fhatdef},~\rf{Ghatdef}) 
\begin{align}\lbl{Ihatdefs}
\widehat{I}^{K}_{jF}(n,w)= & \dfrac{1}{\pi}{\displaystyle\int_\skpi^\infty} 
\dfrac{\widehat{F}^{K}_j(w')\,\sin\delta^{K}_j(w')}
{(w')^n\,(w'-w) \vert\Omega^{K}_j(w')\vert} \,dw'\nonumber\\
\widehat{I}^{K}_{jG}(n,t)= & \dfrac{1}{\pi}{\displaystyle\int_\spipi^\infty} 
\dfrac{\widehat{G}^{K}_j(t')\,\sin\delta^{K}_j(t')}
{(t')^n\,(t'-t) \vert\Omega^{K}_j(t')\vert} \,dt'\nonumber\ .\\
\end{align}
such that eqs.~\rf{MOrepres} form a closed system of integral equations for
the six $F$-functions.
The $\widehat{I}$ integrals contain final-state interaction effects which
involve three particles induced by the two-body interactions described by the
various phase-shifts. These integrals are complex functions, their imaginary
parts being induced both by the $1/(w'-w)$ and $1/(t'-t)$ denominators when
$w$, $t$ approach the real axis from above and from the fact that the
hat-functions are complex. The associated subtracted constants $c_i$ must thus
be complex as well and this generates deviations from Watson's theorem for the
phases of the $F$-functions.

In the case of the $D^0$ amplitudes now, the six additional functions
which are needed satisfy the following equations 
\be\lbl{MOrepresH}
\ba{l}
H_0^{3/2}(w)=\Omega_0^{3/2}(w)\, w^2\,\widehat{I}_{0H}^{3/2}(2,w)\\[0.2cm]
H_0^{1/2}(w)=\Omega_0^{1/2}(w)\,\Big[d_0 +d_1\,w+d_2\,w^2 + 
w^3\,\widehat{I}_{0H}^{1/2}(3,w) \Big]\\[0.2cm]
H_1^{3/2}(w)=\Omega_{1}^{3/2}(w)\,\widehat{I}_{1H}^{3/2}(0,w)\\[0.2cm]
H_1^{1/2}(w)=\Omega_1^{1/2}(w)\Big[d_3+ 
w\,\widehat{I}_{1H}^{1/2}(1,w) \Big]\\[0.2cm]
G_0^0(t)= \Omega_0^0(t)\,\Big[d_4\,t^2 
+ t^3\,\widehat{I}^0_{0G}(3,t)\Big]\\[0.2cm]
\TG(t)= \Omega_1^1(t)\Big[d_5+ d_6\,t
+ t^2\widehat{I}^1_{1\tilde{G}}(2,t)
  \Big]\ .\\
\ea\en
They involve 7 complex polynomial parameters $d_i$.  The
integrals $\widehat{I}^K_{jH}$ have the same form as eqs.~\rf{Ihatdefs}
replacing the $\widehat{F}^K_j$ functions by $\widehat{H}^K_j$ in the
integrands. The expressions of the hat-functions $\widehat{H}^K_j$,
$\widehat{G}_0^0$, $\TTG$ in terms of angular integrals are given in
eqs.~\rf{Hhatfunc},~\rf{Ghatfunc2} in appendix~\sect{Hat-functions}.
They are linear combinations of the twelve one-variable functions.

The above equations have been derived starting from unitarity
equations~\rf{unitarity-s}~\rf{unitarity-t} in which a single channel has been
included (elastic approximation). Experimental studies of $\pi\pi$ and
$\pi{K}$ scattering suggest that this approximation should be valid in energy
regions which include the prominent
$P$-wave resonances $K^*(890)$ and $\rho(770)$ as well as the scalar resonance
$K^*_0(1430)$ but not the $I=0$ scalar $f_0(980)$.  Obviously, as the energy
increases, we expect that more channels should be taken into account. In cases
where inelasticity sets in smoothly one can hope that their effects can be
absorbed, to some extent, into the subtraction polynomials. This is not the
case for the attractive $S$-waves in $\pi\pi$ or $\pi{K}$ scattering for which
inelastic scattering sets in sharply under the effect of the $f_0(980)$ and
$K^*_0(1950)$ resonances respectively. We will
describe below (sec.~\sect{inelasticapprox}) an approximate procedure for
taking this feature into account.
\begin{figure}
\centering
\includegraphics[width=0.5\linewidth]{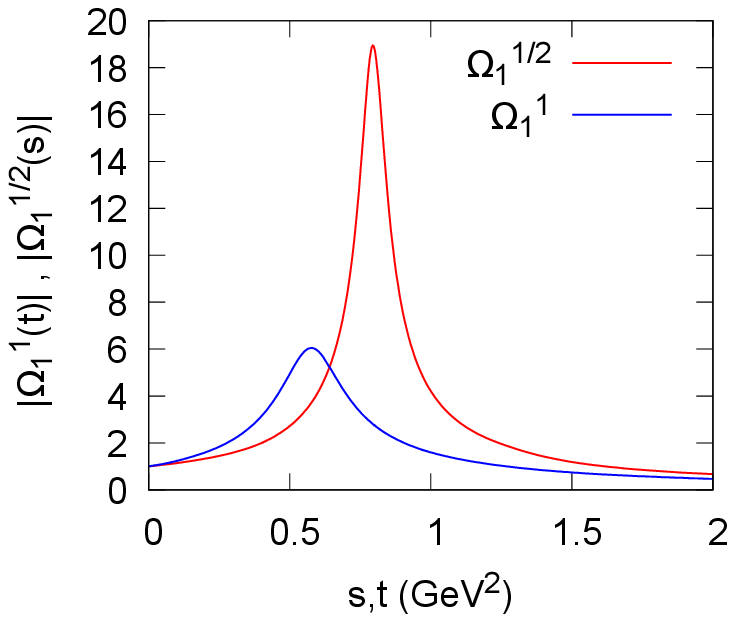}%
\includegraphics[width=0.5\linewidth]{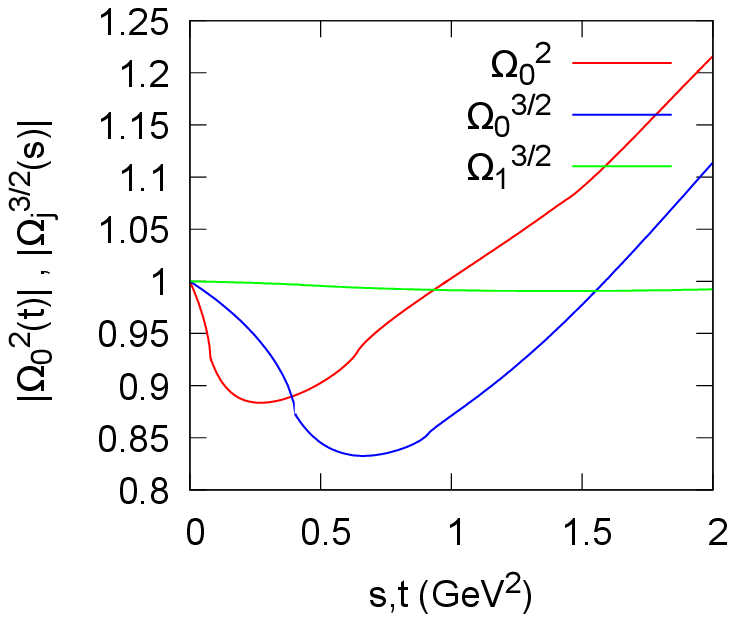}
\caption{\small Left: Omn\`es functions (absolute values) for the $\pi\pi$
  and $\pi{K}$ $P$-waves. Right: Omn\`es functions 
  with isospin $K=3/2$, $I=2$.}
\label{fig:OMfigP+rep}
\end{figure}
\section{Solutions of the KT equations and fits to the data}
\subsection{$\pi{K}$ and $\pi\pi$ phase-shifts and resonances}
The $K^*(892)$ resonance plays a dominant role in all $D\to\Kbar\pi\pi$
decays. Let us start by recalling some points concerning its properties. In
the case of the $K^*(892)^0$ there is good agreement between the values
obtained from semi-leptonic $D$
decays~\cite{FOCUS:2005iqy,delAmoSanchez:2010fd,BESIII:2015hty} (which are the
cleanest channels, in principle) and those obtained from hadroproduction or
other decays and the PDG~\cite{ParticleDataGroup:2022pth} gives
$M_{K^{*0}},\Gamma_{K^{*0}}=(895.6\pm0.2,47.3\pm0.5)$ MeV. For the
$K^*(892)^+$ the most reliable determinations are from $\tau$ decays and the
Belle collaboration~\cite{Epifanov:2007rf} has obtained\footnote{A compatible
  preliminary result was quoted by the Babar
  collaboration~\cite{Adametz:2011zz}
  $M_{K^{*+}},\Gamma_{K^{*+}}=(894.57\pm0.27, 45.56\pm0.71)$ MeV.}
$M_{K^{*+}},\Gamma_{K^{*+}}= (895.47\pm0.77,46.2\pm1.34)$. One sees that
isospin breaking effects are very small for the
$K^*(892)$ which is a favourable point for our isospin symmetric
approach. Determinations of $\pi{K}$ scattering phase-shifts have been
performed in a number of production experiments since the 70's (see
e.g.~\cite{Pelaez:2016tgi} for a recent dispersive analysis and a list of
references). A new set of measurements of this type using a $K_L$ beam is
planned at JLab~\cite{KLF:2020gai}. We describe the $I=1/2$ $P$-wave
phase-shift based on a combined fit of the $\tau$ decay data of
Belle~\cite{Epifanov:2007rf} and the $\pi{K}$ phase-shifts measurements of
refs~\cite{Estabrooks:1977xe,Aston:1987ir} using a three-channel $K$-matrix
from which the $\pi{K}$ vector form-factor $f_+^{K\pi}$ as well as the
phase-shifts can be computed~\cite{Moussallam:2007qc}.

Concerning the properties of the $\rho(770)$ resonance in the $\pi\pi$
$P$-wave, precise determinations have been derived from $e^+e^-\to\pip\pim$
and $\tau^-\to \pim\piz \nu_\tau$. Isospin breaking effects on the mass and
the width again turn out to be smaller than 2 MeV. For the $\pi\pi$ $P$-wave
phase-shift we use the Roy equations solutions provided in
ref.~\cite{Ananthanarayan:2000ht} when $E \le 0.8$ GeV, which are compatible
with the $e^+e^-$ and $\tau$ decay results. In the region $E > 0.8$ GeV we use
simple fits to the phase-shift determination from ref.~\cite{Hyams:1973zf}. We
compute the Omn\`es functions for both the $\pi\pi$ and $\pi{K}$ elastic
scattering phase-shifts. Fig.~\fig{OMfigP+rep} (left) shows that they behave
very similarly to Breit-Wigner functions. Scattering ceases to be purely
elastic for energies above 1 GeV, approximately, for both $\pi\pi$ and
$\pi{K}$ in the $P$-wave. The inelastic effects set in smoothly and we assume
that they can be partly absorbed into the polynomial parameters. Eventually,
we will introduce a modification for $\pi{K}$ by hand in the region of the
$K^*(1680)$ resonance.

The channels with isospin
$I=3/2$ (for $\pi{K}$) and $I=2$ for (for $\pi\pi$) are repulsive. For the $I=2$
$S$-wave phase-shift we use the Roy equations based parametrisation from
ref.~\cite{Ananthanarayan:2000ht} below 0.8 GeV and a simple fit to the
experimental data~\cite{Losty:1973et,Hoogland:1977kt} above, such that
the phase shift goes to zero when $E\to\infty$. For the $I=3/2$ $S$-wave a
parametrisation with a simple rational function of the momentum as proposed
in~\cite{Jamin:2000wn} provides a good fit to the experimental phase
shift. Below 1 GeV we use a parametrisation constrained by the Roy-Steiner
equations~\cite{Buettiker:2003pp}. The $I=3/2$ $P$-wave phase-shift is very
small but it has been measured~\cite{Estabrooks:1977xe}, we have used the
parametrisation proposed in ref.~\cite{Pelaez:2016tgi}. The Omn\`es functions
for these $I=2$ and $I=3/2$ channels are shown in fig.~\fig{OMfigP+rep} (right).  

\begin{figure}[htb]
\centering
\includegraphics[width=0.35\linewidth]{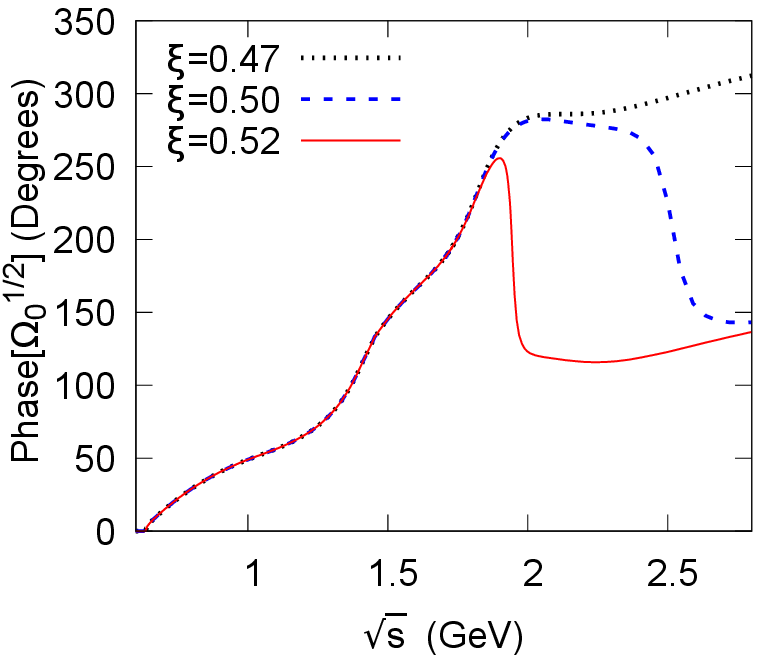}%
\includegraphics[width=0.35\linewidth]{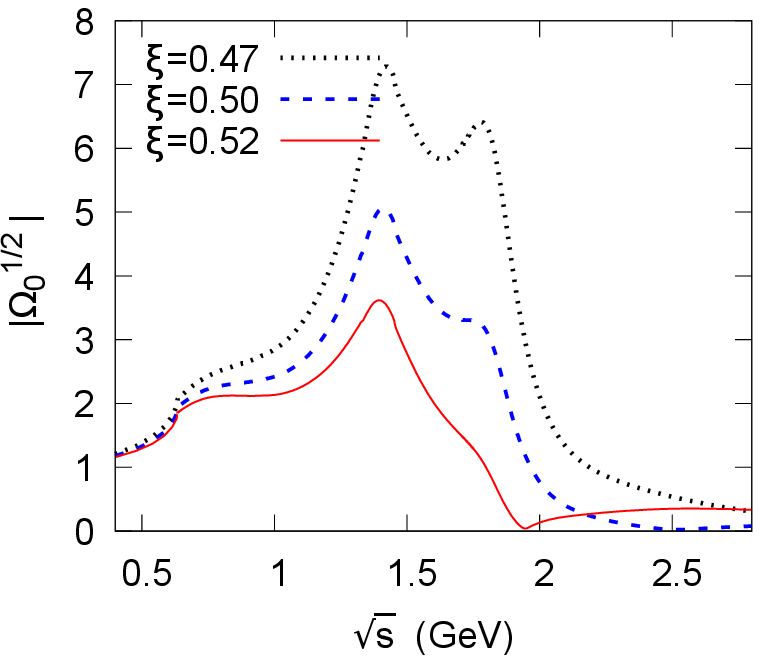}\\
\includegraphics[width=0.35\linewidth]{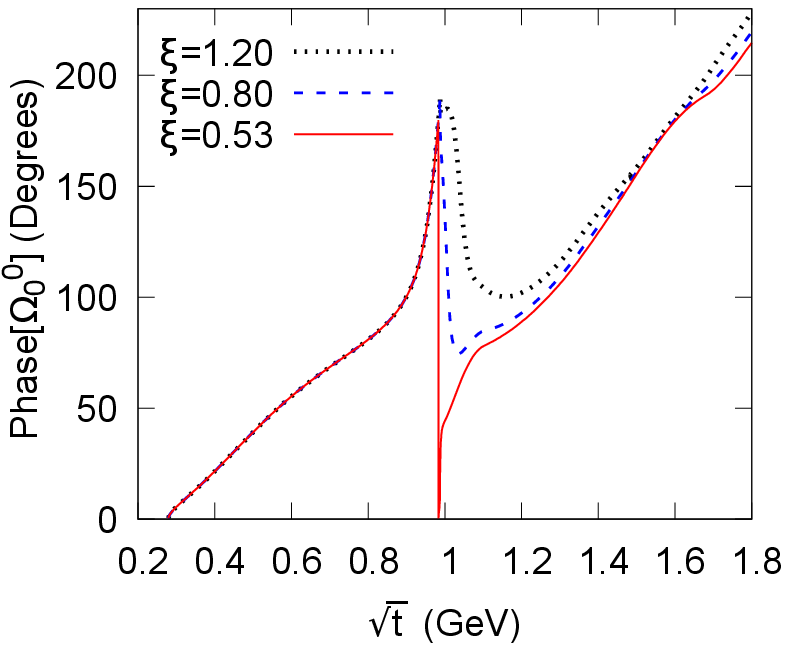}%
\includegraphics[width=0.35\linewidth]{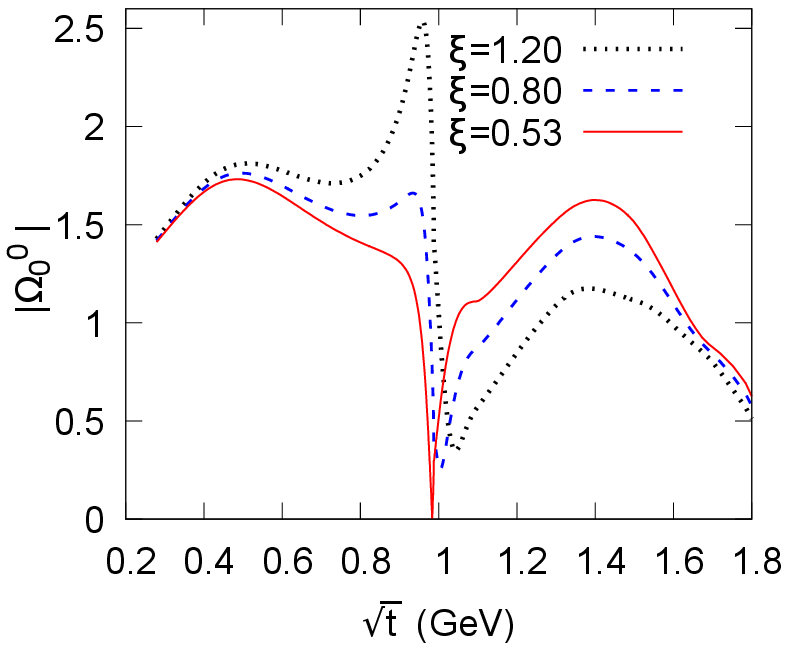}
\caption{\small  Omn\`es functions for the $\pi{K}$ (upper plots)
and $\pi\pi$ (lower plots) attractive $S$-waves generated from 
using two-channel MO matrices (see eq.~\rf{effOMdef}).
The phases are displayed on the left plots and the absolute
values on the right plots.}
\label{fig:effOmnes}
\end{figure}  
\subsection{$S$-wave modelling in inelastic energy regions}\label{sec:inelasticapprox}
Let us now turn to the $S$-waves with $I=0$ and $I=1/2$. For $\pi\pi$
scattering, a distinctive feature is that inelasticity remains
negligibly small below 1 GeV and sets in sharply at the $K\Kbar$
threshold, under the influence of the $f_0(980)$ resonance
(e.g.~\cite{Hyams:1973zf}). A somewhat similar feature is seen in the
$\pi{K}$ experimental data~\cite{Estabrooks:1977xe,Aston:1987ir}. In
that case, inelasticity remains very small below the $\eta'{K}$
threshold and then increases rapidly due to the $K^*_0(1950)$
resonance.
Concerning the $D\to \Kbar\pi\pi$ amplitudes, the physical effect of
$\pi\pi\to K\Kbar$ inelastic rescattering, for instance, is to
introduce a coupling with the $D\to \Kbar K\Kbar$ amplitudes. A proper
treatment is possible, in principle, in which the equation for the
$I=0$ function $G_0^0$ would be replaced by a system of two equations
relating $G_0^0$ and the analogous function $G_{0,K\Kbar}^0$ in the
$D\to \Kbar K\Kbar$ amplitude.  These equations would involve a
$2\times2$ MO matrix instead of the one channel MO function. Here, we
will use $2\times2$ MO matrices simply for generating a plausible
phase to be used in a single channel function in the inelastic region.

We have determined two-channel $T$-matrices for both $\pi\pi$ and
$\pi{K}$ $S$-waves assuming the dominance of a single inelastic
channel and using the experimental scattering data up to 2 GeV. Taking
appropriate asymptotic conditions, one can then compute corresponding
two-channel MO matrices (see~\cite{Babelon:1976ww,Donoghue:1990xh} for
early calculations of this type). Watson's theorem constrains the phases of
the matrix elements of the first line, $\Omega_{1i}$, to be equal to the
elastic scattering phase in the elastic energy region. Forming
linear combinations
\be\lbl{effOMdef}
\Omega_{eff}(z)= \Omega_{11}(z)+\xi \Omega_{12}(z) 
\en
(with $\xi$ real) one obtains a family of one-channel Omn\`es functions which
differ in the behaviour of the phase in the inelastic region. We have
considered values of $\xi$ such that the asymptotic value of the phase is
equal to $\pi$ for both $\pi\pi$ and $\pi{K}$, as assumed in
eqs.~\rf{asymptoticphases}.  The phase of $\Omega_{eff}$ is illustrated in
fig.~\fig{effOmnes} for $\pi\pi$ and $\pi{K}$ scattering.  This procedure
eventually generates a sharp drop of the phase above the inelastic
threshold. For $\pi{K}$, a similar drop was experimentally observed in the
phase of the $\pi{K}$ scalar component of the $D^+\to\Km\pip\pip$
amplitude~\cite{Aitala:2005yh,Bonvicini:2008jw,Link:2009ng}. Correspondingly,
the peak of the $K^*(1950)$ resonance can be strongly reduced (see
fig.~\fig{effOmnes}). We will argue in appendix~\sect{softpions} that the
soft pion theorems for the $D\to\Kbar\pi\pi$ amplitudes support such a
reduction. 

\subsection{Matrix approximations to the KT equations}
We will formulate the KT equations such as to determine the
one-variable functions on a discrete set of points,
\be
(m_K+m_\pi)^2 \le s_1 < s_2 < ... <s_N
\en
for the variables $s,u$, where $s_N$ is a sufficiently large value,
and
\be
4\mpid \le t_1 < t_2 < ... <t_M
\en
for the $t$ variable. We consider first the $F$-functions and introduce two
vectors of size $4N+2M$ containing the values of the one-variable functions at
the points $s_i$, $t_i$ and the values of the corresponding hat-functions,
\be
\mathbb{F}\equiv\left(\ba{c}
F_0^\trdemi(s_i)\\
F_0^\undemi(s_i)\\
F_1^\trdemi(s_i)\\
F_1^\undemi(s_i)\\
G_0^2(t_i)\\
G_1^1(t_i)
\ea\right),\quad
\widehat{\mathbb{F}}\equiv\left(\ba{c}
\widehat{F}_0^\trdemi(s_i)\\
\widehat{F}_0^\undemi(s_i)\\
\widehat{F}_1^\trdemi(s_i)\\
\widehat{F}_1^\undemi(s_i)\\
\widehat{G}_0^2(t_i)\\
\widehat{G}_1^1(t_i)
\ea\right)\ .
\en
We now write the KT integral equations~\rf{MOrepres}~\rf{MOrepresH} using an
approximate evaluation of the $\widehat{I}$ integrals. The integrands
have singularities but it can be shown that the integrals are well
defined and finite by deforming the integration contour in the complex
plane~\cite{Kambor:1995yc,Gasser:2018qtg}. Our method uses real axis
integration, which is fast and convenient, and we have checked that it
correctly reproduces the high-accuracy results derived for $\eta\to3\pi$
in ref.~\cite{Gasser:2018qtg}.  Let us illustrate our
procedure in the case of the integral $\widehat{I}^\trdemi_{0F}$.
We wish to approximate  this integral by a discrete
sum as a function of $\widehat{F}_0^\trdemi(s_{n'})$. The integrand has a Cauchy
singularity $1/(s'-s)$ and, in addition, the function $\widehat{F}_0^\trdemi(s')$
has a square-root singularity when $s'\to \sDmin=(m_D-m_\pi)^2$
(see~\cite{Kambor:1995yc} and appendix~\sect{kernelang}). We multiply and
divide the integrand by $\sqrt{\sDmin-s'}$ and define 
\be
\psi_0^\trdemi(s')\equiv \frac{\sin\delta_0^\trdemi(s')}{(s')^2
  \vert\Omega_0^\trdemi(s') \vert}\sqrt{\sDmin-s'}\widehat{F}_0^\trdemi(s')
\en
which is a finite and continuous function of $s'$ in the whole integration
range. Therefore it can be accurately approximated by a linear function in
the small range $[s_{n'},s_{n'+1}]$. The integration in this range can be
performed using the analytical evaluation of the singular integrals
\be
\int_{s_a}^{s_b} \frac{ds'}{\sqrt{\sDmin-s'}(s'-z)}=I_0(s_b,z)-I_0(s_a,z)
\en
with
\be\lbl{I_0func}
I_0(s',z)=\frac{1}{\sqrt{\sDmin-z}}\left[
 \log(z-s')-\log \left(\sqrt{\sDmin-s'}+ \sqrt{\sDmin-z}\right)^2 \right]\ .
\en
Summing the integrals over the ranges $[s_{n'},s_{n'+1}]$ we obtain an
approximation of $\widehat{I}^\trdemi_{0F}(2,s)$ as a discrete sum
\be\lbl{Ihat_discr}
\widehat{I}^\trdemi_{0F}(2,s)\simeq\frac{1}{\pi}\sum_{n'=1}^{N_s}
\widehat{W}_0^{(n')}(s) 
\psi_0^\trdemi(s_{n'}) \widehat{F}^\trdemi_0(s_{n'}) 
\en
in which the integration weights have the following 
expression\footnote{This formula is for generic values of $n'$. For 
  $n'=1$ or $n'=N_s$ it has to be modified.} 
\begin{align}\lbl{weightsfact}
\widehat{W}^{(n')}_0(s)= &
 \frac{s_{n'+1} -s}{s_{n'+1} -s_{n'}}     \,I_0(s_{n'+1},s)
+\frac{s_{n'-1} -s}{s_{n'} -s_{n'-1}}     \,I_0(s_{n'-1},s)\nonumber\\ \ & 
+\frac{(s_{n'+1}-s_{n'-1})(s      -s_{n'})}
      {(s_{n'}  -s_{n'-1})(s_{n'+1}-s_{n'})}\,I_0(s_{n'},s) \nonumber\\ \ & 
+2\Big(
 \frac{1}{\sqrt{\sDmin-s_{n'}} + \sqrt{\sDmin-s_{n'-1}}}
 -\frac{1}{\sqrt{\sDmin-s_{n'}} + \sqrt{\sDmin-s_{n'+1}}}\Big)\ .
\end{align}
Replacing all the $\hat{I}$ integrals in the set of integral
equations~\rf{MOrepres} by their analogous discretised approximations one
obtains a matrix approximation to these equations 
\be\lbl{matrixF}
\mathbb{F}=\mathbb{F}_{(0)}+\bm{\mathcal{W}}_I^F\!\times\!\widehat{\mathbb{F}}
\en
where $\mathbb{F}_{(0)}$ contains the dependence on the $c_i$ parameters
\be
\mathbb{F}_{(0)}=\left(\ba{c}
0\\
\Omega_0^\undemi(s_i)(c_0+c_1 s_i +c_2 s^2_i)\\
0\\
\Omega_1^\undemi(s_i) c_3\\
0\\
\Omega_1^1(t_i)(c_4+ c_5 t_i)\ea\right)\ .
\en

A similar procedure can be used in order to generate discretised
approximations to the angular integrals. Consider, for example,
$\braque{F_0^\trdemi(u)}_s$. At first, we express the angular integral
as an integral over the right-hand cut (using the ordinary dispersive
representations, see appendix~\ref{sec:kernelang})
\be\lbl{F032av}
\braque{F_0^\trdemi(u)}_s=-\frac{1}{\pi}\int_{(m_K+\mpi)^2}^\infty
K_u^0(s,u')\disc[F_0^\trdemi(u')] du'\ 
\en
this will allow us to use the same grid points $s_i$ and $t_i$ as above.
The integrand in eq.~\rf{F032av} has logarithmic singularities coming from
kernel $K_u^0(s,u')$ at $u'= u^\pm(s)$ (see~\rf{K0u}) and a square-root
singularity from  
$\disc[F_0^\trdemi(u')]$ at $u'=s_D^-$. As before, we multiply and
divide by $\sqrt{\sDmin-u'}$ and, using the unitarity relation to
express the discontinuity, we get
\be\lbl{integexpl}
\braque{F_0^\trdemi(u)}_s=-\frac{1}{\pi}\int_{(m_K+\mpi)^2}^\infty
\frac{K_u^0(s,u')}{\sqrt{\sDmin-u'}}\left[
  \phi_0^\trdemi (u')\left(F_0^\trdemi(u')+\widehat{F}_0^\trdemi(u')
  \right)\right] 
\en
with
\be
\phi_0^\trdemi (u')=\exp\big(-i\delta_0^\trdemi(u')\big)
\sin\big(\delta_0^\trdemi(u')\big)\sqrt{\sDmin-u'}\ .
\en
The quantity inside the square brackets in the integrand of eq.~\rf{integexpl}
is finite and continuous and one can use linear approximations in the
small ranges $[s_{n'},s_{n'+1}]$. The singular integration part which involves
logarithms can be written using the same function $I_0$  which appeared
above~\rf{I_0func} 
\be
\int_{s_a}^{s_b} \frac{\log(u'-u_\pm(s))}{\sqrt{\sDmin-u'}}du'=
\left. -2\sqrt{\sDmin-u'}\log(u'-u_\pm(s))
+2(\sDmin-u')I_0(u',u_\pm(s))\right\vert_{s_s}^{s_b} \ .
\en
It is then not difficult to generate a set of weights which enable one to
express the angular average $\braque{F_0^\trdemi(u)}_s$ in terms of
$F_0^\trdemi(s_{n'})+\widehat{F}_0^\trdemi(s_{n'})$. Repeating this procedure
for the other angular averages and then forming the linear combinations which
give the hat-functions we obtain a second set of linear equations which have
the following form
\be\lbl{matrixFhat}
\widehat{\mathbb{F}}=  \widehat{\mathbb{F}}_{(0)} 
+\bm{\mathcal{W}}_K^F\!\times\!(\mathbb{F}+ \widehat{\mathbb{F}})\ 
\en
where the first term on the right-hand side is generated from the
subtraction constants in the ordinary dispersive representations 
\be
\widehat{\mathbb{F}}_{(0)} =
\begin{pmatrix}
\frac{1}{3}(c_0-c_4\,s_i)+\frac{1}{6}(c'_1+c_4)(\Sigma-s_i+\frac{\Delta}{s_i})
\\[0.2cm]
\frac{1}{3}(2c_0+c_4\,s_i)+\frac{1}{6}(2c'_1-c_4)(\Sigma-s_i+\frac{\Delta}{s_i})
\\[0.2cm]
-\frac{1}{6\,s_i}(c'_1+c_4) 
\\[0.2cm]
-\frac{1}{6\,s_i}(2c'_1-c_4) 
\\[0.2cm]
-\sqrt2(2c_0+c'_1(\Sigma-t_i))
\\[0.2cm]
\frac{1}{2}c'_1\\
\end{pmatrix}
\en
and contains the dependence on the parameters $c_i$.  Those appear only in the
terms ${\mathbb{F}}_{(0)}$ and $\widehat{\mathbb{F}}_{(0)}$, it is then
obvious that the solutions of the system of
equations~\rf{matrixF},~\rf{matrixFhat} are linear functions of
$c_i$. A set of six independent solutions can be determined setting one of the
$c_i$ parameters equal to 1 and the others equal to 0.

Let us now consider the $H$-functions and introduce the corresponding vectors
\be
\mathbb{H}\equiv\left(\ba{c}
H_0^\trdemi(s_i)\\
H_0^\undemi(s_i)\\
H_1^\trdemi(s_i)\\
H_1^\undemi(s_i)\\
G_0^0(t_i)\\
\TG(t_i)
\ea\right),\quad
\widehat{\mathbb{H}}\equiv\left(\ba{c}
\widehat{H}_0^\trdemi(s_i)\\
\widehat{H}_0^\undemi(s_i)\\
\widehat{H}_1^\trdemi(s_i)\\
\widehat{H}_1^\undemi(s_i)\\
\widehat{G}_0^0(t_i)\\
\TTG(t_i) \ea\right)\ .
\en
The first matrix equation is analogous to~\rf{matrixF}
\be\lbl{matrixH}
\mathbb{H}=\mathbb{H}_{(0)} 
+ \bm{\mathcal{W}}_I^H\!\times\! \widehat{\mathbb{H}}
\en
where $\mathbb{H}_{(0)}$ contains the parameters $d_i$
\be
\mathbb{H}_{(0)}=\begin{pmatrix}
0\\[0.2cm]
\Omega_0^\undemi(s_i)(d_0+d_1s_i+d_2s_i^2)\\[0.2cm]
0\\[0.2cm]
\Omega_1^\undemi(s_i)\,d_3\\[0.2cm]
\Omega_0^0(t_i)(d_4 t_i^2)\\[0.2cm]
\Omega_1^1(t_i)(d_5+d_6 t_i)\\
\end{pmatrix}\ .
\en
The extra set of equations generated from the expressions of the $\widehat{H}$
function are similar to eqs.~\rf{matrixFhat}
\be\lbl{matrixHhat}
\widehat{\mathbb{H}}=\widehat{\mathbb{H}}_{(0)} + 
 \bm{\mathcal{W}}_K^{HF}\!\times\! (\mathbb{F}+ \widehat{\mathbb{F}})
+\bm{\mathcal{W}}_K^{HH}\!\times\! (\mathbb{H}+
\widehat{\mathbb{H}})\ .
\en
albeit slightly more complicated due to the contributions from both the $F$
and the $H$ functions (see eqs.~\rf{Hhatfunc}~\rf{Ghatfunc2}). The vector
$\widehat{\mathbb{H}}_{(0)}$ depends on both the $c_i$ and the $d_i$
parameters.
\be
\widehat{\mathbb{H}}_{(0)} =\begin{pmatrix}
\frac{1}{3}(d_0-\sqrt2 d_5\,s_i)
+\frac{1}{6}(d'_1+\sqrt2 d_5)(\Sigma-s_i+\frac{\Delta}{s_i}) 
\\[0.2cm]
\frac{1}{3}(-d_0 +\sqrt2(c_0 -(c_4-4d_5)s_i )
+\frac{1}{6}(-d'_1+\sqrt2(c'_1+c_4-4d_5))(\Sigma-s_i+\frac{\Delta}{s_i})
\\[0.2cm]
-\frac{1}{6\,s_i}(d'_1+\sqrt2 d_5)                  
\\[0.2cm]
-\frac{1}{6\,s_i}(-d'_1+\sqrt2(c'_1+c_4-4d_5) )      
\\[0.2cm]
\frac{1}{3}(-\sqrt2 c_0+3d_0)
+\frac{1}{6}(-\sqrt2 c'_1+3d'_1)(\Sigma-t_i)
\\[0.2cm]
\frac{\sqrt2}{4} d'_1
\end{pmatrix}\ .
\en
The $H$-functions are linear in the $c_i$ and $d_i$ parameters, 
a set of 6+7 independent solutions is generated by setting one of these
parameters equal to 1 and the others to 0 in the RS equations.

The numerical results presented below use values for the number of
points $N\sim M\sim 700-800$ and, for the upper integration ranges,
$s_N\sim t_M =10^3$ $\hbox{GeV}^2$. The matrices which appear in the
discretised form of the RS equations~\rf{matrixF}~\rf{matrixH} are
then rather large but they turn out to be well conditioned and solving
the linear equations does not cause any problem.  As a check, we have
verified that we could reproduce the results of
ref.~\cite{Niecknig:2015ija} on the independent solutions when using
the same input phase-shifts. We have also verified that the solutions
obey ordinary dispersion relations.  Once one has determined the set
of independent solutions, using them to perform fits is just as easy
as with a simple isobar model because of the linearity in the
parameters $c_i$, $d_i$. Such fits will be discussed below.
\subsection{$D^+$ fits}
We will first test our model against the measurements performed by the BESIII
collaboration~\cite{Ablikim:2014cea} on the $D^+\to K_S\piz\pip$ mode. The
Dalitz plot is divided into 1342 equal-size bins and the data consist of a set
of values for the number of events in each bin: $N_i, \Delta{N}_i$. These
numbers are corrected for the background, detector acceptance and energy
resolution effects, such that they can be compared directly to a given
theoretical model.

\begin{figure}
\centering
\includegraphics[width=0.50\linewidth]{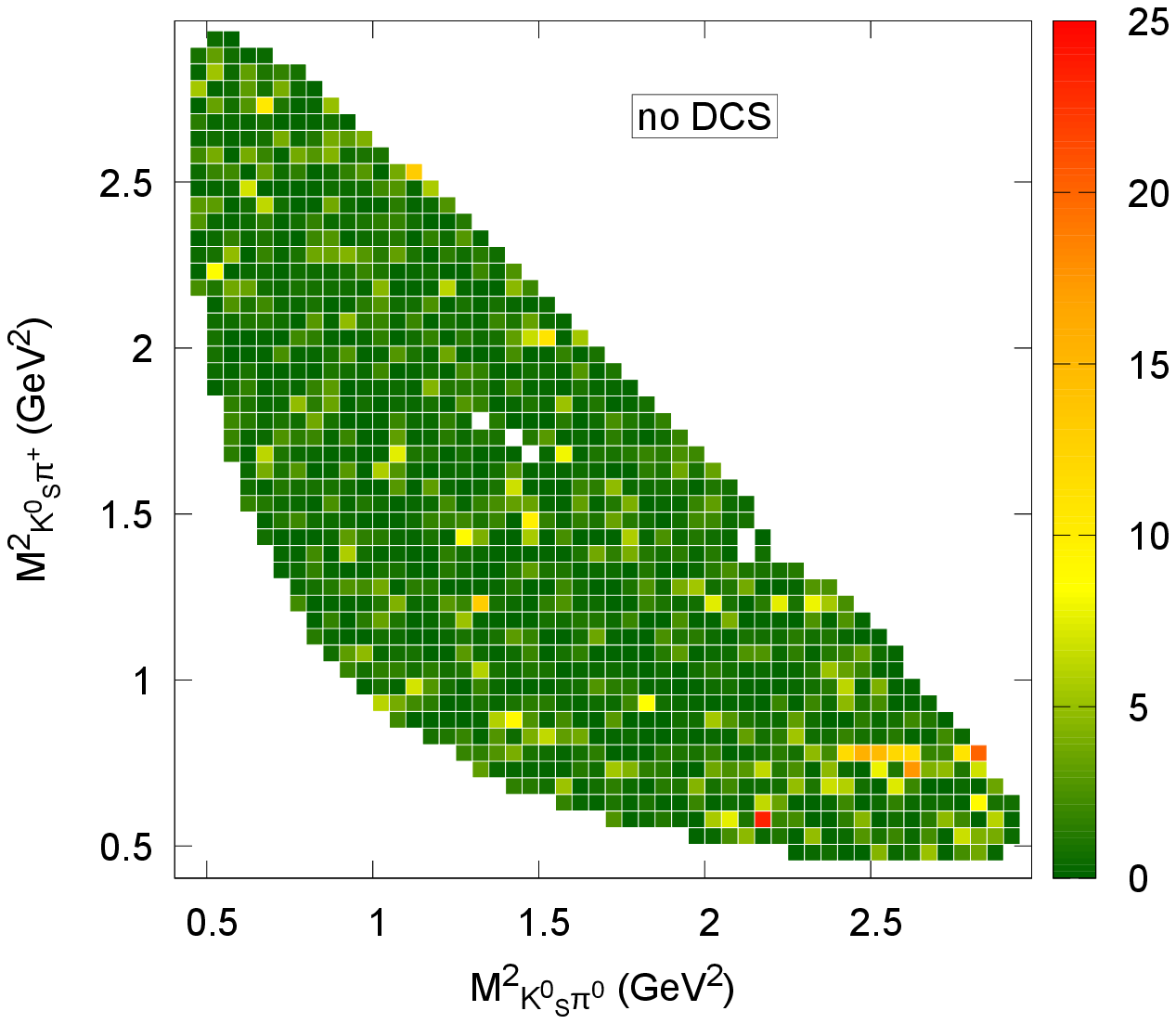}%
\includegraphics[width=0.50\linewidth]{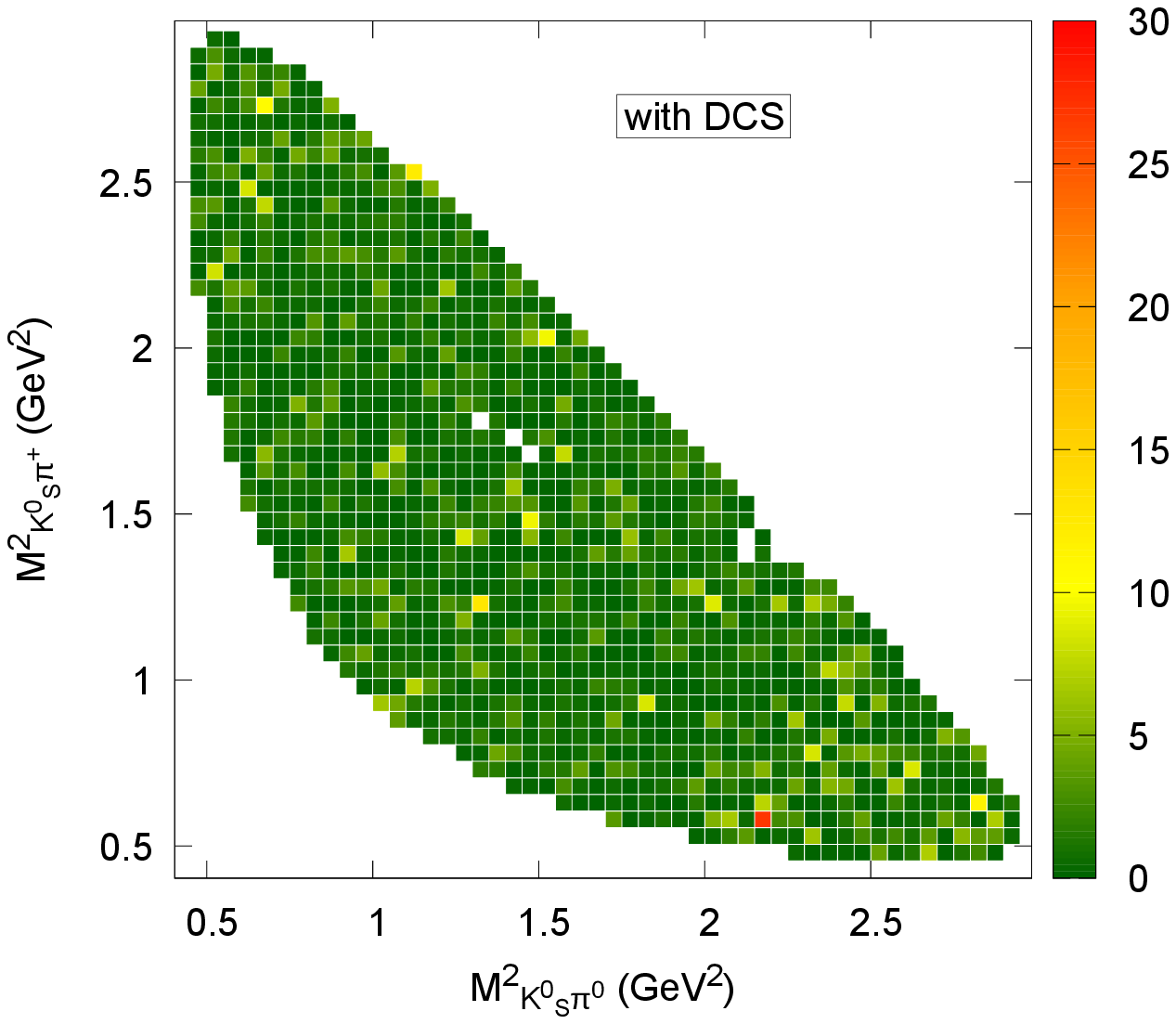}
\caption{\small Distributions of the $\chi^2$ values in the $D^+\to
  K_S\piz\pip$ fits. Left plot: no DCS contribution, right-plot: the DCS
  contribution is included in the fit.}
\label{fig:dalitzchi2_nodcs}
\end{figure} 
We will perform fits over the complete Dalitz plot, i.e. going beyond
the region in which $2\to2$ rescattering can be considered as
elastic. This will be necessary later on for addressing the available
$D^0$ data, because the sizes and shapes of the bins, in that case, do
not allow to separate the elastic from the inelastic regions.  A
further advantage is that this makes it possible to compute decay
widths. As already mentioned the important vector and scalar
resonances, with the exception of the $f_0(980)$, fall within the
region of validity of elastic rescattering. The few smaller
contributions which we add are described with a simple isobar model
approach. At first, since a $K_S$ is detected, a DCS amplitude must be
considered which we assume to be dominated by the $K^*(892)^+$
resonance contribution. We also include the effect of the $D$-wave
resonance $K^*_2(1430)$ and that of the $P$-wave resonance
$K^*(1680)$.
Our model for the $D^+\to K_S \piz\pip$ amplitude, finally, is as
follows\footnote{We take $\ket{K_{S,L}}=(\ket{\Kz}\pm\ket{\Kzb})/\sqrt2$,
  ignoring CP violation.}
\be\lbl{A2+reso}
\ba{l}
\CA_{D^+\to K_S\piz\pip}=\dfrac{F_{norm}}{\sqrt2}\Big\{
-2[F_0^\trdemi(s)+Z_s F_1^\trdemi(s)]+ F_0^\undemi(s)\\
+Z_s[ F_1^\undemi(s)+C_{K^*(1680)} BW_{K^*(1680)}(s)]
+Z_{2s} C_{K^*_2(1430)} BW_{K^*_2(1430)}(s) \\
+3[F_0^\trdemi(u)+Z_u F_1^\trdemi(u)]-\dfrac{\sqrt2}{4} G_0^2(t)
+(s-u)\,G_1^1(t)+  Z_u C_{DCS}BW_{K^*(892)}(u)\Big\}\ 
\ea
\en
where the Breit-Wigner functions, called $BW$, are described in
appendix~\sect{BreitWigner}.
\begin{figure}[t]
\centering
\includegraphics[width=0.50\linewidth]{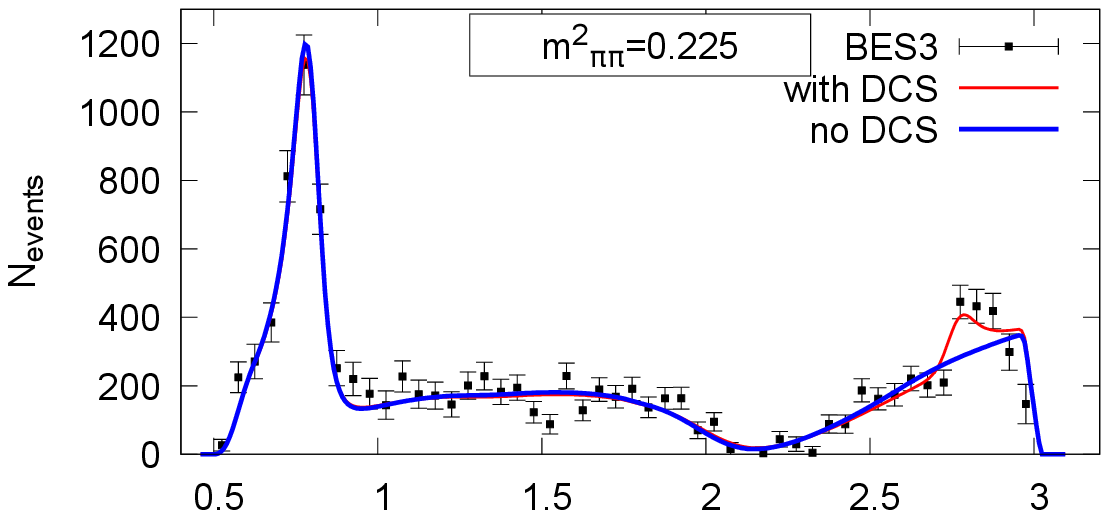}%
\includegraphics[width=0.50\linewidth]{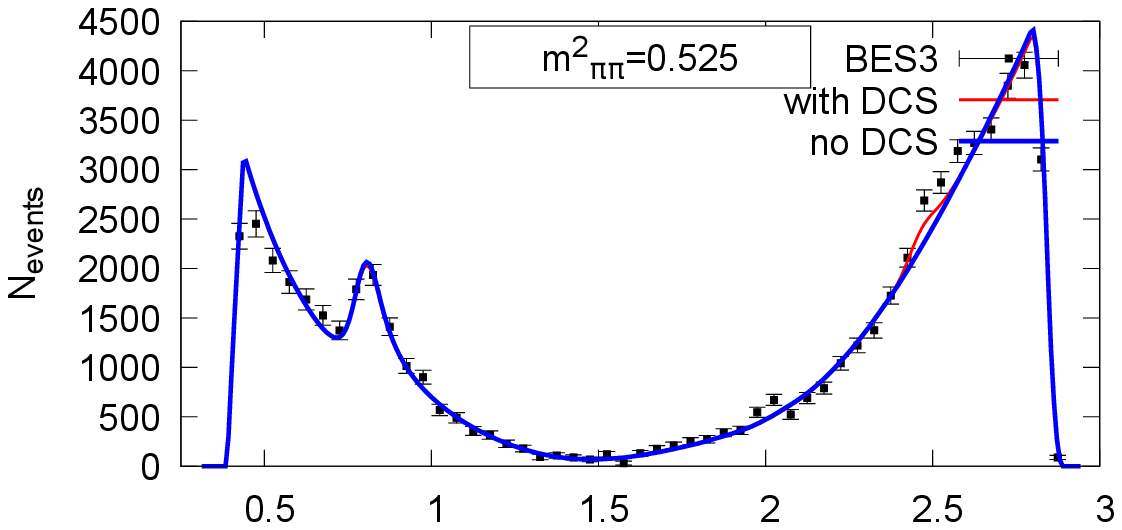}\\
\includegraphics[width=0.50\linewidth]{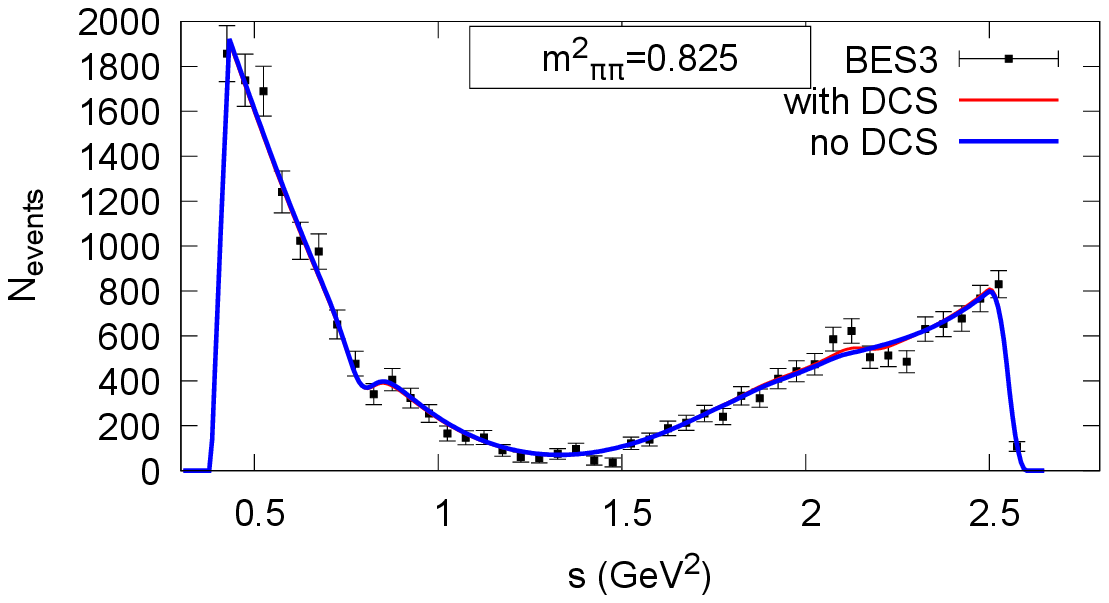}%
\includegraphics[width=0.50\linewidth]{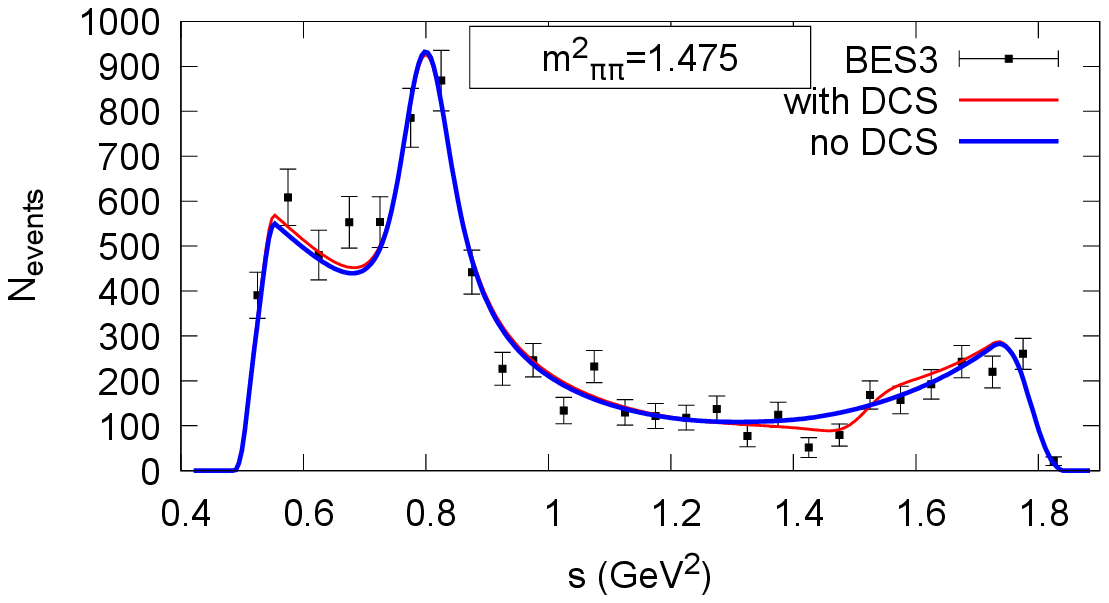}
\caption{\small Comparison of the fitted amplitude with the
  experimental data along slices of Dalitz plot with fixed values of
  $t=m^2_{\pi\pi}$.}
\label{fig:dalitzslices}
\end{figure} 
In addition to the polynomial parameters $c_i$ which drive the KT
amplitudes, it involves the three complex resonance couplings
$C_{K^*(1680)}$, $C_{K^*_2(1430)}$ and $C_{DCS}$ also to be determined from
the fit.  The angular function $Z_{2s}$ associated with the spin 2
resonance $K^*_2(1430)$, proportional to the Legendre polynomial
$P_2(z_s)$, is given in eq.~\rf{Z2s}.
%

\begin{figure}
\centering
\includegraphics[width=0.49\linewidth]{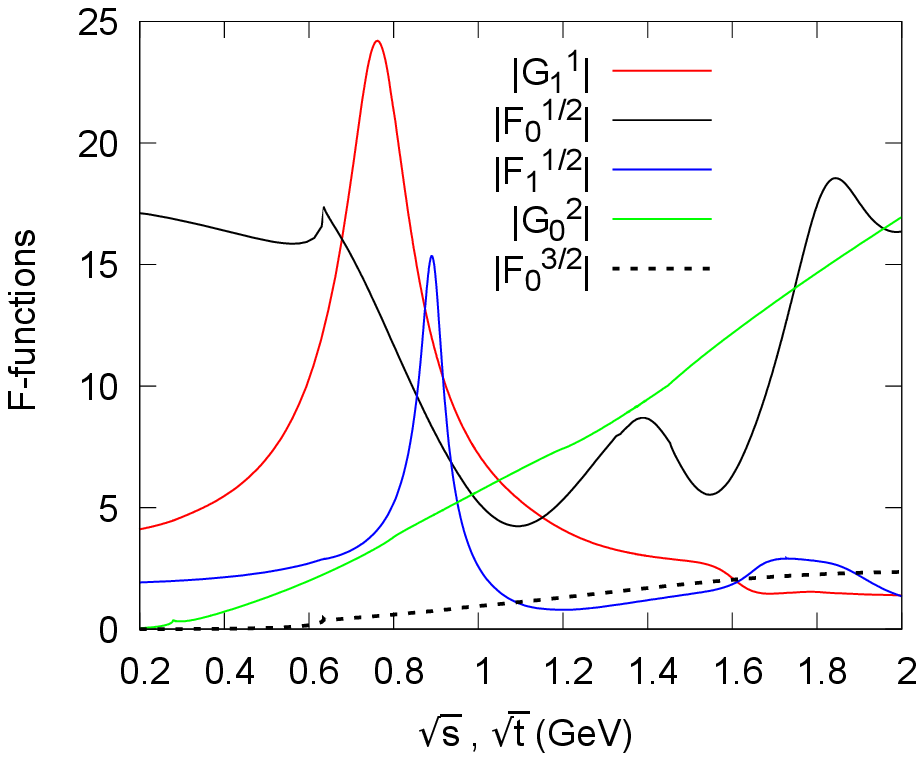}%
\includegraphics[width=0.49\linewidth]{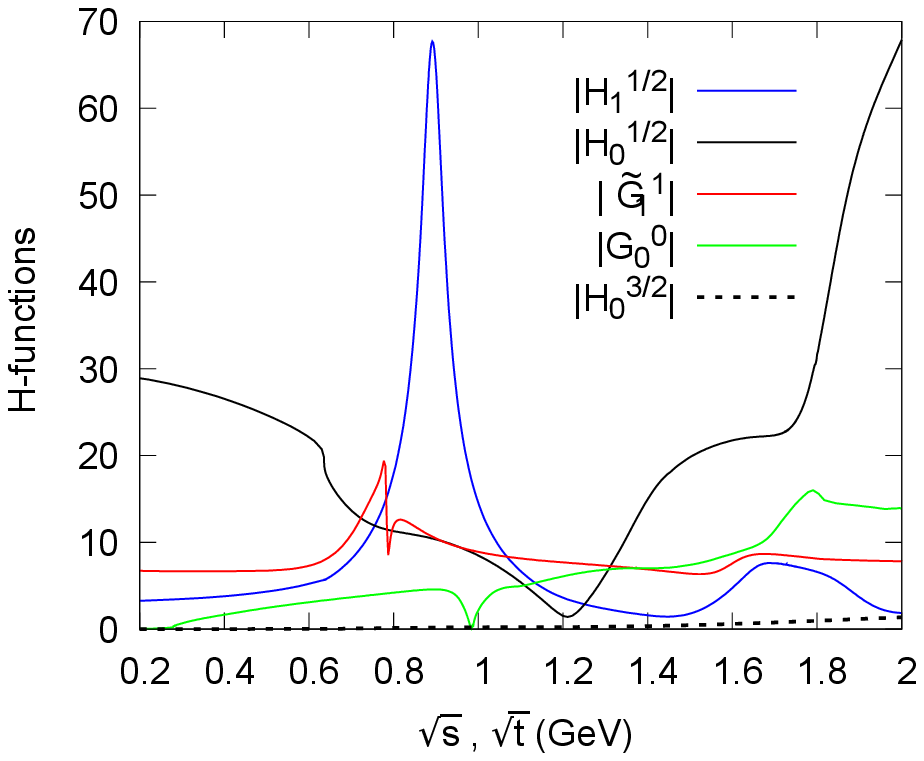}
\caption{\small Left plot: absolute values of the $F$-functions
  resulting from the fit (central values). Right plot:
the corresponding results for the $H$-functions. The small amplitudes
$|F_1^\trdemi|\lapprox0.04$, $|H_1^\trdemi|\lapprox0.008$ are not shown.}
\label{fig:FFplot}
\end{figure} 

When performing the fit, it is convenient to fix one of the
polynomial parameters $c_i$ (we will set $c_3=-1$) and
introduce instead an overall normalisation factor $F_{norm}$. In the fitting
process $F_{norm}$ is adjusted to the total number of experimental
events while the physical value of $F_{norm}$ can be determined from the
experimental value of the width $\Gamma_{D^+\to K_S\piz\pip}$ (see below).
In the calculation of the $\chi^2$ we integrate the absolute value of the
amplitude squared over each bin, dropping the bins which overlap with
the boundary of the Dalitz plot, which leaves $N_{bins}=1182$ .  
At first, we ignore the DCS amplitude i.e. we set $C_{DCS}=0$. We then have
$N_{par}=15$ parameters to fit. An additional parameter is $\xi$ which, via
eq.~\rf{effOMdef}, controls the behaviour of the $\pi{K}$
$S$-wave phase in the inelastic region. It is not included in the fit but its
value is tuned to optimise the result, which gives $\xi\simeq0.515$.
Minimising the  $\chi^2$ we obtain
\be
\chi^2=1576,\quad \chi^2/N_{dof}=1.35
\en
which is comparable to the result of the fit reported in
ref.~\cite{Ablikim:2014cea} ($\chi^2/N_{dof}=1.41$ with model A) using a
substantially larger number of parameters $N_{par}=25$.
Fig.~\fig{dalitzchi2_nodcs} (left) shows the distribution of the $\chi^2$
values across the Dalitz plot. One can clearly see that there is a cluster of
high $\chi^2$ values corresponding to the mass of the
$K^{*}(892)^+$. Including now the DCS contribution, the fit gives a
significantly lower $\chi^2$
\be
\chi^2=1422,\quad \chi^2/N_{dof}=1.22
\en
and fig.~\fig{dalitzchi2_nodcs} (right) shows that, now, the high
$\chi^2$ values are isolated and randomly distributed across the
Dalitz plot. There are obviously other DCS contributions which must be
present: for instance $D^+\to K^*(892)^0\pip$ or $D^+\to K^0\rho^+$
but they cannot be isolated using only data on the 
$D^+\to K_S\piz\pip$ mode. In principle, they could be determined thanks to
the isospin symmetry, using an additional set of experimental data on
the mode $\Dp\to\Km\pip\pip$ (which has no DCS contributions) or using data on
$\Dp\to K_L\piz\pip$ decay.
Further resonances like $K^*_3(1780)$ or $\rho(1450)$ have
been considered in ref.~\cite{Ablikim:2014cea}, but we seem to find no need
for them. One can also see from fig.~\fig{dalitzchi2_nodcs} that the quality
of the fit is similar in the regions where rescattering is elastic and in the
regions where it is not. The quality of the fit and the role of the DCS
amplitudes are illustrated on fig.~\fig{dalitzslices} which shows slices of
the Dalitz plot with fixed values of the $\pi\pi$ invariant mass.  The values
of the 17 parameters which result from our fit are collected in
table~\ref{tab:Dpparamvals}. We note that the Cabibbo suppressed
parameter $|C_{DCS}/c_3|\simeq0.13$ is consistent  in size with
the value of the ratio of CKM parameters,
$|V_{cd}V_{us}/V_{cs}V_{ud}|=0.052\pm0.001$ .

\begin{table}[hbt]
\centering
\bt{lccc}\hline\hline
Parameter &   Modulus &  & Phase(radians) \\ \hline
$ F_{norm} $ & $8365.8\pm91.3$ &          &    -    \\
$ c_0      $ & $17.15 \pm0.78$ &          & $-2.63\pm 0.04$\\
$ c_1      $ & $21.95 \pm1.12$ & GeV$^{-2}$& $ 0.29\pm 0.05$\\
$ c_2      $ & $4.69  \pm0.27$ & GeV$^{-4}$& $-2.80\pm 0.06$\\
$ c_3      $ & $1$             & GeV$^{-4}$&   $\pi$         \\ 
$ c_4      $ & $3.64\pm0.27$   & GeV$^{-2}$& $-2.06\pm 0.07$ \\ 
$ c_5      $ & $1.90 \pm0.45$   & GeV$^{-4}$& $-0.72\pm 0.27$ \\ 
$ C_{K^*_2(1430)}$ & $0.031\pm0.003$ & GeV$^{-6}$& $0.13\pm 0.15$\\
$ C_{K^*(1680)}  $ & $1.91 \pm0.09$  & GeV$^{-2}$& $2.52\pm 0.03$\\
$ C_{DCS}        $ & $0.13 \pm0.01$ & GeV$^{-2}$ & $1.24\pm 0.09$\\ \hline\hline
\et
\caption{\small Values of the parameters entering the $D^+\to\bar{K}\pi\pi$
amplitudes resulting from the fit to the data from
  ref.~\cite{Ablikim:2014cea}.} 
\label{tab:Dpparamvals}
\end{table}

Fig.~\fig{FFplot} (left) illustrates the magnitudes of the $F$-functions
corresponding to the central values of the fitted parameters. The definitions
of these functions, we recall, assume the fixing conditions~\rf{condFH}. One
notices the rather large size of the $I=2$ amplitude $G_0^2$. These results
are similar to those obtained in ref.~\cite{Niecknig:2017ylb}. In particular,
the function $F_0^\undemi$ displays a large cusp at the $\pi{K}$ threshold and
has a first minimum close to 1 GeV followed by a maximum around 1.4 GeV. 
Using the parameters from table~\ref{tab:Dpparamvals} one can
evaluate the integral
\be
I_{A2}=\frac{1}{256\pi^3m^3_D}\int dsdt\,
\vert\CA_{D^+\to K_S\piz\pip}(s,t,u)\vert^2\ =(9.583\pm0.029)\cdot10^{-3}\ \hbox{GeV}
\en
from which the physical value of the normalisation factor $F_{norm}$
can be determined using~\cite{ParticleDataGroup:2022pth}
\be
|F_{norm}|^2
I_{A2}=\Gamma^{exp}_{D^+\to K_S\piz\pip}=(4.69\pm0.14)\cdot10^{-14}\ \hbox{GeV}\ ,
\en
which gives
\be\lbl{Fnormphys}
F_{norm}(phys.)=(2.212\pm0.031)\cdot10^{-6}\ .
\en
This determination of $F_{norm}(phys)$ is useful for discussing the soft pion
limits relations (see appendix~\sect{softpions}). A prediction can be
made for the decay width of the mode $\Dp\to\Km\pip\pip$. Introducing
the ratio of the widths
\be
{\cal R}\equiv\frac{\Gamma(D^+\to\Km\pip\pip)}{\Gamma(D^+\to K_S\piz\pip)}
\en
the model predicts
\be
{\cal R}^{model}= 1.235\pm 0.013
\en
which is compatible with the experimental value~\cite{ParticleDataGroup:2022pth}
\be
{\cal R}^{exp}= 1.274\pm0.042\ .
\en

\subsection{$D^0$ fits}
\begin{figure}[ht]
\centering
\includegraphics[width=0.45\linewidth]{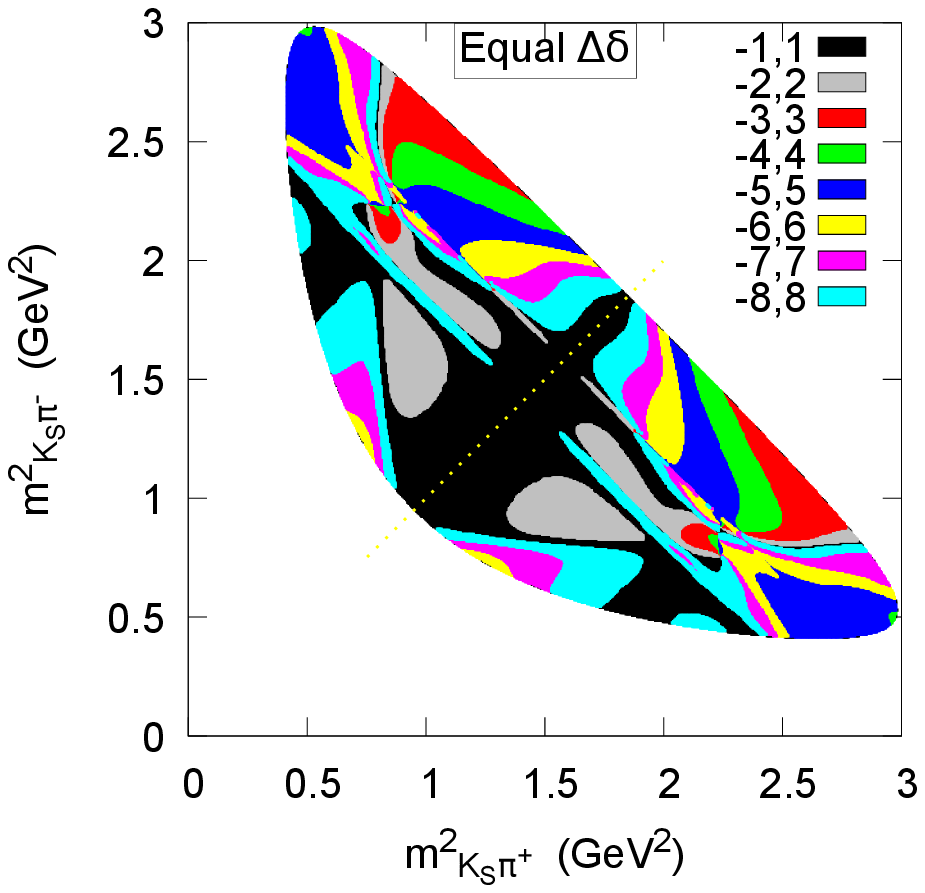}%
\includegraphics[width=0.45\linewidth]{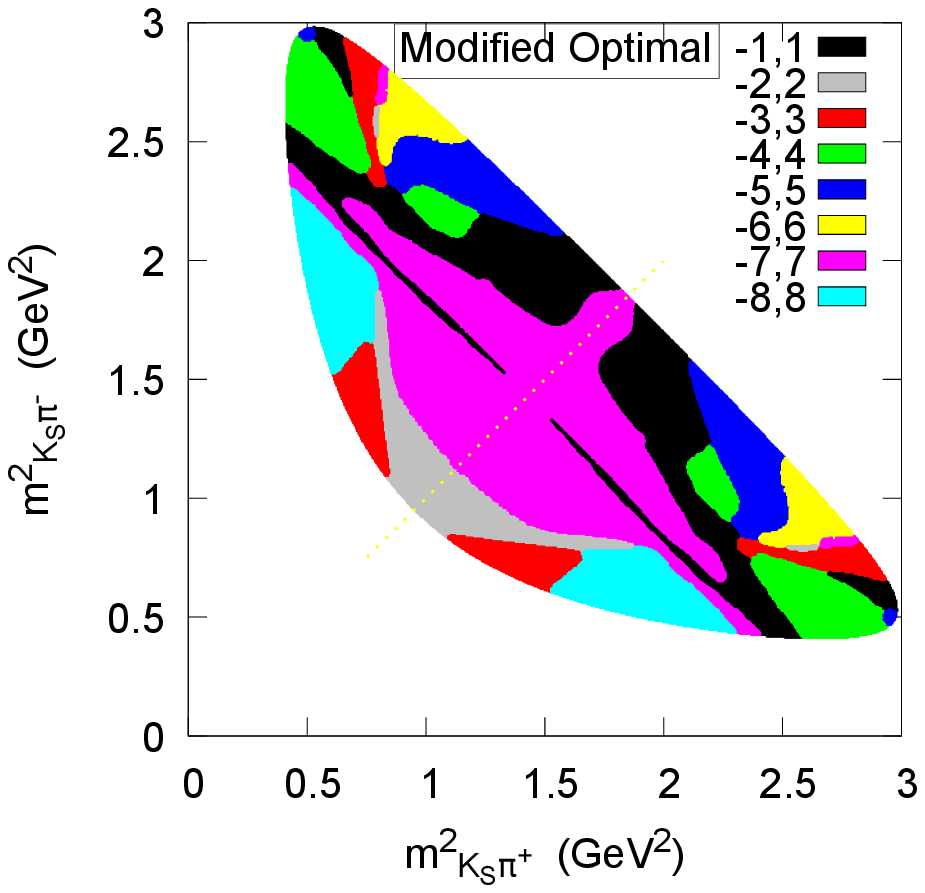}%
\caption{\small Illustration of two of the binnings used in the measurements
  of refs.~\cite{Libby:2010nu,BESIII:2020khq}. }
\label{fig:3binnings}
\end{figure}

From the $D^+$ fits discussed above, we have determined the parameters of the
$F$-functions, we now want to determine the analogous parameters of the
$H$-functions using $D^0$ decay inputs. Of the three possible decay modes, the
mode $D^0\to K_S\pim\pip$ has attracted much interest because of its
connection to the BPGGSZ~\cite{Giri:2003ty} approach for measuring the angle
$\gamma/\phi_3$.  Dalitz plot data for $D^0\to K_S\pim\pip$ on three different
binnings have been published~\cite{Libby:2010nu,BESIII:2020khq}, which we can
use for our purposes. The first binning is called ``Equal $\Delta\delta_D$'',
it is defined such that for $s,u$ in bin $i$ the phase difference
$\Delta\delta_D(s,u)$ (defined in eq.~\rf{Deltadeltadef}) lies between the two
values
\be\lbl{equaldeltaD}
\frac{2\pi}{8}(i-\frac{3}{2})\le \Delta\delta_D(s,u)<
\frac{2\pi}{8}(i-\frac{1}{2})\ ,
\en
with $i=1,\cdots,8$.
Fig.~\fig{3binnings} shows the shapes of the bins computed using the
Babar amplitude model~\cite{Aubert:2008bd}. The other two binnings are
called ``Optimal'' and ``Modified Optimal'' they have been generated
in ref.~\cite{Bondar:2007ir} from variations of the ``Equal
$\Delta\delta_D$'', binning such as to further optimise the determination of
$\gamma/\phi_3$ in the BPGGSZ method.\footnote{The description of these binnings
  is provided in the form of ``lookup tables'' containing 53842 equally-spaced
  points in which each point is attributed an index label. } The ``Modified
Optimal'' binning is also shown in fig.~\fig{3binnings}.

Refs.~\cite{Libby:2010nu,BESIII:2020khq} provide measures of three
different observables, $F_i$, $\bar{c}_i$, $\bar{s}_i$, in each bin. $F_i$, at
first, is the normalised number of events in bin $i$,
\be
F_i=\frac{N_i}{\sum_{-8}^8 N_i},\quad
N_i=\int_{bin_i} ds du\, \vert {\cal A}_D(s,t,u)\vert^2
\en
One notes that the sums
\be\lbl{S+S-}
S_-\equiv \sum_{i<0}F_i\simeq0.25,\quad
S_+\equiv \sum_{i>0}F_i\simeq0.75
\en
refer to halves of the Dalitz plot and are thus the same for the three
binnings.  The numerical values in eq.~\rf{S+S-}, taken
from~\cite{BESIII:2020khq}, indicate that the $K^*(892)^-$ resonance is
located in the region where the bins have a positive index $i>0$,
which must therefore satisfy $m^2_{K_S\pim} < m^2_{K_S\pip}$
, i.e. $s<u$ (contrary to what is stated in~\cite{BESIII:2020khq}).
The quantities $\bar{c}_i$, $\bar{s}_i$ are averages involving the cosine and the
sine of the phase difference $\Delta\delta_D$,
\be
\ba{l}
\bar{c}_i=\dfrac{\displaystyle{\int_{bin_i}} ds du\, \vert
{\cal A}_D(s,t,u) {\cal A}_D(u,t,s)\vert\,\cos\Delta\delta_D(s,u)}
{\sqrt{N_i N_{-i}}}\\[0.3cm]
\bar{s}_i=\dfrac{\displaystyle{\int_{bin_i}} ds du\, \vert
{\cal A}_D(s,t,u) {\cal A}_D(u,t,s)\vert\,\sin\Delta\delta_D(s,u)}
{\sqrt{N_i N_{-i}}}\ .
\ea\en
Using the Schwartz inequality
\be
\ba{l}
\left({\displaystyle\int_{bin_i}} ds du\, \vert
{\cal A}_D(s,t,u) {\cal A}_D(u,t,s)\vert\,\cos\Delta\delta_D(s,u)\right)^2
\le \\[0.3cm]
\quad\quad{\displaystyle\int_{bin_i}} ds du\, \vert {\cal A}_D(s,t,u)\vert^2
{\displaystyle\int_{bin_i}ds du} \vert {\cal A}_D(u,t,s)\vert^2
\cos^2\Delta\delta_D(s,u)\ ,
\ea\en
one easily deduces that $\bar{c}_i$ and $\bar{s}_i$ must lie inside a
circle~\cite{Bondar:2008hh} 
\be\lbl{cisibound}
\bar{c}_i^2 +\bar{s}_i^2 \le 1\ .
\en

As one can see from fig~\fig{3binnings}, each bin has an intricate
structure, extending over several regions of the Dalitz plot. In order
to address the experimental results it is necessary for the model to
be applicable in the whole Dalitz plot. As we did above, inelasticity
effects are simulated in the $\pi{K}$ and $\pi\pi$ attractive
$S$-waves by generating a phase in the inelastic region using a
two-channel Omn\`es matrix (see eq.~\rf{effOMdef}). We also add a few
resonance contributions to the KT amplitudes. In addition to the
$K^*(1680)$ and $K^*_2(1430)$ which were considered above, we now also
include the $f_2(1270)$ which couples to $(\pi\pi)_{I=0}$. One also
has to account for the isospin violating contribution from the
$\omega(782)$ coupling to $\pip\pim$.  We finally describe the DCS
amplitude as before via a $K^*(892)^+$ Breit-Wigner function. There are
clear indications that further contributions are required in this
amplitude (see below) but it is necessary to keep the number of
parameters to be as small as possible, due to the limited number of bins.
Not including them obviously introduces some bias in the
determination of the parameters of the CA amplitude. The $D^0\to K_S\pim\pip$
amplitude, finally, is described as follows,
\be\lbl{A6+reso}
\ba{l}
\CA_{D^0\to   K_S \pim \pip}(s,t,u)=  \dfrac{F_{norm}}{\sqrt2}\Big\{
\sqrt2(F_0^\trdemi(s)+Z_sF_1^\trdemi(s))\\
\quad-(H_0^\trdemi(s)+Z_sH_1^\trdemi(s))-3(H_0^\trdemi(u)+Z_uH_1^\trdemi(u))\\
\quad-\big[H_0^\undemi(s)+Z_s(H_1^\undemi(s)+D_{K^*(1680)}BW_{K^*(1680)}(s))\\
\quad
+Z_{2s} D_{K^*_2(1430)} BW_{K^*_2(1430)}(s)  \big]  
+\dfrac{1}{6}G_0^2(t)-G_0^0(t)-Z_{2t} D_{f_2(1270)} BW_{f_2(1270)}(t)\\
\quad-(s-u)\big[\sqrt2\tilde{G}_1^1(t)+ D_{\omega(892)} BW_{\omega(892)}(t)\big]\
+D_{DCS} Z_u BW_{K^*(892)}(u)  \Big\}\ 
\ea\en
where the angular function $Z_{2t}$ is given in eq.~\rf{Z2s}.
The physical value of $F_{norm}$ has been determined from $D^+$ decays and is
given in eq.~\rf{Fnormphys}.  One observes that the $F$-functions play a
relatively minor role in this amplitude since only 
$F_0^\trdemi$, $F_1^\trdemi$ and $G_0^2$ appear, which do not involve
resonances. For this reason, it is interesting to also consider other $D^0$
amplitudes which probe different interferences between the $F$ and the
$H$-functions. We will use below the experimental decay widths for the three
modes $K_S\pim\pip$, $\Km\piz\pip$ and $K_S\piz\piz$.

\begin{figure}[ht]
\centering
\includegraphics[width=0.45\linewidth]{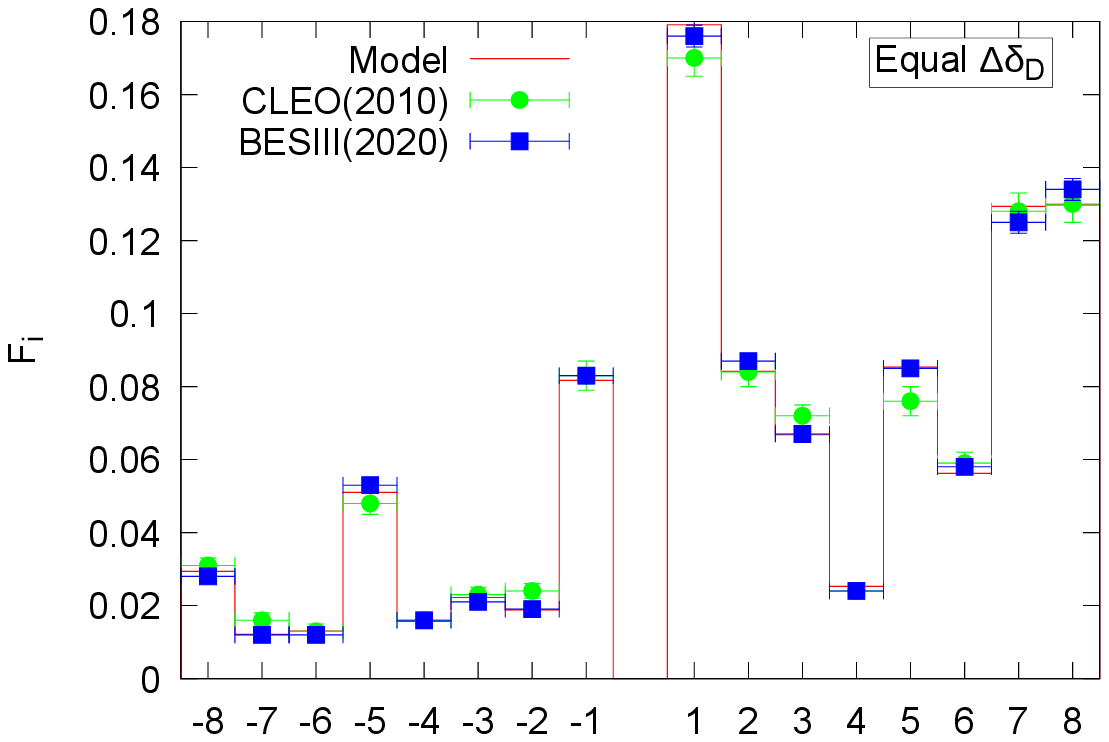}%
\includegraphics[width=0.45\linewidth]{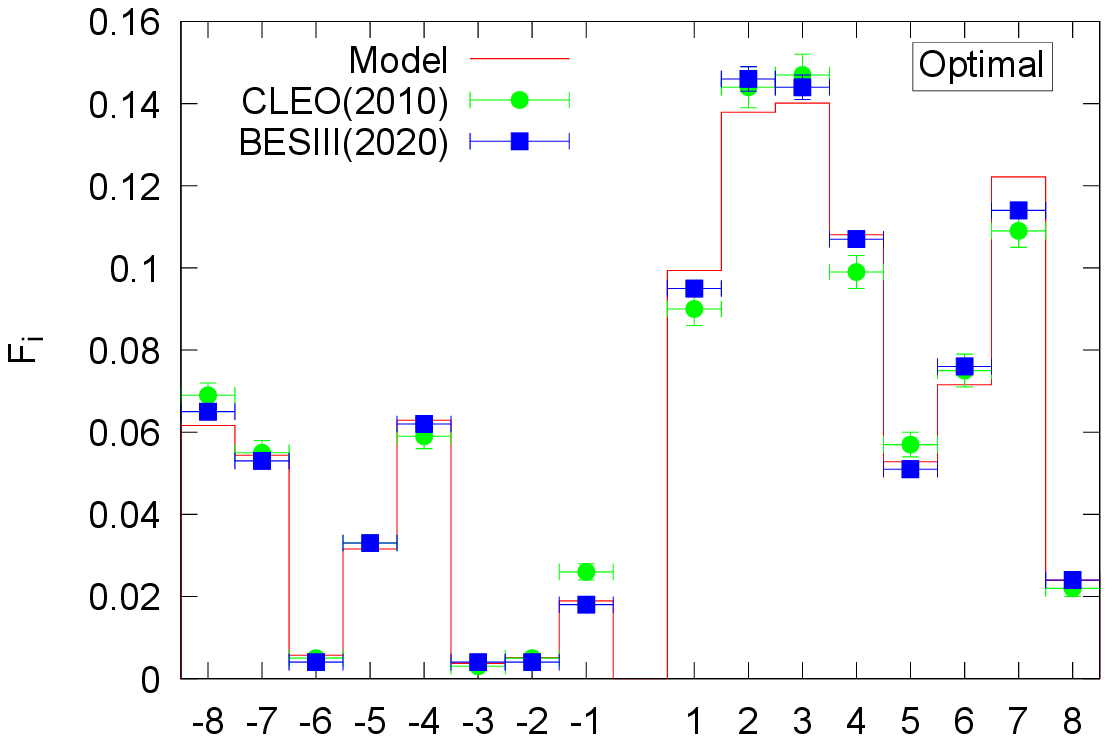}
\includegraphics[width=0.45\linewidth]{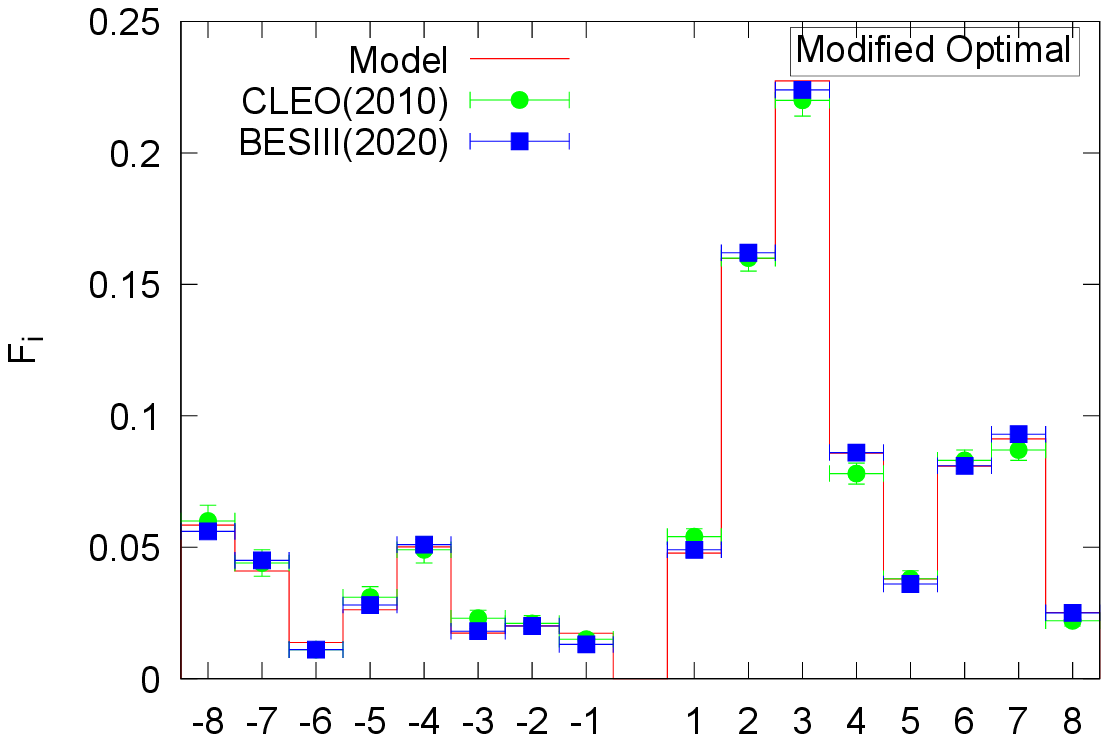}
\caption{\small Normalised number of events $F_i$ in each bin of three
  binnings. The experimental results are from
  ref.~\cite{BESIII:2020khq} (blue squares) and ref.~\cite{Libby:2010nu}
  (green circles). The horizontal error bars are only to show the
  extent of a bin. The results from our amplitude after fitting the 24
  parameters is represented by the red curve. }
\label{fig:compFi}
\end{figure}
\begin{figure}[htb]
\centering
\includegraphics[width=0.40\linewidth]{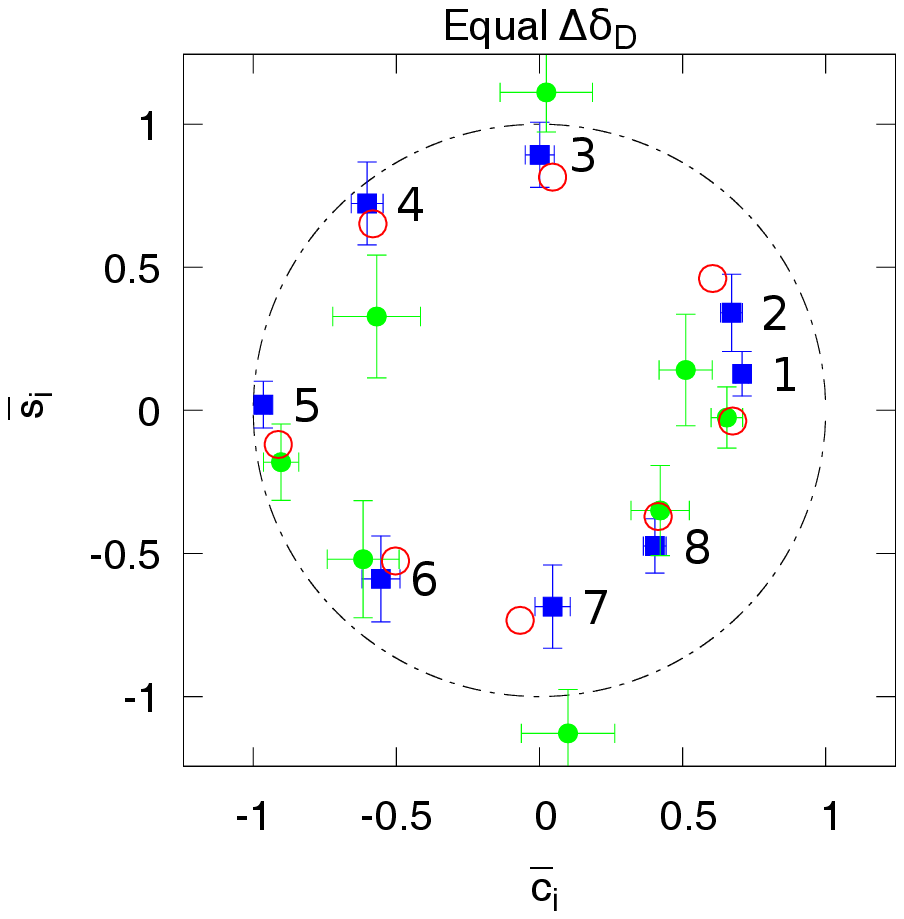}%
\includegraphics[width=0.40\linewidth]{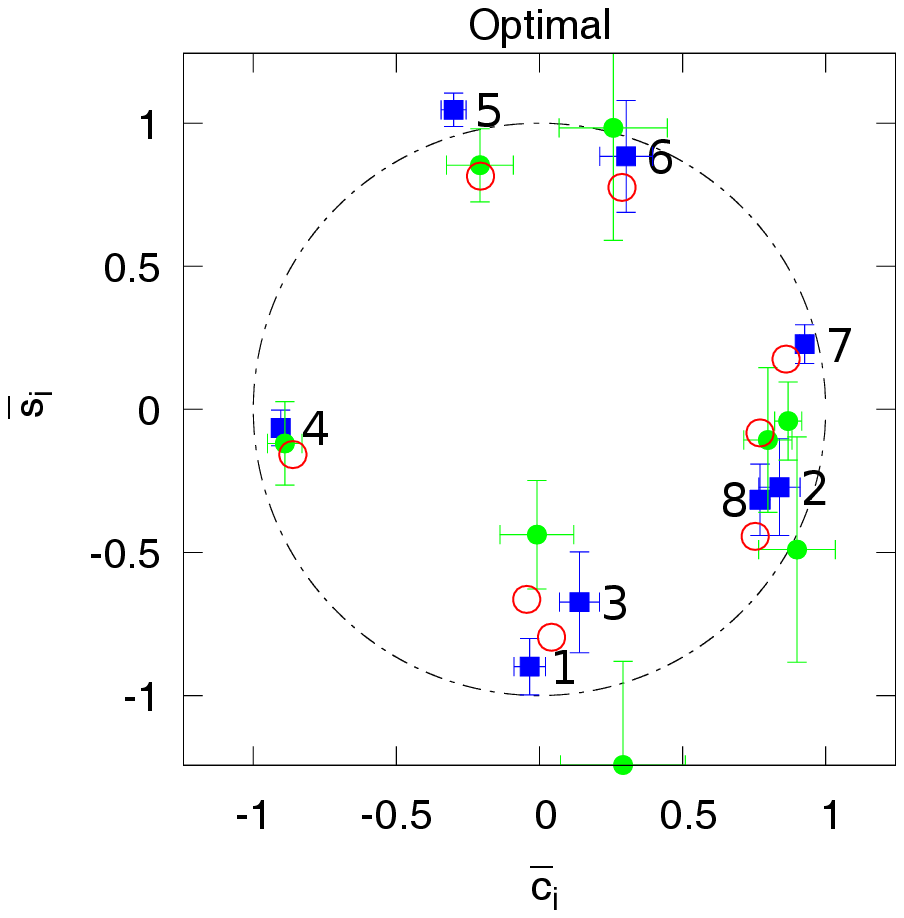}\\
\includegraphics[width=0.40\linewidth]{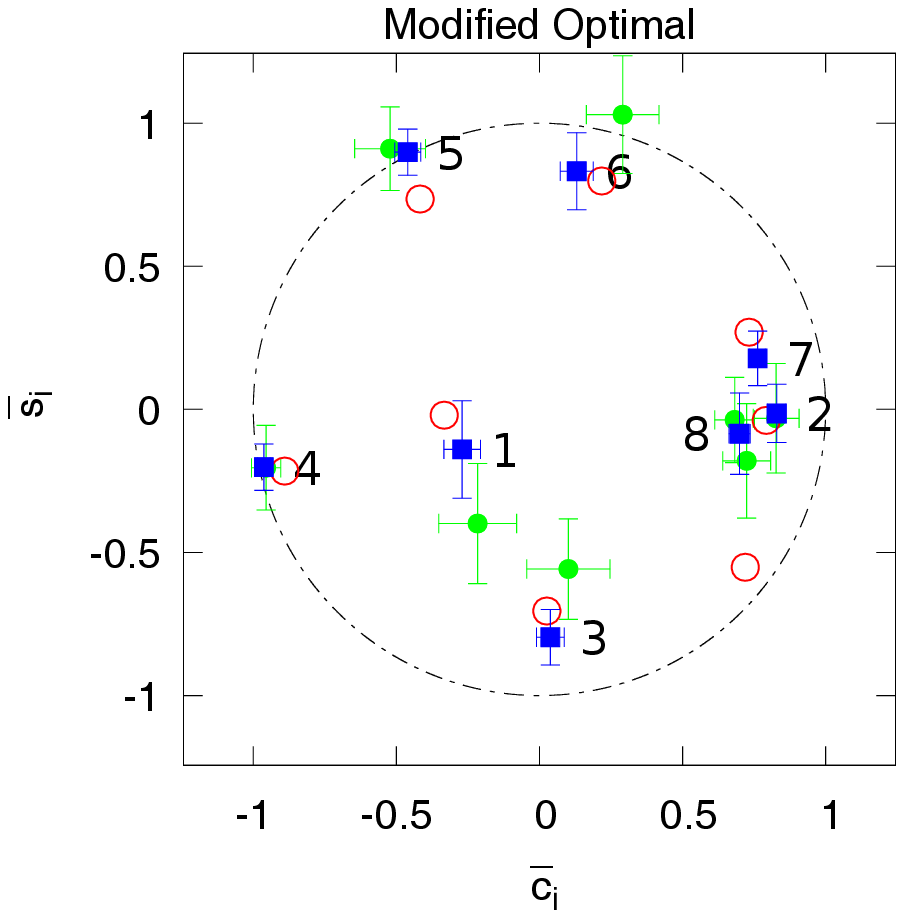}
\caption{\small Experimental values of the cosine and
  sine averages $\bar{c}_i$, $\bar{s}_i$
  (ref.~\cite{BESIII:2020khq}: green circles, ref.~\cite{Libby:2010nu}:
  blue squares) compared with the results from our amplitude (red
  circles). 
}
\label{fig:compcisi}
\end{figure}

The free parameters in the $D^0$ amplitudes are: the polynomial
coefficients $d_0,\cdots d_6$, the coefficients of the Breit-Wigner
functions $D_{K^*_2(1430)}$, $D_{K^*(1680)}$, $D_{f_2(1270)}$,
$D_{\omega(892)}$ and $D_{DCS}$. In total, we have 24 real parameters
to determine. For comparison, the Babar model~\cite{Aubert:2008bd} for
the $D^0\to K_S\pim\pip$ amplitude involves 43 parameters, while the
model used by Belle~\cite{Abe:2007rd} involves 40. The model proposed
in ref.~\cite{Dedonder:2014xpa}, which assumes a structure based on
naive factorisation, involves 33 free parameters. The parameters $\xi$
which influence the values of the $S$-wave phases in the inelastic
region are also tuned to improve the results. For $\pi{K}$ we take
$\xi_{\pi{K}}=0.515$ for the $F$-functions and $\xi_{\pi{K}}=0.5293$ for
the $H$-functions, while for $\pi\pi$ we take $\xi_{\pi\pi}=0.53025$.
We have included only the results from the BESIII
experiment~\cite{BESIII:2020khq} in the fit since they are compatible
with the previous results from CLEO~\cite{Libby:2010nu} and have
significantly smaller statistical uncertainties. 
We perform a fit by minimising a $\chi^2$ which includes:
\begin{itemize}
\item[a)] The 48 values $F_i$ of the
  number of events in  each bin $i$,
\item[b)] The 16 pairs of values $\bar{c}_i$, $\bar{s}_i$ of
the cosine and sine averages in the two binnings ``Equal $\Delta\delta_D$''
and ``Modified Optimal''. The ``Optimal'' binning was not
included in order not to give too much weight to these phase integrals and
because one pair $\bar{c}_i$, $\bar{s}_i$ fails to satisfy~\rf{cisibound}. 
\item[c)] The 3 experimental values of the $D^0$ decay
widths~\cite{ParticleDataGroup:2022pth},  
\be\lbl{D0largeurs}
\ba{l}
\Gamma^{exp}_{D^0\to K_S\pim\pip}=(4.49\pm0.29)\cdot10^{-14}\ \hbox{GeV}\\
\Gamma^{exp}_{D^0\to \Km\piz\pip}=(23.10\pm0.80)\cdot10^{-14}\ \hbox{GeV}\\
\Gamma^{exp}_{D^0\to K_S\piz\piz}=(1.46\pm 0.18)\cdot10^{-14}\ \hbox{GeV}\ .
\ea\en
\end{itemize}

\begin{figure}[htb]
\centering
\includegraphics[width=0.55\linewidth]{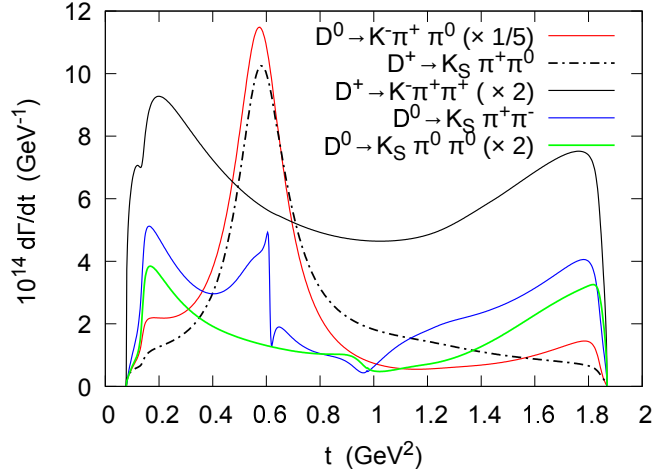}
\caption{\small Distributions of the widths as a function of the
  $\pi\pi$ energy squared, $d\Gamma/dt$, for the various 
  $D\to\Kbar\pi\pi$ decay modes as predicted from isospin symmetry and our
  fits to the two modes $\Dp\to K_S\piz\pip$ and $\Dz\to K_S\pim\pip$.}
\label{fig:Itn}
\end{figure}
The input experimental data are described in a qualitatively correct, but not
highly precise way, with a $\chi^2$ per-point value $\chi^2/N= 1.99$. These
results are illustrated in figs.~\fig{compFi} and~\fig{compcisi}
which show respectively the number of events $F_i$ and the averages
$\bar{c}_i$,$\bar{s}_i$. The agreement between the data and the model
is best in the case of the ``Equal $\Delta\delta_D$'' binning. The
fitted amplitude also reproduces the experimental values of the three
$D^0$ widths~\rf{D0largeurs} within their errors. The various
$D\to\Kbar\pi\pi$ widths show remarkable differences, e.g. the width of 
the $D^0\to K^-\piz\pip$ mode is 16 times larger than that of the 
$D^0\to K_S\piz\piz$ mode. These differences reflect the interference
patterns of the one-variable amplitudes as encoded in
eqs.~\rf{A1A2express}-\rf{A7express}. 
The reason for the large value of $\Gamma_{D^0\to K^-\piz\pip}$ is partly due
to the fact that two $I=1$ amplitudes add coherently in the combination
$G_1^1+2\TG$ which appears in the corresponding amplitude, generating a large
$\rho(770)$ peak. This is illustrated in fig.~\fig{Itn} which shows the
distribution $d\Gamma_{\Km\piz\pip}/dt$ (divided by 5) compared to the other
distributions. It is in qualitative agreement with the measurement by the CLEO
collaboration~\cite{Kopp:2000gv}.
The smallest width is that of
the $D^0\to K_S\piz\piz$ mode which is four times smaller than the
width of the mode $D^+\to \Km\pip\pip$. The main difference between
these two amplitudes is a significant suppression of the $I=1/2$
$S$-wave in the former due to a cancellation in the combination
$\sqrt2F_0^\undemi-H_0^\undemi$ which appears in eq.~\rf{A7express}. The
isobar-type model proposed in ref.~\cite{Lowrey:2011yd} for the
$K_S\piz\piz$ mode includes no $I=1/2$ $S$-wave contribution at all,
which is probably an oversimplification.  The distribution
$d\Gamma_{K_S\piz\piz}/dt$ (ignoring the DCS contribution as it cannot
be deduced from isospin symmetry) is also shown in fig.~\fig{Itn}. It
is qualitatively similar to the CLEO measurement~\cite{Lowrey:2011yd}.

\begin{table}[hbt]
\centering
\bt{lccc}\hline\hline
Parameter     &Modulus &           & Phase(radians) \\ \hline
$d_0$         &  $ 29.20\pm 4.25$     &           &  $ -1.96\pm 0.16$       \\
$d_1$         &  $ 42.10\pm 6.17$     & GeV$^{-2}$  &  $  1.07\pm 0.15$       \\
$d_2$         &  $  7.90\pm 1.03$     & GeV$^{-4}$  &  $ -1.58\pm 0.19$        \\
$d_3$         &  $  3.08\pm 0.09$     & GeV$^{-4}$  &  $ -1.36\pm 0.14$       \\
$d_4$         &  $ 12.85\pm 3.35$     & GeV$^{-4}$  &  $  2.63\pm 0.24$      \\
$d_5$         &  $  6.49\pm 0.91$     & GeV$^{-2}$  &  $ -2.34\pm 0.15$     \\
$d_6$         &  $ 12.80\pm 1.64$     & GeV$^{-4}$  &  $  0.98\pm 0.16$   \\
$D_{\omega(892)}$ &  $0.12\pm 0.04$      & GeV$^{-2}$  &  $-2.06 \pm 0.36$     \\
$D_{K^*_2(1430)}$ &  $0.13 \pm 0.02$     & GeV$^{-6}$  &  $ -0.78\pm 0.21$      \\
$D_{K^*(1680)}$  &  $ 5.41\pm 0.30$     & GeV$^{-2}$  &  $  0.17\pm 0.17$     \\
$D_{f_2(1270)} $ &  $ 1.95\pm 0.11$     & GeV$^{-4}$  &  $ -2.69\pm 0.17$      \\
$D_{DCS}$      &  $ 0.25\pm 0.03$      & GeV$^{-2}$  &  $ -1.33\pm 0.20$    \\ \hline\hline
\et
\caption{\small Numerical parameters in the amplitude $D^0\to
  K_S\pim\pip$ (see eq.~\rf{A6+reso}) resulting from the fit.} 
\label{table:D0params}
\end{table}
\subsection{Comparison of the $F$- and $H$-functions}
\begin{figure}[hbt]
\centering
\includegraphics[width=0.50\linewidth]{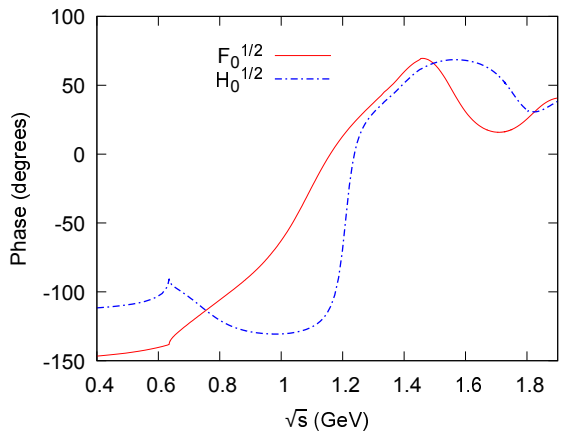}
\caption{\small Comparison of the phases of the $\pi{K}$ $S$-wave
  amplitudes $F_0^\undemi$ and $H_0^\undemi$.}
\label{fig:compFH2}
\end{figure}

The magnitudes of the $H$-functions
resulting from the fit (the numerical values of the corresponding
parameters are given in table~\ref{table:D0params}) are shown in
fig.~\fig{FFplot} (right). They can be compared to the magnitudes of the
fitted $F$-functions which are shown in the same figure (left). 
The $P$-wave function $H_1^\undemi$ is larger than the corresponding
$F_1^\undemi$ by a factor of four while the $S$-waves $F_0^\undemi$,
$H_0^\undemi$ are roughly comparable in magnitude in the physical region but
rather different in their shapes, in particular concerning the shapes of the
cusps at the $\pi{K}$ threshold.
Fig.~\fig{compFH2} compares
the phases of the $F_0^\undemi$ and $H_0^\undemi$ functions. While the phase
of $F_0^\undemi$ increases steadily in the low energy region (in agreement
with the phases determined through bin-by-bin measurements in
refs.~\cite{Aitala:2005yh,Bonvicini:2008jw,Link:2009ng}\footnote{The value
  $c_3=-1$ (see table~\ref{tab:Dpparamvals}) for the parameter appearing in
  the $P$-wave amplitude $F_1^{1/2}$ was chosen such as to correspond to the
  normalisation of the $K^*(892)$ Breit-Wigner function in these references.})
the phase of $H_0^\undemi$ is flat and slowly decreasing. Interestingly, these
different behaviours in the threshold region cannot be generated by the
polynomial terms in the KT representation (since they are smooth
functions). They must be caused by the rescattering integrals $\widehat{I}$
which have a cusp at the $\pi{K}$ threshold which combines with the cusp
present in the Omn\`es function $\Omega_0^\undemi$. This provides an
interpretation to the fact that, within isobar model descriptions, a strong
contribution from the broad $\kappa$ resonance is required in $D^+$
amplitudes~\cite{E791:2002xlc,FOCUS:2007mcb,Bonvicini:2008jw,Ablikim:2014cea}
but not in $D^0$ amplitudes (e.g.~\cite{CLEO:2002uvu}).

A remark is in order, finally, concerning the DCS amplitude for which we have
used a very crude description involving simply the $K^*(892)^+$
resonance. While the DCS amplitude is small (compared to the Cabibbo favoured
one) it can significantly affect the phase of the overall amplitude. Previous
fits~\cite{Poluektov:2006ia,Aubert:2008bd,BaBar:2018cka} of the $K_S\pim\pip$
mode have included several additional contributions from $K_0^*(1430)^+$,
$K_2^*(1430)^+$, $K^*(1680)^+$,... resonances. The DCS operator can also
couple to states like $\Kz\rho^0(770)$, $\Kz f_0(980)$... which cannot be
distinguished from $\Kzb\rho^0(770)$, $\Kzb f_0^0(980)$ unless measurements of
both $K_S\pim\pip$ and $K_L\pim\pip$ are performed.  We find clear indications
that additional contributions are necessary since, with our crude DCS model,
we cannot reproduce the experimental results of both the $K_S\pip\pim$ and the
$K_L\pip\pim$ modes which are reported in
refs.~\cite{Libby:2010nu,BESIII:2020khq} and we have thus focused only on the
$K_S\pim\pip$ mode. As mentioned above in the case of the $D^+$, making use of
the isospin symmetry and inputs from a mode without DCS contributions
(i.e. the $\Km\piz\pip$ mode) would allow a more complete determination of the
DCS amplitude in the $K_S\pim\pip$ mode.

\section{Conclusions}
The main result of our work is contained in the
formulas~\rf{A1A2express} to~\rf{A7express} which express all the
$D\to\Kbar\pi\pi$ amplitudes in terms of 6 $F$-functions plus 6
$H$-functions which depend on one of the Mandelstam variables. These formulas
derive from the fact that the relevant weak Hamiltonian for these
decays is a pure isospin 1 operator and from the reconstruction theorem
limited to $j=0,1$ partial-waves. The formulas apply to isobar-type models as
well and, in that framework, can be easily generalised to contain $j\ge 2$
resonances. Based on this representation, we have derived two sets of coupled
Khuri-Treiman equations, which generalise the previous results of
refs.~\cite{Niecknig:2015ija,Niecknig:2017ylb}.  Solving (numerically) these
integral equations, the two sets of $F$- and $H$-functions get determined as
linear combinations in terms of respectively 6 and 13 complex parameters.

In order to enlarge the region of applicability of the results two additional
parameters are introduced which control the behaviour of the $\pi{K}$ and
$\pi\pi$ $S$-wave phases in the region where scattering is no longer elastic,
allowing to simulate e.g. some cusp effects. As a check of this procedure, we
have considered the soft pion relations which constrain the amplitudes at the
two points $s,u=(\mkd,\mdd)$ and $(\mdd,\mkd)$ and are sensitive to these
parameters. A description of the complete Dalitz plot region requires further
small contributions from higher spin or mass resonances and, eventually, from
Cabibbo suppressed amplitudes: we describe these in a simple isobar-type
model.

We have illustrated some applications of these results by performing a
combined fit of a set of data for the $\Dp\to K_S\piz\pip$ mode and a set of
data for the $D^0\to K_S\pim\pip$ mode.
A rather good description of the $D^+$ data can be obtained in
this manner, clearly showing the presence of  the small DCS
contribution. Concerning the $D^0$, the available data refer to binnings with
only eight bins. We have then used a limited number of parameters in
particular in the description of the DCS amplitude. These data are
nevertheless rather constraining as they concern the behaviour of both the
modulus and the phase of the amplitude in these bins. The obtained
$\chi^2/N\simeq1.99$ is moderately good but the estimates of the $H$-functions
seem already reasonable. This is illustrated in the result for the energy
distributions of the widths, shown in  fig.~\fig{Itn}, which indicate correct
interference patterns between the $F$ and the $H$ functions.
An interesting difference between the two $S$-wave functions $F_0^{1/2}$ and
$H_0^{1/2}$ can be observed near the threshold, see fig.~\fig{compFH2} which
shows the phases. This difference illustrates a clear three-body
rescattering effect, which is accounted for in the KT formalism.

Our results allow one to introduce constraints in phenomenological models by
performing combined fits of several $D$ decay modes. Such fits have been
performed in the past~\cite{MARK-III:1987qok,E691:1992rwf,E687:1994wlh}
without, however, making use of the isospin symmetry relations. The modes with
a $K_S$ (or a $K_L$) receive a small DCS contribution. Making use of isospin
symmetry in the evaluation of the Cabibbo favoured component would enable one
to determine this DCS part more completely than has been possible so far. As
more data become available for DCS
amplitudes~\cite{LHCb:2018xff,BESIII:2022chs} one could also make use of
isospin symmetry in this sector. 

\vspace{0.8cm}
\noindent{\Large\bf Acknowledgements:} We would like to thank
Prof. Chengdong Fu for sending us the experimental results of the
BESIII publication~\cite{Ablikim:2014cea} and Prof. Andrzej Kupsc for
his efficient help. We thank Prof. Jim Libby for correspondence and
explanations and for sending us the lookup tables of the Babar model.
This work is supported in part by the European Union's Horizon2020
research and innovation programme (HADRON-2020) under the grant
agreement $\hbox{N}^\circ$ 824093.

\appendix
\section{Soft pion limits}\label{sec:softpions}
Application of the soft pion theorems (e.g.~\cite{Donoghue:1992dd},
chap. IV) to the $D\to\Kbar\pi\pi$ amplitudes relates these to $D\to
\Kbar\pi$ amplitudes at the kinematical points $s,u=(m_D^2,m_K^2)$ or
$(m_K^2,m_D^2)$ which provides qualitative but interesting constraints
on the scalar $\Kbar\pi$ component of the $D\to\Kbar\pi\pi$
amplitudes~\cite{Gronau:1999zt}.  The soft pion
relations have the following form
\be\lbl{genericsoftpi}
\lim_{p=0}\braque{\Kbar\pi^b\pi^a(p)\vert H_W(0)\vert D}=
-\frac{i}{\fpi}\braque{\Kbar\pi^b\vert [Q_5^a,H_W(0)]\vert D}\ 
\en
(analogous to those for $K\to 3\pi$~\cite{Suzuki:1966zz}) where $a,b$
are isospin indices, $Q_5^a$ is an axial charge and $F_\pi\simeq
92.2$ MeV is the pion decay constant. Isospin symmetry allows one to
express the three physical $D\to\Kbar\pi$ amplitudes in terms of two
isospin amplitudes $A^{1/2}$, $A^{3/2}$~\cite{Bauer:1986bm},
\be\lbl{DKpiisodecomp}
\ba{ll}
A(D^+ \to \Kzb\pip)&=   A^{3/2}\\[0.1cm]
A(D^0 \to \Km\pip) &=  \frac{1}{3}\left(A^{3/2}+ 2A^{1/2}\right)\\[0.1cm]
A(D^0 \to \Kzb\piz)&=  \frac{\sqrt2}{3}\left(A^{3/2}-A^{1/2}\right)\ 
\ea\en
and using the experimental results~\cite{ParticleDataGroup:2022pth}
\be
\ba{l}
\Gamma(D^+ \to \Kzb\pip)=(1.93\pm0.04)\cdot10^{-14},\\
\Gamma(D^0 \to \Km\pip) =(6.33\pm0.05)\cdot10^{-14},\\
\Gamma(D^0 \to \Kzb\piz)=(3.59\pm0.12)\cdot10^{-14}
\ea\en
(all in GeV) one deduces from~\rf{DKpiisodecomp} the absolute values of the
two  isospin amplitudes 
\be
|A^{3/2}|=(1.40\pm0.01)\,10^{-6}\ \hbox{(GeV)},\quad
|A^{1/2}|=(3.76\pm0.03)\,10^{-6}\ \hbox{(GeV)}
\en
and the value of the phase difference
\be
\phi^{1/2}-\phi^{3/2}=\pm(91.8\pm2.0)^{\circ}\ .
\en
Considering all the $D\to\Kbar\pi_1\pi_2$ amplitudes, one can write three
relations of the type~\rf{genericsoftpi} corresponding to a soft $\piz$, four
relations corresponding to a soft $\pip$ and one relation with a soft
$\pim$. The latter relation is special because the combination which appears
on the right-hand side
\be\lbl{AdlerZero}
\ba{ll}
\lim_{p=0}\braque{\Kzb\pim(p)\pip\vert H_W\vert D^0}= &
\dfrac{1}{\sqrt2\fpi}\Big(A(D^+\to \Kzb\pip)\\[0.2cm]
\ & -A(D^0\to\Km\pip)-\sqrt2 A(D^0\to\Kzb\piz)\Big)=0\ 
\ea\en
vanishes upon using the isospin relations~\rf{genericsoftpi}. This
corresponds to a zero of the amplitude $A_{\Kzb\pim\pip}$ at the point $s=\mkd$,
$u=\mdd$. This is the only soft pion zero. For instance, for the same
amplitude at the symmetric point $s=\mdd$, $u=\mkd$ one obtains
\be
\lim _{p=0}\braque{\Kzb\pim\pip(p)\vert H_W\vert D^0}= \frac{i}{F_\pi}
A(D^0\to \Kzb\piz)\ .
\en
\begin{table}
\centering  
\bt{c|ccc}\hline\hline
\rule{0pt}{4ex} & $10^5\dfrac{|A^\undemi|}{\fpi}$ & $10^5\dfrac{|A^\trdemi|}{\fpi}$ &
$\phi^\undemi-\phi^\trdemi$ \\[1.5ex] \hline
\TT Exp.                & 4.08(3)   & 1.52(1)    & $\pm91.8(2.0)^\circ $  \\ \hline
\TT\BB$\xi=0.515$ & 2.62     & 0.98      & $+98.5^\circ $ \\ \hline
\TT\BB$\xi=0.5293$& 3.99     & 1.29      & $-96.5^\circ $ \\ \hline\hline
\et
\caption{\small  Test of the two soft pion relations~\rf{softrels1} as
a function of the parameter $\xi$ which tunes the
value of the $\pi{K}$ phase in the inelastic region according
to~\rf{effOMdef}. For a given value of $\xi$, the polynomial
parameters $c_i$ are derived from a best fit to the data. }
\label{table:softpions}
\end{table} 

Only four (out of the eight) soft-pion relations are independent. One can
first write two relations which involve only the $F$-functions
\be\lbl{softrels1}
\ba{l}
\dfrac{iA^\undemi}{\fpi}= -3F_0^\undemi(\mdd)-2F_0^\undemi(\mkd)
-5F_0^\trdemi(\mkd)+\dfrac{5\sqrt2}{4}G_0^2(2\mpid)
-\Delta_{DK}G_1^1(2\mpid)\\[0.3cm]
\dfrac{iA^\trdemi}{\fpi}=  3F_0^{3/2}(\mdd) 
-2F_0^{3/2}(\mkd)+F_0^{1/2}(\mkd) -\dfrac{\sqrt2}{4}G_0^2(2\mpid)
-\Delta_{DK}G_1^1(2\mpid)\ .
\ea\en
In writing these relations, we have dropped all contributions which are
$O(\mpid)$, in particular, those involving the angular factors $Z_s$, $Z_u$.
The right-hand sides of the relations~\rf{softrels1} are sensitive to the
value of the parameter $\xi$ which we introduced for tuning the $I=1/2$ $\pi
K$ $S$-wave phase above the inelastic threshold.  Table~\ref{table:softpions}
illustrates that taking a value of the $\xi$ parameter slightly larger than
0.5, the right orders of magnitudes can be obtained for the soft pion
relations~\rf{softrels1}. It is possible to fine-tune the value of $\xi$ such
that they are satisfied within 15\%.  The last two independent soft pion
relations can be written as
\be\lbl{softrels2}
\ba{ll}
H_0^{1/2}(\mdd)= & -\sqrt2\left(F_0^{1/2}(\mdd) 
                  +F_0^{1/2}(\mkd)
                  +\frac{4}{3}F_0^{3/2}(\mkd)\right)
                  +\frac{1}{3}H_0^{1/2}(\mkd)\\[0.2cm]
\ &              -\frac{8}{3}H_0^{3/2}(\mkd)       
                 +\frac{10}{9}G_0^2(2\mpid)  
             -\frac{2}{3}G_0^0(2\mpid)
             -\frac{4\sqrt2}{3}\Delta_{DK}\widetilde{G}_1^1(2\mpid)\\[0.2cm]
H_0^{3/2}(\mdd)= &\frac{\sqrt2}{3}F_0^{3/2}(\mkd)    
             -\frac{1}{3}H_0^{1/2}(\mkd)-\frac{1}{3}H_0^{3/2}(\mkd)\\[0.2cm]
\ &          +\frac{1}{18}G_0^2(2\mpid)-\frac{1}{3}G_0^0(2\mpid)
+\frac{\sqrt2}{3}\Delta_{DK}\widetilde{G}_1^1(2\mpid)\ .
\ea\en
Setting the value of $\xi$ as in the last line of table~\ref{table:softpions}
the first relation in eq.~\rf{softrels2} is satisfied within $\simeq50\%$, the
left-hand side equalling $(8.6+i5.9)\,10^{-5}$ and the right-hand side:
$(5.1+i3.8)\,10^{-5}$.  The second relation fails to be satisfied even
qualitatively. In that case, the left-hand side equals $(2.4-i0.6)\,10^{-6}$ and
the right-hand side $(-9.8+i1.8)\,10^{-6}$, differing in their signs and by a
factor of 4 in magnitude. This indicates that our evaluation of $H_0^{3/2}$ in
the inelastic region is not quite correct. That relation on $H_0^{3/2}(m_D^2)$
also insures the presence of the zero in $A_{\Kzb\pim\pip}$ at $s=\mkd$,
$u=m_D^2$ (see eq.\rf{AdlerZero}). Our solution amplitude is thus not
sufficiently suppressed at this point.

\section{Details of the expansions with isospin
  amplitudes}\label{sec:isodetails} 
\begin{table}
\centering  
\bt{c|cccc}
\       & $\CF^\trtr$ & $\CF^\untr$ & $\CF^\trun$ & $\CF^\unun$ \\ \hline
$\CA_1$ &$-\frac{2\sqrt2}{3\sqrt{15}}$ & $\frac{1}{3\sqrt3}$   &
           $ -\frac{\sqrt2}{3\sqrt3}$  & $\frac{2\sqrt2}{3\sqrt3}$   \\[0.15cm]
$\CA_2$&  $-\frac{4}{3\sqrt{15}}$      & $\frac{2}{3\sqrt6}$ &
         +$\frac{1}{3\sqrt3}$  & $-\frac{2}{3\sqrt3}$ \\[0.15cm]
$\CA_3$& $\frac{2}{\sqrt{15}}$ & $-\frac{1}{\sqrt6}$   & 0 & 0 \\[0.15cm]
$\CA_4$& $-\frac{2}{\sqrt{15}}$ & 0 & $-\frac{1}{\sqrt3}$ & 0 \\[0.15cm]
$\CA_5$& $\frac{4}{3\sqrt{15}}$ & +$\frac{1}{3\sqrt6}$ &
         +$\frac{2}{3\sqrt3}$ &   +$\frac{2}{3\sqrt3}$ \\[0.15cm]
$\CA_6$& $\frac{\sqrt2}{\sqrt{15}}$ & 0 & $-\frac{\sqrt2}{\sqrt3}$ & 0\\[0.15cm]
$\CA_7$& $\frac{4\sqrt2}{3\sqrt{15}}$ & +$\frac{1}{3\sqrt3}$ &
      $-\frac{\sqrt2}{3\sqrt3}$ & $-\frac{\sqrt2}{3\sqrt3}$ \\[0.15cm]
$\CA_8$& $\frac{\sqrt2}{\sqrt{15}}$ & +$\frac{1}{\sqrt3}$ & 0 & 0 \\ \hline
\et\hspace{1cm}  
\bt{c|cccc}
\ &   $\CG^{10}$ & $\CG^{12}$ & $\CG^{01}$ & $\CG^{11}$ \\ \hline
$\tilde{\CA}_1$  & 0 & $\frac{\sqrt3}{\sqrt5}$ & 0 & 0 \\[0.15cm]           
$\tilde{\CA}_2$ & 0 & $-\frac{\sqrt3}{2\sqrt{10}}$ &
                      $-\frac{1}{2\sqrt3}$        &
                      $+\frac{1}{2\sqrt2}$         \\[0.15cm]         
$\tilde{\CA}_3$ & 0 &$-\frac{\sqrt3}{2\sqrt{10}}$ &
                      $+\frac{1}{2\sqrt3}$        &
                      $-\frac{1}{2\sqrt2}$        \\[0.15cm]         
$\tilde{\CA}_4$ & 0 &$\frac{\sqrt3}{2\sqrt{10}}$  &
                      $-\frac{1}{2\sqrt3}$        &
                      $-\frac{1}{2\sqrt2}$        \\[0.15cm]     
$\tilde{\CA}_5$ & 0 &$\frac{\sqrt3}{2\sqrt{10}}$  &
                     $+\frac{1}{2\sqrt3}$         &
                     $+\frac{1}{2\sqrt2}$         \\[0.15cm]     
$\tilde{\CA}_6$ &   $\frac{1}{\sqrt3}$            &
             $+\frac{1}{\sqrt{60}}$   & 0         &
                    $ -\frac{1}{2}$               \\[0.15cm]     
$\tilde{\CA}_7$ &$\frac{1}{\sqrt3}$   & $-\frac{1}{\sqrt{15}}$
                                    & 0 & 0\\[0.15cm]     
$\tilde{\CA}_8$ &   $\frac{1}{\sqrt3}$      &
             $+\frac{1}{\sqrt{60}}$ & 0 &
                     $+\frac{1}{2}$          \\  \hline
\et
\caption{\small Coefficients of the linear expansions of the
physical amplitudes $\CA_i$ and $\tilde{\CA}_i$ (defined
in~\rf{calAidef}~\rf{tildecalAidef}) in terms of isospin amplitudes.}          
\label{table:isoexpress}
\end{table}  
\subsection{General relations}
Let us consider $2\to 2$ scattering amplitudes induced by the weak Hamiltonian
$H_W$: $\CA(s,t,u)=\braque{cd\vert H_W\vert ab}$ with $s=(p_a+p_b)^2$,
$t=(p_b-p_d)^2$, $u=(p_a-p_d)^2$.  We start by forming 8 scattering amplitudes
of the type $\braque{D\pi_2\vert H_W\vert \bar{K}\pi_1}$ with all possible
charges of $\pi_1$, $\pi_2$ and $\Kbar$
\be\lbl{calAidef}
\ba{ll}
\CA_1=\braque{D^+\pim \vert H_W\vert \Km\pip}  &
             \CA_5=\braque{D^0\piz \vert H_W\vert \Km\pip}  \\
\CA_2=\braque{D^+\pim \vert H_W\vert \Kzb\piz} &
             \CA_6=\braque{D^0\pim \vert H_W\vert \Kzb\pim}\\
\CA_3=\braque{D^+\piz \vert H_W\vert \Kzb\pip} &
             \CA_7=\braque{D^0\piz \vert H_W\vert \Kzb\piz}\\
\CA_4=\braque{D^0\pim \vert H_W\vert \Km\piz}  &     
             \CA_8=\braque{D^0\pip \vert H_W\vert \Kzb\pip}\\
\ea\en
and then 8 scattering amplitudes of the type
$\braque{D K\vert H_W\vert \pi_1 \pi_2}$
\be\lbl{tildecalAidef}
\ba{ll}
\tilde{\CA}_1=\braque{D^+ \Kp \vert H_W\vert \pip\pip}  &
             \tilde{\CA}_5=\braque{D^0\Kp\vert H_W\vert\piz\pip}  \\
\tilde{\CA}_2=\braque{D^+ \Kz \vert H_W\vert\pip \piz} &
             \tilde{\CA}_6=\braque{D^0\Kz \vert H_W\vert\pip\pim}\\
\tilde{\CA}_3=\braque{D^+\Kz \vert H_W\vert\piz \pip} &
             \tilde{\CA}_7=\braque{D^0\Kz\vert H_W\vert\piz\piz}\\
\tilde{\CA}_4=\braque{D^0\Kp \vert H_W\vert\pip \piz}  &     
             \tilde{\CA}_8=\braque{D^0\Kz \vert H_W\vert\pim\pip}\ .
\ea\en
Next, using the following conventional isospin assignments
\be
\begin{pmatrix}
\Kp\\
\Kz\\
\end{pmatrix}\sim 
\begin{pmatrix}
\ket{\unun}\\[0.2cm]
\ket{\unmun}\\
\end{pmatrix},\ 
\begin{pmatrix}
\Kzb\\
\Km\\
\end{pmatrix}\sim 
\begin{pmatrix}
\ket{\unun}\\[0.2cm]
-\ket{\unmun}\\
\end{pmatrix},\ 
\begin{pmatrix}
\Dp\\
\Dz\\
\end{pmatrix}\sim 
\begin{pmatrix}
\ket{\unun}\\[0.2cm]
-\ket{\unmun}\\
\end{pmatrix}
\en
and
\be
\begin{pmatrix}
\pip\\
\piz\\
\pim\\
\end{pmatrix}\sim
\begin{pmatrix}
-\ket{11}\\
\ket{10}\\
\ket{1,\!-\!1}\\
\end{pmatrix}
\en
one can expand the two particle states in terms of isospin states and then
apply the Wigner-Eckart theorem~\rf{wignereckart}. The resulting coefficients
of the expressions of the physical amplitudes $\CA_i$ in terms of the isospin
amplitudes ${\cal F}^{KK'}$ and those of $\tilde{\CA}_i$ in terms of ${\cal
  G}^{II'}$ (see~\rf{isoamplit}) are collected in
table~\ref{table:isoexpress}.
\begin{table}
\centering  
\bt{c|cccc}
\    & $\CF^\trdemi$ & $\CF^\undemi$ & $\CH^\trdemi$ & $\CH^\undemi$ \\ \hline
\TT$\CA_1$ & $-\sqrt2$ & $-\sqrt2$ & 0  & 0 \\[0.15cm]
$\CA_2$ & $-2$      & $1$       & 0  & 0 \\[0.15cm]
$\CA_3$ & 3         & 0         & 0  & 0 \\[0.15cm]
$\CA_4$ & $-2$      & 0         & $\sqrt2$  & $-\frac{1}{\sqrt2}$\\[0.15cm]
$\CA_5$ & 1         &$-1$       & $-\sqrt2$ & $\frac{1}{\sqrt2}$\\[0.15cm]
$\CA_6$ & $\sqrt2$  &0          &$-1$       & $-1$  \\[0.15cm]
$\CA_7$ & $\sqrt2$  &$\frac{1}{\sqrt2}$&$-2$& $-\frac{1}{2}$ \\[0.15cm]
$\CA_8$ & 0         &0          &$-3$       & 0     \\ \hline
\et\hspace{1cm}  
\bt{c|cccc}
\    & $\CG^0$   & $\CG^2$    & $\CG^1$    & $\tilde{\CG}^1$ \\ \hline
\TT$\tilde{\CA}_1$ & 0 & 1 & 0 & 0 \\[0.15cm]
$\tilde{\CA}_2$ & 0 & $-\frac{\sqrt2}{4}$ & 1 & 0  \\[0.15cm]
$\tilde{\CA}_3$ & 0 & $-\frac{\sqrt2}{4}$ & $-1$ & 0  \\[0.15cm]
$\tilde{\CA}_4$ & 0 & $-\frac{\sqrt2}{4}$ & 1 & $-2$  \\[0.15cm]
$\tilde{\CA}_5$ & 0 & $\frac{\sqrt2}{4}$ & $-1$ & $2$ \\[0.15cm]
$\tilde{\CA}_6$ &$-1$& $\frac{1}{6}$  & 0 & $-\sqrt2$\\[0.15cm]  
$\tilde{\CA}_7$ &$-1$& $-\frac{1}{3}$ & 0 & 0        \\[0.15cm]  
$\tilde{\CA}_8$ &$-1$& $\frac{1}{6}$  & 0 & $\sqrt2$ \\ \hline
\et
\caption{\small  Expansion coefficients of the physical amplitudes $\CA_i$ and
$\tilde{\CA}_i$ in terms of the new isospin amplitudes.}
\label{table:newisoexpress}
\end{table}  

The expressions of the $D^+$ amplitudes can be further simplified if we assume
that the effects of the $D\pi$ or $DK$ interactions can be neglected in the
energy region of interest. This allows us to combine amplitudes with different
isospins of $D\pi$ or $DK$. In the channels $\braque{D\pi_i\vert
  H_W\vert\Kbar\pi_j}$ we introduce the following linear combinations and keep
a unique isospin label referring to the $K\pi$ system
\be
\bp
\CF^\trdemi\\
\CF^\undemi\\
\CH^\trdemi\\
\CH^\undemi\\
\ep\equiv\bp
\frac{2}{3\sqrt{15}}& -\frac{1}{3\sqrt6} & 0 & 0\\
 0                  & 0   & \frac{1}{3\sqrt3} & -\frac{2}{3\sqrt3}\\
-\frac{\sqrt2}{3\sqrt{15}}&-\frac{1}{3\sqrt3} & 0 & 0\\
0 & 0 & \frac{\sqrt2}{\sqrt3} & 0 \\   
\ep
\bp
\CF^{\trtr}\\
\CF^{\untr}\\
\CF^{\trun}\\
\CF^{\unun}\\
\ep\ .
\en
Similarly, in the channels $\braque{DK\vert H_W\vert\pi_i\pi_j}$ we introduce
four combinations labelled with the isospin of the pion pair,
\be
\bp
\CG^0\\
\CG^2\\
\CG^1\\
\tilde{\CG}^1\\
\ep\equiv\bp
-\frac{1}{\sqrt3} & 0 & 0 & 0 \\
0 &\frac{\sqrt3}{\sqrt5} & 0 & 0\\
0 & 0 &-\frac{1}{2\sqrt3} & \frac{1}{2\sqrt2} \\
0 & 0 & 0 & \frac{1}{2\sqrt2} \\
\ep\bp
\CG^{10}\\
\CG^{12}\\
\CG^{01}\\
\CG^{11}\\
\ep\ .
\en
The coefficients of the expansions of the physical amplitudes in terms
of these new isospin amplitudes are given in
table~\ref{table:newisoexpress}. One notes, in particular, that the
expansions of the $D^+$ amplitudes (first three lines in the tables)
now involve only two isospin amplitudes instead of four.

\subsection{Crossing symmetries}\label{sec:crossingsym}
The amplitudes $\CA_i$, $\tilde{\CA}_i$ obey a number of crossing
symmetry relations. Under $s-t$ crossing one has the 8 relations
\be\lbl{stcross}
\tilde{\CA}_i(s,t,u)=\CA_i(t,s,u)
\en
and under $s-u$ crossing one has 5 relations
\be\lbl{uscross}
\ba{l}
\CA_1(s,t,u)=\CA_1(u,t,s),\\ 
\CA_2(s,t,u)=\CA_3(u,t,s),\\
\CA_4(s,t,u)=\CA_5(u,t,s),\\
\CA_6(s,t,u)=\CA_8(u,t,s),\\
\CA_7(s,t,u)=\CA_7(u,t,s)\ .\\
\ea\en
The crossing-symmetry relations~\rf{stcross}~\rf{uscross} induce
corresponding linear relations among the isospin
amplitudes. Collecting the isospin amplitudes into vectors
$\vec{\CF}(s,t,u)$ and $\vec{\CG}(s,t,u)$ (see eq.~\rf{vecFvecG}) one
has the crossing relations given in eq.~\rf{isocrossrels} in which the
crossing matrices $\bm{C_{st}}$, $\bm{C_{us}}$ read
\be\lbl{crossmat}
\bm{C_{st}}=\bp
-\frac{2\sqrt2}{3} & -\frac{\sqrt2}{6} & 2 & \frac{1}{2} \\[0.2cm]
-\sqrt2            & -\sqrt2           & 0 & 0\\[0.2cm]
-\frac{5}{2}       &\frac{1}{2}        & 0 & 0\\[0.2cm]
-\frac{1}{2}       & 0                 &-\frac{\sqrt2}{2} & \frac{\sqrt2}{4}\\
\ep
\ ,\quad
\bm{C_{us}}=\bp
-\frac{2}{3}  & \frac{1}{3} & 0 & 0 \\[0.2cm]
\frac{5}{3}   & \frac{2}{3} & 0 & 0 \\[0.2cm]
-\frac{\sqrt2}{3} & 0 & \frac{1}{3} & \frac{1}{3} \\[0.2cm]
-\frac{\sqrt2}{3} & \frac{\sqrt2}{3}&\frac{8}{3}&-\frac{1}{3}\\ \ .
\ep\ .
\en
We note that they satisfy the following properties
\be\lbl{CusCstprop}
\bm{C_{us}}^2=\bm{1}\ ,\quad  \bm{C_{st}}\bm{C_{us}} \bm{C_{st}}^{-1}=
 diag(1,1,-1,-1) \ .
\en

With the ingredients presented above we can  write the representation of
the isospin amplitudes in terms of single-variable functions which derive from 
the reconstruction theorem~\cite{Stern:1993rg}. Including  $j=0,1$
partial-waves,  we can collect the relevant one-variable functions into vectors
$\vec{F}_0$, $\vec{F}_1$, $\vec{G}_0$, $\vec{G}_1$ as given
in eq.~\rf{onevarvectors}. 
Imposing the crossing-symmetry relations one finds that the representation
with single-variable functions must have the form given in eq.~\rf{isosvrep1}.
Using these relations together with table~\ref{table:newisoexpress} one
obtains the expressions of the physical amplitudes in terms of single-variable
isospin amplitudes.

\section{Angular integrations}\label{sec:angularint}

\subsection{Hat-functions}\label{sec:Hat-functions}
We give below the expressions of the hat-functions in terms of
angular integrals. For the $z_s$ integrals we use the
notation~\cite{Anisovich:1996tx} 
\be\lbl{zsintegr}
\ba{l}
\braque{z_s^n \Phi(u)}_s\equiv\frac{1}{2}\displaystyle\int_{-1}^1 dz_s\,
z_s^n\, \Phi(u(s,z_s))\\[0.3cm]
\braque{z_s^n \Phi(t)}_s\equiv\frac{1}{2}\displaystyle\int_{-1}^1 dz_s\,
z_s^n\, \Phi(t(s,z_s))\ .
\ea\en
The hat-functions $\widehat{F}_j^K$ have the following expressions
\be\lbl{Fhatfunc}
\ba{ll}
\widehat{F}_0^\trdemi(s)= &
-\frac{2}{3}\,\braques{F^\trdemi_0(u)}
+\frac{1}{3}\,\braques{F^\undemi_0(u)}
-\frac{2}{3}\,\braques{\Zu\,F^\trdemi_1(u)}
\\[0.2cm] &
+\frac{1}{3}\,\braques{\Zu\,F^\undemi_1(u)}
-\frac{\sqrt2}{12}\,\,\braques{G_0^2(t)}
-\frac{1}{3}\,\braques{\su\,G_1^1(t)}
\\[0.3cm]
\widehat{F}_0^\undemi(s)= &
\frac{5}{3}\,\braques{F^\trdemi_0(u)}
+\frac{2}{3}\,\braques{F^\undemi_0(u)}+\frac{5}{3}\,\braques{\Zu\,F^\trdemi_1(u)}
\\[0.2cm] &
+\frac{2}{3}\,\braques{\Zu\,F^\undemi_1(u)}
-\frac{5\sqrt2}{12}\,\,\braques{G_0^2(t)}
+\frac{1}{3}\,\braques{\su\,G_1^1(t)}
\\[0.3cm]
\widehat{F}_1^\trdemi(s)= & \dfrac{1}{s\kappa_s(s)}\Big[
-2\,\braques{z_s\,F^\trdemi_0(u)}
+\braques{z_s\,F^\undemi_0(u)}
-2\,\braques{z_s\,\Zu\,F^\trdemi_1(u)}
\\[0.1cm] &
+\braques{z_s\,\Zu\,F^\undemi_1(u)}
-\frac{1}{4}\,\sqrt2\,\braques{z_s\,G_0^2(t)}
-\braques{z_s\,\su\,G_1^1(t)} \Big]
\\[0.3cm]
\widehat{F}_1^\undemi(s)= & \dfrac{1}{s\kappa_s(s)}\Big[
 5\,\braques{z_s\,F^\trdemi_0(u)}
+2\,\braques{z_s\,F^\undemi_0(u)}
+5\,\braques{z_s\,\Zu\,F^\trdemi_1(u)}
\\[0.1cm] &
+2\,\braques{z_s\,\Zu\,F^\undemi_1(u)}
-\frac{5}{4}\,\sqrt2\,\braques{z_s\,G_0^2(t)}
+\braques{z_s\,\su\,G_1^1(t)} \Big]\ .
\ea
\en
Where $\zu\equiv -u(u+2s-\Sigma)+\Delta$, $Z_t\equiv
t+2s-\Sigma$. Next, the hat-functions $\widehat{H}_j^K$ read
\be\lbl{Hhatfunc}
\ba{ll}
\widehat{H}_0^\trdemi(s)= &
-\frac{\sqrt2}{3}\,\,\braques{F^\trdemi_0(u)}
+\frac{1}{3}\,\braques{H^\trdemi_0(u)}
+\frac{1}{3}\,\braques{H^\undemi_0(u)}
\\[0.2cm] &
-\frac{\sqrt2}{3}\,\,\braques{\Zu\,F^\trdemi_1(u)}
+\frac{1}{3}\,\braques{\Zu\,H^\trdemi_1(u)}
+\frac{1}{3}\,\braques{\Zu\,H^\undemi_1(u)}
\\[0.2cm] &
+\frac{1}{3}\,\braques{G_0^0(t)}-\frac{1}{18}\,\braques{G_0^2(t)}
-\frac{\sqrt2}{3}\,\,\braques{\su\,\widetilde{G}_1^1(t)}
\\[0.3cm]
\widehat{H}_0^\undemi(s)= &
-\frac{\sqrt2}{3}\,\,\braques{F^\trdemi_0(u)}
+\frac{\sqrt2}{3}\,\,\braques{F^\undemi_0(u)}+\frac{8}{3}\,\braques{H^\trdemi_0(u)}
-\frac{1}{3}\,\braques{H^\undemi_0(u)}
\\[0.2cm] &
-\frac{\sqrt2}{3}\,\,\braques{\Zu\,F^\trdemi_1(u)}
+\frac{\sqrt2}{3}\,\,\braques{\Zu\,F^\undemi_1(u)}
\\[0.2cm] &
+\frac{8}{3}\,\braques{\Zu\,H^\trdemi_1(u)}
-\frac{1}{3}\,\braques{\Zu\,H^\undemi_1(u)}
\\[0.2cm] &
+\frac{2}{3}\,\braques{G_0^0(t)}-\frac{5}{18}\,\braques{G_0^2(t)}
-\frac{\sqrt2}{3}\,\,\braques{\su\,G_1^1(t)}
+\frac{4\sqrt2}{3}\,\,\braques{\su\,\widetilde{G}_1^1(t)}\\[0.2cm]
\widehat{H}_1^\trdemi(s)= & \dfrac{1}{s\kappa_s(s)}\Big[
-\sqrt2\,\braques{z_s\,F^\trdemi_0(u)}
+\braques{z_s\,H^\trdemi_0(u)}
+\braques{z_s\,H^\undemi_0(u)}
\\[0.1cm] &
-\sqrt2\,\braques{z_s\,\zu\,F^\trdemi_1(u)}
+\braques{z_s\,\zu\,H^\trdemi_1(u)}
+\braques{z_s\,\zu\,H^\undemi_1(u)}
\\[0.1cm] &
+\braques{z_s\,G_0^0(t)}
-\frac{1}{6}\,\braques{z_s\,G_0^2(t)}
-\sqrt2\,\braques{z_s\,\su\,\widetilde{G}_1^1(t)} \Big]
\\[0.3cm]
\widehat{H}_1^\undemi(s)= & \dfrac{1}{s\kappa_s(s)}\Big[
-\sqrt2\,\braques{z_s\,F^\trdemi_0(u)}
+\sqrt2\,\braques{z_s\,F^\undemi_0(u)}
+8\,\braques{z_s\,H^\trdemi_0(u)}
\\[0.1cm] &
-\braques{z_s\,H^\undemi_0(u)}
-\sqrt2\,\braques{z_s\,\Zu\,F^\trdemi_1(u)}
+\sqrt2\,\braques{z_s\,\Zu\,F^\undemi_1(u)}
\\[0.1cm] &
+8\,\braques{z_s\,\Zu\,H^\trdemi_1(u)}
-\braques{z_s\,\Zu\,H^\undemi_1(u)}
+2\,\braques{z_s\,G_0^0(t)}
\\[0.1cm] &
-\frac{5}{6}\,\braques{z_s\,G_0^2(t)}
-\sqrt2\,\braques{z_s\,\su\,G_1^1(t)}
+4\,\sqrt2\,\braques{z_s\,\su\,\widetilde{G}_1^1(t)}\Big]\ .
\ea\en
Concerning the $t$-channel hat-functions, we will use the notation
\be\lbl{ztintegr}
\braque{z_t^n \Phi(s)}_t\equiv \frac{1}{2}\int_{-1}^1
dz_t\, z_t^n \Phi(s(t,z_t))\ .
\en
The analogous integrals involving $\Phi(u)$ can be replaced using the
relation, 
\be
\braque{z_t^n \Phi(u)}_t =(-1)^n \braque{z_t^n \Phi(s)}_t\ .
\en
The hat-functions $\widehat{G}_0^2$, $\widehat{G}_1^1(t)$ are given by
\be\lbl{Ghatfunc1}
\ba{ll}
\widehat{G}_0^2(t)= &
-2\,\sqrt2\,\braquet{F^\trdemi_0(s)}
-2\,\sqrt2\,\braquet{F^\undemi_0(s)}\\[0.2cm] &
-2\,\sqrt2\,\braquet{\Zs\,F^\trdemi_1(s)}
-2\,\sqrt2\,\braquet{\Zs\,F^\undemi_1(s)}\\[0.2cm] 
\widehat{G}_1^1(t)= & \dfrac{1}{\kappa_t(t)}\Big[
-15\,\braquet{z_t\,F^\trdemi_0(s)}
+3\,\braquet{z_t\,F^\undemi_0(s)}\\[0.2cm] &
-15\,\braquet{z_t\,\Zs\,F^\trdemi_1(s)}
+3\,\braquet{z_t\,\Zs\,F^\undemi_1(s)}\Big]
\\[0.2cm]
\ea
\en
where $\zs=s(s+2t-\Sigma)+\Delta$.
The hat-functions $\widehat{G}_0^0$, $\TTG$, finally, are given by 
\be\lbl{Ghatfunc2}
\ba{ll}
\widehat{G}_0^0(t)= &
-\frac{4\sqrt2}{3}\,\braquet{F^\trdemi_0(s)}
-\frac{\sqrt2}{3}\,\braquet{F^\undemi_0(s)}
+4\,\braquet{H^\trdemi_0(s)} 
+\braquet{H^\undemi_0(s)}\\[0.2cm] &
-\frac{4\sqrt2}{3}\braquet{\Zs\,F^\trdemi_1(s)}
-\frac{\sqrt2}{3}\,\braquet{\Zs\,F^\undemi_1(s)}\\[0.2cm] &
+4\,\braquet{\Zs\,H^\trdemi_1(s)}
+\braquet{\Zs\,H^\undemi_1(s)}\\[0.2cm] 
\TTG(t)= &\dfrac{1}{\kappa_t(t)}\Big[
-3\,\braquet{z_t\,F^\trdemi_0(s)}
-3\,\sqrt2\,\braquet{z_t\,H^\trdemi_0(s)}
+\frac{3}{2}\,\sqrt2\,\braquet{z_t\,H^\undemi_0(s)}\\[0.2cm] &
-3\,\braquet{z_t\,\Zs\,F^\trdemi_1(s)}
-3\,\sqrt2\,\braquet{z_t\,\Zs\,H^\trdemi_1(s)}
+\frac{3}{2}\,\sqrt2\,\braquet{z_t\,\Zs\,H^\undemi_1(s)}\Big]\ .
\ea\en

\subsection{Angular integrals expressed in terms of kernels}\label{sec:kernelang}
The one-variable functions satisfy ordinary dispersion relations. We
will use these in order to re-express the angular integrations over
$z_s$ and $z_t$. The dispersion relations are written below, taking
into account the assumed asymptotic behaviour. Firstly, one has
\begin{align}\lbl{ordinaryDRF}
F_0^\undemi(w)= & c_0 +c'_1\,w+\frac{w^2}{\pi}\int_\skpi^\infty 
\frac{\disc[F_0^\undemi(w')]}{(w')^2(w'-w)} \,dw'\nonumber\\
F_0^\trdemi(w)= & \frac{w^2}{\pi}\int_\skpi^\infty 
\frac{\disc[F_0^\trdemi(w')]}{(w')^2(w'-w)}\,dw'\nonumber\\ 
F_1^\undemi(w)= & \frac{1}{\pi}\int_\skpi^\infty 
\frac{\disc[F_1^\undemi(w')]}{(w'-w)}\,dw'\nonumber\\ 
F_1^\trdemi(w)= & \frac{1}{\pi}\int_\skpi^\infty 
\frac{\disc[F_1^\trdemi(w')]}{(w'-w)}\,dw'\\
G_0^2(t)= & \frac{t^2}{\pi}\int_\spipi^\infty 
\frac{\disc[G_0^2(t')]}{(t')^2(t'-t)} \,dt'\nonumber\\
G_1^1(t)=& c_4+\frac{t}{\pi}\int_\spipi^\infty 
\frac{\disc[G_1^1(t')]}{(t')(t'-t)}\,dt'\nonumber
\end{align}
with
\be
c'_1=c_1+ c_0\,\dot{\Omega}_0^\undemi(0)\ 
\en
and the remaining set reads
 \begin{align}\lbl{ordinaryDRH}
H_0^\undemi(w)= & d_0 +d'_1\,w+\frac{w^2}{\pi}\int_\skpi^\infty 
\frac{\disc[H_0^\undemi(w')]}{(w')^2(w'-w)} \,dw'\nonumber\\
H_0^\trdemi(w)= & \frac{w^2}{\pi}\int_\skpi^\infty 
\frac{\disc[H_0^\trdemi(w')]}{(w')^2(w'-w)}\,dw'\nonumber\\ 
H_1^\undemi(w)= & \frac{1}{\pi}\int_\skpi^\infty 
\frac{\disc[H_1^\undemi(w')]}{(w'-w)}\,dw'\nonumber\\ 
H_1^\trdemi(w)= & \frac{1}{\pi}\int_\skpi^\infty 
\frac{\disc[H_1^\trdemi(w')]}{(w'-w)}\,dw'\\
G_0^0(t)= & \frac{t^2}{\pi}\int_\spipi^\infty 
\frac{\disc[G_0^0(t')]}{(t')^2(t'-t)}\,dt' \nonumber\\
\TG(t)=& d_5+\frac{t}{\pi}\int_\spipi^\infty 
\frac{\disc[\TG(t')]}{(t')(t'-t)}\,dt'\nonumber
\end{align}
with
\be
d'_1=d_1+d_0\,\dot{\Omega}_0^\undemi(0)\ .
\en
Computing angular integrals using eqs.~\rf{ordinaryDRF}~\rf{ordinaryDRH}
one can invert the integrations order and perform the $z_s$, $z_t$
integrations first, obtaining a set of kernels.  These can be written
in analytical form and encode the contour prescriptions for performing
the angular integrals.  Representations of the angular averages are
then derived involving integrals of these kernels multiplied by
discontinuities of the one-variable functions. For the $F$-functions,
the $z_s$ angular integrals can be expressed as follows 
\begin{align}\lbl{kernint-zs}
\braques{F_0^K(u)}=&\delta_{2K,1}\big[c_0
+\frac{1}{2}c'_1 (\Sigma-s+\frac{\Delta}{s})\big]
-\frac{1}{\pi}\int_\skpi^\infty
K^0_u(s,u')\,\disc[F_0^K(u')]\,du'\nonumber\\
\braques{z_sF_0^K(u)}= & \delta_{2K,1}\big[-\frac{1}{6}c'_1 \kappa_s(s)\big]
-\frac{1}{\pi}\int_\skpi^\infty  K^1_u(s,u')
\,\disc[F_0^K(u')]\,du'\nonumber\\
\braques{\zu F_1^K(u)}=& \frac{-1}{\pi}\int_\skpi^\infty  
K^0_{\zu}(s,u')\,\disc[ F_1^K(u')]\,du'\nonumber\\
\braques{z_s\zu F_1^K(u)}=& \frac{-1}{\pi}\int_\skpi^\infty  
K^1_{\zu}(s,u')\,\disc[ F_1^K(u')]\, du'\nonumber\\
\braques{G_0^2(t)}=& \frac{-1}{\pi}\int_\spipi^\infty 
K^0_t(s,t')\,\disc[G_0^2(t')]\,dt'\nonumber\\
\braques{z_sG_0^2(t)}=& \frac{-1}{\pi}\int_\spipi^\infty 
K^1_t(s,t')\,\disc[G_0^2(t')]\,dt'\nonumber\\
\braques{Z_tG_1^1(t)}=& c_4\frac{1}{2}(3s-\Sigma-\frac{\Delta}{s})
-\frac{1}{\pi}\int_\spipi^\infty
K^0_{Z_t}(s,t')\, \disc[G_1^1(t')]\,dt'\nonumber\\
\braques{z_sZ_tG_1^1(t)}=& c_4\frac{1}{6}\kappa_s(s)
-\frac{1}{\pi}\int_\spipi^\infty
K^1_{Z_t}(s,t')\, \disc[G_1^1(t')]\,dt'\ .  
\end{align}
The representation of $z_s$ integrals of the functions $H^K_j$
and $G_0^0$ are analogous to those of $F_j^K$ and $G_0^2$
respectively, eventually replacing the polynomial parameters $c_i$ by $d_i$.
The representation for $\TG$ is analogous to that of
$G_1^1$ replacing the constant $c_4$ by $d_4$.
We express in a similar way the  $z_t$ angular integrals 
\begin{align}\lbl{kernint-zt}
\braquet{F_0^K(s)}=& \delta_{2K,1}\big[c_0+\frac{1}{2}c'_1(\Sigma-t)\big]
                          -\frac{1}{\pi}\int_\skpi^\infty K^0_s(t,s')
                        \disc[F_0^K(s')]\,ds'\nonumber\\
\braquet{z_t F_0^K(s)}=& \delta_{2K,1}\big[\frac{1}{6}c'_1\kappa_t(t)\big]
                         -\frac{1}{\pi}\int_\skpi^\infty K^1_s(t,s')
                          \disc[F_0^K(s')]\,ds'\nonumber\\
\braquet{\zs F_1^K(s)}=&     \frac{-1}{\pi}\int_\skpi^\infty K^0_{\zs}(t,s')
                       \disc[F_1^K(s')]\,ds'\nonumber\\
\braquet{z_t \zs F_1^K(s)}=& \frac{-1}{\pi}\int_\skpi^\infty K^1_{\zs}(t,s')
                       \disc[F_1^K(s')]\,ds'\ .
\end{align}
The twelve kernels which appear in eqs.~\rf{kernint-zs},~\rf{kernint-zt} can
be written as follows, after making a change of variables
\begin{align}\lbl{kernellist}
K^0_u(s,u') =& \frac{1}{\kappa_s(s)}\int_\umin^\uplus 
\frac{z^2}{(u')^2 (z-u')}dz\nonumber\\
K^1_u(s,u') =& \frac{1}{\kappa^2_s(s)}\int_\umin^\uplus 
\frac{z^2(-2z+\uplus+\umin)}{(u')^2 (z-u')}dz\nonumber\\
K^0_{\zu}(s,u')= & \frac{1}{\kappa_s(s)}\int_\umin^\uplus
\frac{-z^2+z(\Sigma-2s)+\Delta}{z-u'}\,dz\nonumber\\
K^1_{\zu}(s,u')= & \frac{1}{\kappa^2_s(s)}\int_\umin^\uplus
\frac{(-2z+\uplus+\umin)(-z^2+z(\Sigma-2s)+\Delta)}{z-u'}
\,dz\nonumber\\
K^0_t(s,t')= & \frac{1}{\kappa_s(s)}\int_\tmin^\tplus
\frac{z^2}{(t')^2(z-t')}\,dz\nonumber\\
K^1_t(s,t')= & \frac{1}{\kappa^2_s(s)}\int_\tmin^\tplus
\frac{z^2(2z-\tplus-\tmin)}{(t')^2(z-t')}\,dz\nonumber\\
K^0_{Z_t}(s,t')= &\frac{1}{\kappa_s(s)}\int_\tmin^\tplus
\frac{z(z+2s-\Sigma)}{t'(z-t')}\,dz\nonumber\\
K^1_{Z_t}(s,t')= &\frac{1}{\kappa^2_s(s)}\int_\tmin^\tplus
\frac{z(z+2s-\Sigma)(2z-\tplus-\tmin) }{t'(z-t')}\,dz\nonumber\\
K^0_s(t,s')= & \frac{1}{\kappa_t(t)} \int_\smin^\splus
              \frac{z^2}{(s')^2 (z-s')}\,dz\nonumber\\
K^1_s(t,s')= & \frac{1}{\kappa^2_t(t)}\int_\smin^\splus
              \frac{z^2(2z-\splus-\smin)}{(s')^2 (z-s')}\,dz\nonumber\\
K^0_{\zs}(t,s')= & \frac{1}{\kappa_t(t)} \int_\smin^\splus
               \frac{(z^2+z(2t-\Sigma)+\Delta)}{z-s'}\,dz\nonumber\\
K^1_{\zs}(t,s')= & \frac{1}{\kappa^2_t(t)} \int_\smin^\splus
               \frac{(z^2+z(2t-\Sigma)+\Delta)(2z-\splus-\smin)}
              {z-s'}\,dz\ 
\end{align}
where
\be
u_\pm(s)=\frac{1}{2}\big(\Sigma-s+\frac{\Delta}{s}
\pm\kappa_s(s)\Big),\quad
t_\pm(s)=\frac{1}{2}\big(\Sigma-s-\frac{\Delta}{s}
\pm\kappa_s(s)\Big),\quad
\en
and
\be
s_\pm(t)=\frac{1}{2}(\Sigma-t \pm \kappa_t(t))\ .
\en
When these integration endpoints are real and overlapping with the unitarity
cut, they must be shifted infinitesimally from the real axis by using the 
prescription $m_D^2\to m_D^2+i\epsilon$.  The twelve kernels~\rf{kernellist}
can easily be expressed in terms of three basic logarithms,
\be
L_u(s,u')=\int_{u_-(s)}^{u_+(s)} \frac{dz}{z-u'},\quad
L_t(s,t')=\int_{t_-(s)}^{t_+(s)} \frac{dz}{z-t'},\quad
\en
and
\be
L_s(t,s')=\int_{s_-(t)}^{s_+(t)}\frac{dz}{z-s'}
\en
which can be written in the following simple way in terms of the usual complex
log function
\be
\ba{l}
L_u(s,u')=\log(u'-u_+(s))-\log(u'-u_-(s))\\[0.1cm] 
L_t(s,t')=\log(t'-t_+(s))-\log(t'-t_-(s))\\[0.1cm] 
L_s(t,s')=\log(s'-s_+(t))-\log(s'-s_-(t))\ .
\ea\en
The angular averages have singularities on the real axis when the
variable $s$ approaches $(m_D-m_\pi)^2$ (or the variable $t$
approaches $(m_D-m_K)^2$) (e.g.~\cite{Kambor:1995yc}). Let us illustrate
how these appear via the log functions. Consider, for instance, the kernel
$K^0_u(s,u')$ (see~\rf{kernellist}) whose expression is
\be\lbl{K0u}
K^0_u(s,u')=\frac{u'+u_+(s)+u_-(s)}{(u')^2} +\frac{L_u(s,u')}{\kappa_s(s)}\ .
\en
When $s\to (m_D-m_\pi)^2$ the function $\kappa_s(s)$ in the denominator
vanishes while in the numerator $L_u(s,u')$ remains finite for a range of
values of $u'$ 
\be
(m_K+m_\pi)^2\le u'\le u_\pm( (m_D-m_\pi)^2)\ :\ L_u(s,u')\vert_{s\to (m_D-m_\pi)^2}=-2i\pi
\ .
\en
This is because the integration boundaries $u_\pm(s)$ have
equal real parts but opposite infinitesimal imaginary parts when $s\to
(m_D-m_\pi)^2$. The divergence which is induced, for instance, in the angular
average $\braque{F_0^K(u)}_s$ reads
\be
\left.\braque{F_0^K(u)}_s\right\vert^{div}_{s\to (m_D-\mpi)^2} =
\frac{A}{ \sqrt{ ((\md-\mpi)^2-s)}}
\int_{(\mk+\mpi)^2}^{u_+((\md-\mpi)^2)}  \disc[F_0^K(u')] \,du'\ 
\en
with $A= (\md-\mpi)^2/\sqrt{\md\mpi\lambda_{K\pi}((\md-\mpi)^2)}$.

\subsection{Breit-Wigner functions}\label{sec:BreitWigner}
We have used the following form for the Breit-Wigner functions used for
describing some of the resonance contributions with $J=1$ or $J=2$
\be\lbl{breit-wigner}
BW_J(w)=\frac{1}{m_r^2 -w -i m_r \Gamma_r^J(w,\mu_1,\mu_2)} F_r^J(w,m_1,m_2)
\en
where $m_r$ is the mass of the resonance. The energy dependence of the
width assumes the dominance of a two-particle decay mode with masses
$\mu_1,\mu_2$, 
\be\lbl{edepwidth}
\Gamma_r^J(w,\mu_1,\mu_2)=\Gamma_r\times\left(\frac{q(w,\mu_1,\mu_2)}
{q(m_r^2,\mu_1,\mu_2)}\right)^{2J+1} \frac{m_r}{\sqrt{w}}
(F_r^J(w,\mu_1,\mu_2))^2
\en
where $q$ is the centre-of-mass momentum
\be
q(w,\mu_1,\mu_2)=\sqrt{\frac{(w-(\mu_1-\mu_2)^2)(w-(\mu_1+\mu_2)^2)}{4w} }
\en
and $F_r^J$ is a Blatt-Weisskopf penetration
factor~\cite{Blatt:1952ije},
\be
F_r^J(w,\mu_1,\mu_2)=\sqrt{\frac{1+X_r^2}{1+X^2}},\quad
F_r^2(w,\mu_1,\mu_2)=\sqrt{\frac{9+3X_r^2+X_r^4}{9+3X^2+X^4}}
\en
with $X=q(w,\mu_1,\mu_2)R$, $X_r=q(m^2_r,\mu_1,\mu_2)R$ and the value
of the radius is taken as $R=1.5$ $\hbox{GeV}^{-1}$
(following~\cite{E791:2002xlc}). A Blatt-Weisskopf factor is also
included in eq.~\rf{breit-wigner} corresponding to the vertex between
the resonance and the two light pseudoscalar mesons. The values of
the masses $\mu_1,\mu_2$ in eq.~\rf{edepwidth} are taken as
$\mu_1=\mpi$ and $\mu_2=\mpi$ ($=m_K$) for a non-strange (strange)
resonance except for the $K^*(1680)$ for which we choose
$\mu_2=m_{K^*(892)}$. In the case of $\omega(782)$ the width was taken
as constant. The values of the masses and widths parameters $m_r$,
$\Gamma_r$ are collected in table~\ref{table:resparams}.
\begin{table}
\centering  
\bt{c|ccccc}\hline \hline 
\ & $K^*(892)$ & $K^*(1680)$ & $K^*_2(1430)$ & $\omega(782)$ &
$f_2(1270)$ \\ \hline
$m_r$ (MeV)      & 895.55 & 1718.0 & 1432.4 & 782.66 & 1275.5 \\
$\Gamma_r$ (MeV) & 47.33  &  422.0 &  109.0 &  8.68  & 186.7 \\  \hline \hline 
\et
\caption{\small  Masses and widths parameters used in the Breit-Wigner
  functions. They correspond to the central values in the PDG except
  for the $K^*(1680)$ width which is taken at the upper end of the
  allowed values.}
\label{table:resparams}
\end{table}


\providecommand{\href}[2]{#2}\begingroup\raggedright
\endgroup

\end{document}